\documentclass[10pt,journal,compsoc,x11names]{IEEEtran}
\usepackage[switch]{lineno}
\usepackage{amsfonts}
\usepackage{amsmath, bm}
\usepackage{xspace}
\usepackage{enumitem}
\usepackage{amssymb}
\usepackage{booktabs}
\usepackage[hyphens]{url}

\usepackage{ragged2e}
\usepackage{hyperref}
\usepackage{multirow}
\usepackage{graphicx}
\usepackage{xcolor}
\usepackage{color}
\usepackage{colortbl}
\usepackage{tablefootnote}
\usepackage{pifont}
\usepackage{makecell}
\usepackage[most]{tcolorbox}
\usepackage{framed}
\usepackage{mdframed}
\usepackage{subfigure}
\usepackage{caption}
\usepackage{longtable}
\usepackage{float}
\usepackage{booktabs}
\newcommand{\vpara}[1]{\vspace{0.07in}\noindent\textbf{#1}\xspace}

\newcolumntype{H}{>{\setbox0=\hbox\bgroup}c<{\egroup}@{}}

\newcommand{\ignore}[1]{}

\definecolor{gold}{RGB}{205,133,63}
\definecolor{fGreen}{RGB}{34,139,34}
\definecolor{tOrange}{RGB}{255,215,0}
\definecolor{tBlue}{RGB}{135,206,250}
\definecolor{tPink}{RGB}{255,204,204}
\definecolor{tGreen}{RGB}{205,230,199}
\definecolor{tGold}{RGB}{255,215,0}

\usepackage{cite}
\usepackage[numbers,sort&compress]{natbib}

\ifCLASSINFOpdf
\else
\fi

\usepackage{amsmath}
\usepackage{amssymb}  
\usepackage{mathtools}
\usepackage{utils/math_commands}

\usepackage{xcolor}
\usepackage{times}
\usepackage{ragged2e}
\usepackage{bm}
\usepackage{CJKutf8}
\usepackage{latexsym}
\usepackage{subfigure}
\usepackage{graphicx}
\usepackage{float}
\usepackage{longtable}
\usepackage{booktabs} 
\usepackage{multirow}

\usepackage[ruled,vlined]{algorithm2e}
\usepackage{color, soul}

\usepackage{fancyhdr}
\usepackage[T1]{fontenc}

\usepackage[utf8]{inputenc}

\begin{document}
\title{AI-driven inverse design of materials: Past, present and future}

\author{Xiao-Qi Han, Xin-De Wang, Meng-Yuan Xu, Zhen Feng, Bo-Wen Yao, Peng-Jie Guo, Ze-Feng Gao*, Zhong-Yi Lu*
\IEEEcompsocitemizethanks{ 
\IEEEcompsocthanksitem * Corresponding author: Ze-Feng Gao and Zhong-Yi Lu
% \IEEEcompsocthanksitem GitHub link: \url{https://github.com/xxx} 
\IEEEcompsocthanksitem 
The authors are mainly with School of Physics, Renmin University of China, Beijing, China; Meng-Yuan Xu is with School of Physics and Information, Shaanxi Normal University, Xi’an, Shaanxi, China.
\IEEEcompsocthanksitem Contact e-mail: zfgao@ruc.edu.cn; zlu@ruc.edu.cn
}
}

\markboth{}%
{Shell \MakeLowercase{\textit{et al.}}: Bare Advanced Demo of IEEEtran.cls for IEEE Computer Society Journals}

\IEEEtitleabstractindextext{%
\begin{abstract}
\justifying
The discovery of advanced materials is the cornerstone of human technological development and progress. The structures of materials and their corresponding properties are essentially the result of a complex interplay of multiple degrees of freedom such as lattice, charge, spin, symmetry, and topology. This poses significant challenges for the inverse design methods of materials. Humans have long explored new materials through a large number of experiments and proposed corresponding theoretical systems to predict new material properties and structures. With the improvement of computational power, researchers have gradually developed various electronic structure calculation methods, such as the density functional theory and high-throughput computational methods. Recently, the rapid development of artificial intelligence technology in the field of computer science has enabled the effective characterization of the implicit association between material properties and structures, thus opening up an efficient paradigm for the inverse design of functional materials. A significant progress has been made in inverse design of materials based on generative and discriminative models, attracting widespread attention from researchers. Considering this rapid technological progress, in this survey, we look back on the latest advancements in AI-driven inverse design of materials by introducing the background, key findings, and mainstream technological development routes. In addition, we summarize the remaining issues for future directions. This survey provides the latest overview of AI-driven inverse design of materials, which can serve as a useful resource for researchers.
\end{abstract}

}

\maketitle

\IEEEdisplaynontitleabstractindextext

\IEEEpeerreviewmaketitle

\section{Introduction}
Advanced materials form a cornerstone of our modern information society, acting as a key catalyst for technological progress and industrial expansion, advancing at an unprecedented rate~\cite{zunger2018inverse,lee2023machine,wang2022inverse,long2024generative,wu2024physics,chen2021generative}. Their utilization extends across diverse industries such as aerospace, biomedical engineering, energy storage, and information technology, which are all poised to benefit significantly from the integration of innovative materials that can surmount existing constraints~\cite{abu2023inverse}. Undoubtedly, the evolution of novel materials is crucial for fostering technological breakthroughs, invigorating economic prosperity, and elevating the standard of living.

Generally, \emph{materials science} is a major scientific discipline dedicated to the study of advanced functional materials. The search for advanced materials through inverse design of materials is an important research field within materials science. Inverse design of materials essentially involves creating an optimization space based on the desired performance attributes of materials. This process strives to establish a high-dimensional, nonlinear mapping from material properties to structural configurations, while adhering to physical constraints. The development of inverse design of materials has garnered widespread attention in academic works and can be categorized into four major paradigms:

$\bullet$ \emph{Experiment-driven paradigm.} The experimental-driven paradigm is the original method of material discovery, and these methods have played a crucial role in propelling the field of materials science forward. For example, Madame Curie used experiments to discover the new elements radium and polonium, Onnes discovered the superconductivity phenomenon in mercury~\cite{onnes1911further}, and the discovery of the high-temperature superconducting material MgB$_2$~\cite{nagamatsu2001superconductivity}. However, this experimental-driven paradigm heavily relies on trial-and-error experimentation, individual expertise, and phenomenological scientific theories. Moreover, these methods are characterized by iterative cycles of experiments and observations to determine the properties and behaviors of materials, which is not only time-consuming and resource-intensive but also leads to extended research cycles and increased costs. In materials research, a heavy dependence on personal experience is also common. Although experienced researchers can guide experimental design with their intuition and prior knowledge, this method is limited in terms of reproducibility and scalability. The quantification and transfer of personal experience often face challenges, hindering the process of knowledge accumulation and dissemination~\cite{kim2020inverse}. Additionally, personal biases and misunderstandings may arise due to individual backgrounds and cognitive limitations~\cite{lee2023machine}.

$\bullet$ \emph{Theory-driven paradigm.} The theory-driven paradigm emphasizes the key role of theoretical insights and computational models in materials science. This paradigm is characterized by the widespread use of molecular dynamics simulations and thermodynamic models to understand and predict material behavior. These developments have, to a certain extent, simplified material research and enhanced the efficiency of investigations into new materials. There are some very famous materials and states of matter that were first predicted by theory and then verified experimentally, and these discoveries have extremely high scientific significance, even winning Nobel Prizes. In 1930, British physicist Paul Dirac derived the existence of ``antimatter'' from his quantum mechanical equations, suggesting that for every particle there is a corresponding antiparticle~\cite{vidmar2011dirac}. Dirac's equations first predicted the existence of the positron, the antiparticle of the electron. In 1932, Carl Anderson discovered the positron in cosmic rays, confirming Dirac's prediction~\cite{anderson1932apparent}. Anderson was awarded the Nobel Prize in Physics in 1936 for this discovery, and Dirac also received the Nobel Prize in Physics in 1933 for his contributions to quantum mechanics, including the theoretical prediction of antimatter. John Bardeen, Leon Cooper, and John Schrieffer proposed the BCS theory, which explained the cause of superconductivity—the pairing of electrons to form Cooper pairs, leading to zero electrical resistance~\cite{bardeen1957theory,bardeen1955theory,cooper1956bound}. Although the BCS theory itself did not directly predict new materials, it became the foundation for research into low-temperature superconductivity. Bardeen, Cooper, and Schrieffer were awarded the Nobel Prize in Physics in 1972 for their work. In 2005, Charles Kane and Eugene Mele predicted the existence of the quantum spin Hall effect through topological quantum theory, a phenomenon where edge state conduction occurs without an external magnetic field. This concept laid the foundation for topological insulators. In 2007, scientists first realized the quantum spin Hall effect in HgTe quantum wells, verifying the accuracy of the theory. These classic predictions demonstrate the powerful predictive ability of theoretical physics in the exploration of new materials and phenomena, and also promote the development of experimental physics. Many such theoretical breakthroughs ultimately won the Nobel Prize for experimental verification, marking significant advances in physics. Nevertheless, these theoretical frameworks often require complex mathematical models that are demanding in terms of computational resources and expertise. Their applicability may be limited, especially when it comes to material systems that exhibit multi-scale phenomena and complex interactions.

$\bullet$ \emph{Computation-driven paradigm.} Established on the groundwork of theoretical progress, the computation-driven paradigm has risen alongside the surge in computational power and data accessibility. This paradigm leverages computational models to simulate material behaviors and inform the design process. The application of density functional theory (DFT)~\cite{DFT-1,DFT-2,DFT-3} and computational chemistry tools like Hartree-Fock theory has revolutionized our ability to predict and optimize material properties. The DFT has played a crucial role in investigating the electronic structure of graphene. It was through the DFT that the zero-band gap structure of graphene was uncovered, a feature that is essential for its electronic properties and holds significant potential for its application in electronic devices~\cite{novoselov2004electric}. Furthermore, the Hartree-Fock method has been extensively applied in calculating molecular orbitals, such as in studies of water molecules, offering vital insights into their geometric and electronic characteristics~\cite{pople1976theoretical}. The potency of these methods is clear in their capacity to manage complex systems that were once difficult to dissect. However, reliance on computational models introduces challenges, as the accuracy of predictions heavily relies on model quality and available computational resources. Additionally, high-throughput screening (HTP) and combinatorial screening have become significant methodologies for exploring new materials and systems. This approach greatly expedites the discovery and development of novel materials, particularly in drug discovery, catalyst design, and energy storage material development, where it allows for the parallel assessment of vast compound libraries to identify potential new materials~\cite{wang2022inverse}. Despite HTP's achievements in materials science, challenges remain, including substantial resource requirements for physical or computational experiments, constraints by existing material libraries, and insufficient consideration of intricate relationships between material properties~\cite{park2024has}.

\begin{figure*}[t]
		\centering  
		\includegraphics[width=1.0\linewidth]{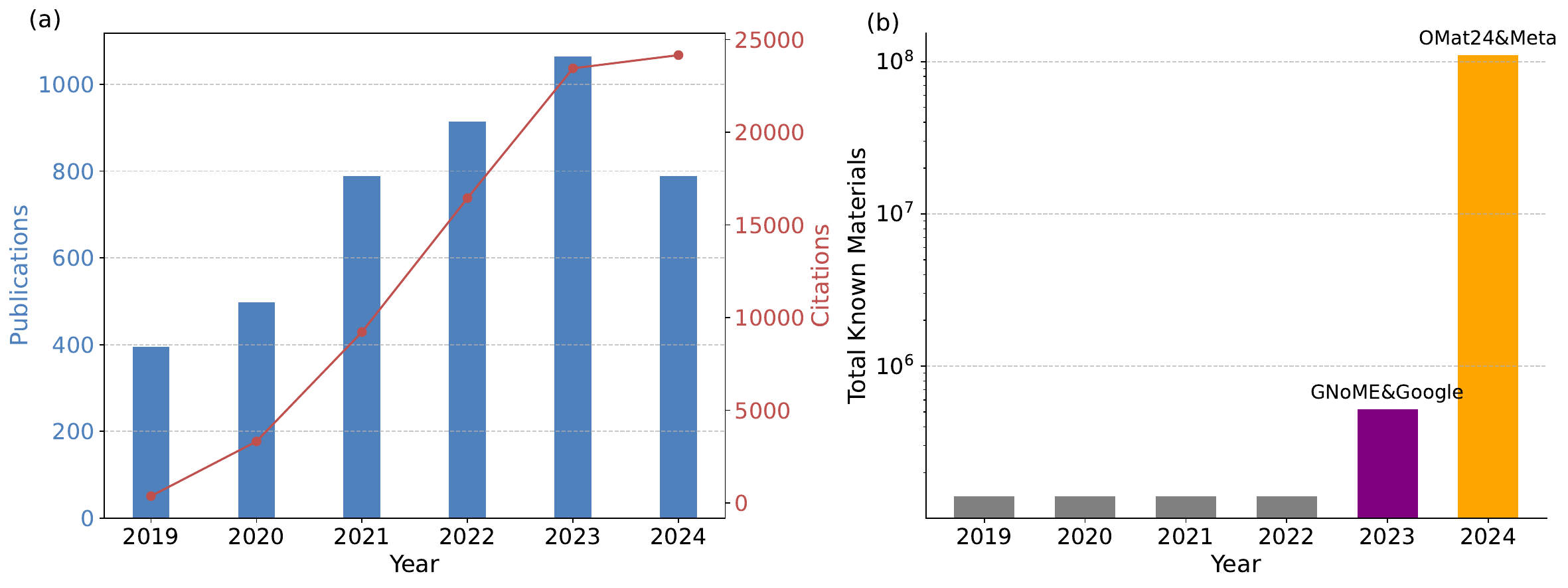}
		\caption{Trends in publications and citations in the field of ``Machine Learning Materials Discovery'' from 2019 to 2024. The left panel (a) illustrates the growth in the number of publications (represented by blue bars) alongside the total citations (depicted by the red line with markers), reflecting a significant increase in both metrics over the past few years. The right panel (b) presents the variation in the total known materials over time on a logarithmic scale, highlighting the acceleration of material discovery processes facilitated by GNoME~\cite{merchant2023scaling} of Google and OMat24~\cite{OMat24} of Meta. This figure underscores the rapid development within the field of machine learning materials discovery}
		\label{fig:pubcit-1} 
\end{figure*}

$\bullet$ \emph{AI-driven paradigm.} With the dawn of the big data era, materials science has transitioned into an AI-driven paradigm. Artificial Intelligence (AI) marks a significant shift within the engineering field, heralding an era defined by enhanced intelligence and automation. Both the widely utilized auto-regressive models based on the Transformer architecture~\cite{vaswani2017attention} and the robust diffusion models~\cite{CDVAE2022,jiao2024crystal} have made substantial contributions to the advancement of inverse design of materials~\cite{chen2024mattergpt,choudhary2024atomgpt,fernandez2024denoising,lyu2024microstructure}. The capacity of AI to discern patterns and formulate predictions from data has sparked notable transformations in materials science. By examining vast experimental data and computational simulation outcomes, AI models reveal intricate correlations between material properties and their underlying crystal structures~\cite{gao2023ai,ansari2024dziner}. Furthermore, a recent survey~\cite{chen2024mattergpt} has integrated inverse design methodologies with machine learning models to forecast the mechanical properties of materials, including elastic modulus and yield strength, thus expediting the discovery and development of innovative materials. These data-driven strategies not only bolster the precision of predictions but also considerably shorten material development cycles. Consequently, ``machine learning materials discovery'' is garnering escalating research interest (See Figure~\ref{fig:pubcit-1} (a)). In parallel, the exponential growth of the total known materials over time highlights the accelerated pace of material discovery processes, with significant contributions from Google's GNoME~\cite{merchant2023scaling} and Meta's OMat24~\cite{OMat24}.

\begin{figure*}[h]
		\centering  
		\includegraphics[width=1.0\linewidth]{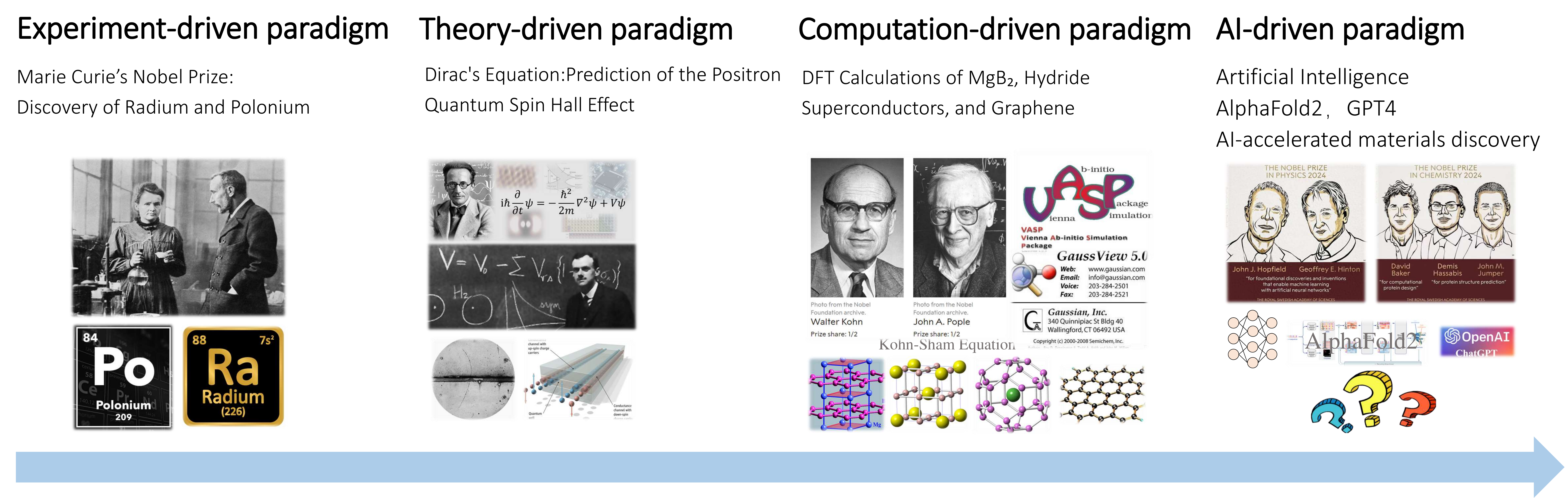}
		\caption{Materials Science Research Paradigms
This figure illustrates the evolution of research paradigms in materials science, emphasizing key milestones across various approaches. It highlights the increasing role of artificial intelligence in driving future materials discovery, with AI becoming the dominant force in shaping the field. The experiment-driven paradigm is exemplified by Marie Curie’s Nobel Prize-winning discovery of radium and polonium. The theory-driven paradigm is represented by Dirac's equation, which predicted the existence of the positron and the quantum spin Hall effect. The computation-driven paradigm is demonstrated through DFT calculations applied to materials such as MgB$_2$, hydride superconductors, and graphene. Finally, the AI-driven paradigm showcases recent breakthroughs in artificial intelligence, including AlphaFold2, GPT-4, and AI-accelerated materials discovery, signaling the frontier of research in the field.}
		\label{fig:research-paradigms} 
\end{figure*}

As discussed before, AI-driven methods are not a new technical concept for inverse design of functional materials, but has evolved with the advance of AI over the decades. The experiment-driven and theory-driven paradigms mainly aim to discover new functional materials based on heavily trial-and-error experiments or theoretical models, while latest AI-driven methods concentrate on constructing the hidden mappings between material functions and crystal structures.  From conventional experiment methods to data-driven methods, it is an important leap in using modern advanced computational methods to design target functional materials. In Figure~\ref{fig:research-paradigms}, we delineate the evolution of materials science discoveries alongside the progression of time and technological advancements. Initially, the confirmation of new materials was primarily achieved through experimentation (such as Madame Curie's discovery of new elements), which entailed substantial costs and numerous trials. Subsequently, theoretical paradigms were introduced, predicting physical properties that were later experimentally verified (for instance, the prediction and validation of semiconductors). With the advancement of computer technology and the enhancement of computational ability, high-throughput computational methods have become a significant avenue for the discovery of new materials. Ultimately, as the latest generation of technology, AI methods can efficiently generate and screen new functional materials by elucidating the hidden correlations between crystal structures and properties, thereby accelerating the discovery of new materials. In summary, throughout this evolutionary process, technological progress has enabled us to employ more sophisticated methods to expedite the discovery of new functional materials.

With the development of technology, a research field has emerged in materials science known as the inverse design of materials. Unlike traditional trial-and-error methods, it utilizes complex computational techniques to design materials with specific properties from the ground up. The process involves several key steps: first, defining the target properties or functionalities; then, selecting an appropriate design space; followed by using modeling and simulation tools (such as DFT and finite element analysis) to simulate material properties; optimizing the material structure through algorithms to minimize the difference between simulated properties and target goals; finally, validating the design through experimental verification, followed by iterative refinement based on experimental results. This approach is widely applied in areas such as optics, electronics, energy storage, catalysis, and composite materials, significantly accelerating the development of new materials. However, inverse design also faces challenges such as computational cost, data quality, and experimental validation. With the advancement of computational power and optimization algorithms, materials inverse design is expected to play an increasingly important role in the future development of materials science.

In the existing literature, the inverse design of materials has been extensively discussed and surveyed~\cite{lee2023machine,wang2022inverse,long2024generative,wu2024physics,chen2021generative}. However, current surveys often focus on specific machine learning algorithms or particular types of materials, leading to a lack of systematic integration of diverse methodologies and a broad range of material systems. Many existing surveys predominantly concentrate on specific categories of materials or application scenarios, such as topological insulators or high-entropy alloys~\cite{chen2022inverse,liu2024machine}, thereby failing to provide a comprehensive examination of AI-driven inverse design of materials. Specifically, there is a significant absence of comparative and analytical studies across different researches about AI-driven inverse design of materials, which hinders a holistic understanding of the advantages, disadvantages, and applicable contexts of various methodologies. These limitations pose challenges for readers attempting to fully grasp the overarching landscape of AI-driven inverse design of materials, particularly with respect to the dynamic evolution of new methods and their applications.

Faced with both opportunities and challenges, it needs more attention on the research and development of AI-driven inverse design of materials. In order to provide a basic understanding of this research filed, this survey aims to address the gaps in the current body of research by comprehensively examining previous studies from two perspectives: the discovery of functional materials based on AI methods and the development of AI methods in materials science. We systematically analyze the latest advancements in AI technologies within the domain of inverse design of materials. We conduct an exhaustive survey of the literature to synthesize the pivotal discoveries, AI methods, and procedural methods in the inverse design of functional materials. We are also aware of several relative survey articles on inverse design of materials~\cite{han2024surveygeometricgraphneural,reiser2022graph}. The reference~\cite{reiser2022graph} provides an overview of the importance and utilization of graph neural networks (GNNs) within the realms of chemistry and materials science. GNNs can directly process the graphical representations of molecules and materials, thereby capturing the essential information required to characterize these materials. The review further outlines the fundamental principles of GNNs, including commonly used datasets and architectures, and emphasizes their critical role throughout the materials development. Another reference~\cite{han2024surveygeometricgraphneural} systematically explores the research advancements of geometric GNNs in the applications of materials discovery and drug discovery. It emphasizes the significance of geometric graphs in scientific fields, particularly their ability to capture geometric features such as the three dimensional coordinates of nodes. The survey further defines geometric graphs and compares them with traditional graphs, elucidating their distinct characteristics. Additionally, it summarizes existing models, including invariant GNNs and equivariant GNNs. Our survey presents the latest advancements in functional materials accelerated by machine learning techniques, with particular emphasis on the robust representational capabilities of geometric GNNs. Furthermore, we provide a comprehensive overview of advanced generative models and large language models, highlighting their pivotal roles in driving materials discovery. In the end, we have gathered standard datasets and benchmarks that are crucial for material discovery, in order to advance the methods of AI-driven inverse design of materials.

``\emph{All great achievements take time.}''--The inverse design of materials has undergone a long development to achieve the latest successes. Our goal is to delve into the historical progression of material directional design, with a particular focus on the innovative approaches to material discovery that have arisen in the era of artificial intelligence. This survey will shed light on the critical issues inherent in inverse design of materials. Subsequently, we discuss the specific classes of materials that are currently at the forefront of materials science~(including superconducting materials, magnetic materials and so on), and detail the AI technologies. We have also compiled a comprehensive overview of the evolution of AI technologies within the materials domain. It is our aspiration that the contributions presented in this paper will propel the field of AI-driven inverse design of materials to new heights of advancement. In what follows, we will introduce the the recent applications of AI-driven method in expediting the inverse design of functional materials in Section~\ref{sec-2}. We then elaborate on the development history of AI technology in the field of materials science in Section~\ref{sec-3}. Finally, we will briefly discuss a series of open problems and promising directions in the future in Section~\ref{sec-future} and conclude in Section~\ref{sec-conclusion}.

\section{The Discovery of Functional Materials Based on AI Methods}
\label{sec-2}
AI-accelerated discovery of functional materials has emerged as a dominant trend in materials science. As the demand for novel materials continues to grow, traditional material discovery methods often prove too slow and labor-intensive to keep up with the pace of technological innovation. Consequently, artificial intelligence—especially machine learning (ML) and deep learning (DL)—has demonstrated significant potential in advancing material discovery. AI facilitates the efficient identification of patterns in data, the prediction of material properties, and the optimization of complex material compositions, offering tools for materials science that were previously unimaginable. This chapter provides a comprehensive review of AI applications across several key categories of functional materials, including superconducting materials, magnetic materials, thermoelectric materials, carbon-based nanomaterials, two-dimensional materials, photovoltaic materials, catalytic materials, high-entropy alloys, and porous materials. These materials are crucial to the development of advanced technologies in areas such as energy storage, electronics, and environmental sustainability.
\begin{figure*}[t]
		\centering  
		\includegraphics[width=1.0\linewidth]{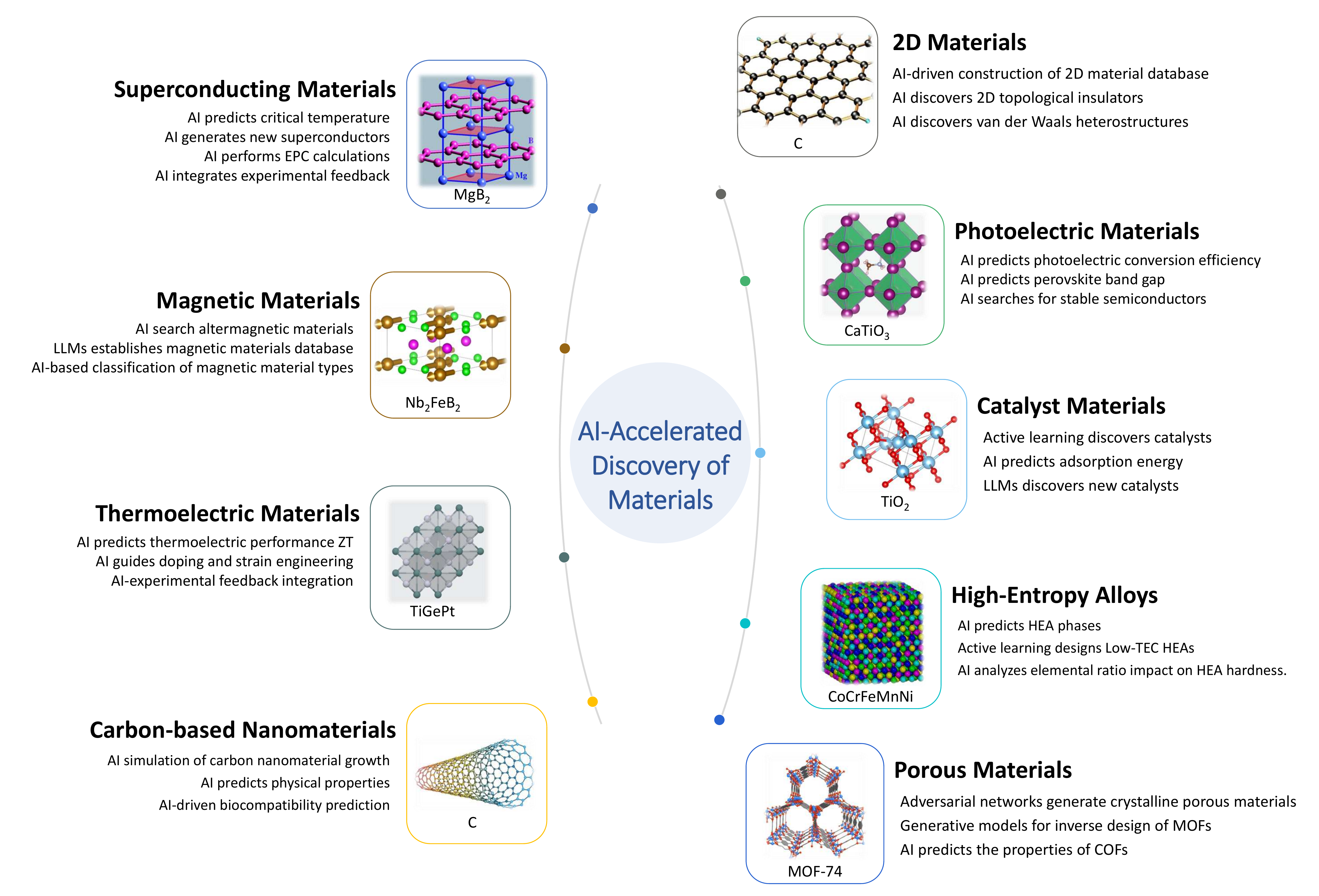}
		\caption{AI-driven discovery of materials.
This figure illustrates the role of artificial intelligence in accelerating the discovery of various types of materials. The examples presented highlight how AI-driven approaches are employed to optimize the identification, design, and property prediction of materials across diverse categories, including superconducting materials, magnetic materials, thermoelectric materials, carbon-based nanomaterials, 2D materials, photovoltaic materials, catalyst materials, high-entropy alloys, and porous materials. By leveraging large datasets and advanced computational techniques, AI methods facilitate more efficient screening and prediction, thereby significantly advancing the pace of material discovery. These examples represent only a subset of the broad potential of AI in transforming materials research, with further discussion provided in the main text.}
		\label{fig:aiacc} 
\end{figure*}

\subsection{Superconducting Materials}
Superconducting materials exhibit zero electrical resistance and complete expulsion of magnetic fields below a critical temperature~($T_c$). These materials have wide applications due to their outstanding physical properties, such as in magnetic resonance imaging~\cite{lvovsky2013novel} and nuclear fusion technology~\cite{bruzzone2018high}. Superconductor-based devices are used in quantum information processors, advanced sensors, and communication systems~\cite{mirhosseini2020superconducting,gambetta2017building,degen2017quantum}. The discovery of superconducting materials back to 1911 when mercury exhibited zero resistance at 4.2K~\cite{onnes1911resistance}. In 1933, the Meissner effect~\cite{Meissner1933} was observed, where superconductors expel magnetic fields entirely. Significant milestones include the 1973 discovery of niobium-germanium alloy~\cite{gavaler1973superconductivity} with a $T_c$ of 23.2K, and the 1986 breakthrough in copper oxide superconductors~\cite{bednorz1986possible} reaching 35K, followed by yttrium-barium-copper-oxide materials~\cite{PhysRevLett.58.908} surpassing 77K in 1987. In 2008, iron-based compounds exceeded 55K~\cite{zhi2008superconductivity}. In recent years, high-$T_c$ superconductors have advanced, with scandium setting a record for elemental superconductors at 36K~\cite{PhysRevLett.130.256002}, and pressed nickelate bilayers reaching liquid-nitrogen temperatures~\cite{Sun-nature}. Hydride superconductors, such as $\text{H}_3\text{S}$, have also been confirmed under high pressure~\cite{drozdov2015conventional}.

The search for new high-$T_c$ superconductors is a significant task in condensed matter physics. With the advancement of AI technologies, there has been substantial exploration in the field of AI-accelerated superconducting material discovery. The work in~\cite{Stanev2018} utilized data extracted from the SuperCon database~\cite{ward2016general}, comprising approximately 16,400 compounds~(without crystal structures). They employed a random forest algorithm to build machine learning models aiming at predicting the critical temperature of superconducting materials. Separate regression models were developed for different superconducting families, such as cuprates, iron-based superconductors, and low-$T_c$ superconductors. Through ablation studies, it was demonstrated that these regression models could not be generalized across different superconducting families, likely due to their distinct superconducting mechanisms. Considering the crucial role of band theory in explaining superconducting mechanisms, recent work~\cite{li2024deeplearningapproachsearch} employed electronic band structure data and applied a Transformer-based model with attention mechanisms to predict the superconducting $T_c$ of materials. Teng-dong Zhang~\emph{et al.} released the SuperBand dataset, comprising electronic band structures suitable for machine learning training, generated through high-throughput DFT calculations~\cite{zhang2024superbandelectronicbandfermisurface}.

Traditional machine learning algorithms struggle to effectively model complex crystal structures, whereas GNNs~\cite{GNN2008,GNNs2020,GCN2016,GraphSAGE2017,GAT2017,CONVgraph2016} have a natural advantage in representing such structures. Kamal Choudhary~\emph{et al.}~\cite{choudhary2022designing} leveraged the atomistic line graph neural network (ALIGNN)~\cite{alignn2021} to accelerate the discovery of superconductors. They employed ALIGNN to predict key physical properties such as Debye temperature, electronic density of states (DOS) at the Fermi level, and the critical temperature of superconducting materials. By screening materials with high Debye temperatures and high electronic densities of states at the Fermi level, they conducted electron-phonon coupling calculations on 1,058 materials, constructing a systematic Bardeen-Cooper-Schrieffer (BCS) superconductivity performance database. Using the McMillan-Allen-Dynes formula, they identified 105 dynamically stable materials with $T_c$ exceeding 5K. Recent work~\cite{bcsHalignn,zhao2024machine,li2024machine,cerqueira2024searching} has demonstrated that ALIGNN can also be used to predict the superconducting $T_c$ of hydride superconductors under \emph{varying pressures condition}, leading to the discovery of 122 dynamically stable structures with $T_c$ higher than that of MgB$_2$ (39K). With the advancement of algorithms, superconducting datasets containing crystal structures, such as 3DSC~\cite{2023_3dsc} and SuperCon3D, have been successively released.

The method of searching for high-$T_c$ superconductors based on existing databases has only touched a very small region of chemical space. In recent years, generative models~\cite{stablediffusion2022,DDPM2020,song2019generative} have gained significant attention due to their ability to produce highly realistic images, and such algorithms have also been widely applied in areas like molecular docking~\cite{alphafold3} and material generation~\cite{jiao2024crystal,CDVAE2022}. In principle, generative models can explore an infinite chemical space. Xiao-Qi Han~\emph{et al.}~\cite{xiaoqiAIsuper} developed an AI workflow for discovering high-$T_c$ superconductors. This workflow first generates new candidate superconducting crystal structures using a diffusion-based generative model. Then, a superconductivity classification model~\cite{xie2018crystal, MatAltMag} assesses the likelihood of the material being a superconductor, followed by a formation energy prediction model~\cite{chen2019graph,Zhang_2023} that evaluates the material’s stability. Subsequently, a DPA-2 model~\cite{zhang2023dpa} optimizes the structure, and finally, ALIGNN is used to predict the $T_c$ of  materials. The results  have been validated through first-principles electronic structure calculations, leading to the identification of 74 dynamically stable superconducting candidates with $T_c$ above 15K, none of which are found in existing datasets.A series of similar studies~\cite{wines2023inverse} have also employed generative models to explore superconducting materials.

When validating the stability of candidate superconducting materials, the DFT requires expensive phonon spectrum calculations. Recent work~\cite{phononprediction_2024} introduced the virtual node graph neural network (VGNN), which can directly predict $\Gamma$-point phonon spectra and full dispersion relations across the entire Brillouin zone, using only atomic coordinates as input. They also developed a $\Gamma$-point phonon database containing over 146,000 materials. Further validation of superconducting materials requires the calculation of electron-phonon coupling (EPC), and Yang Zhong~\emph{et al.}~\cite{Zhong_epc_2024} provided a machine learning approach to accelerate the computation of EPC matrices.

Most of the AI models mentioned above are primarily combined with high-throughput DFT calculations to discover new high-$T_c$ superconductors, where the superconducting mechanisms are well-understood within the framework of BCS theory. However, the superconducting mechanisms for most high-$T_c$ superconductors remain unclear, and these materials are considered unconventional superconductors. There are also empirical rules~\cite{scalingCuO2Tc,lee2012relationship,mizuguchi2010anion,peng2017influence} that describe correlations between $T_c$ and structural features, but these rules are limited to cuprates and iron-based superconductors and cannot be generalized to other materials. Given that chemical bonding and electronic interactions in the lattice are crucial factors for superconductivity, Gu Liang~\emph{et al.}~\cite{gu2024bond} developed a Bond Sensitive Graph Neural Network (BSGNN) to predict the upper limit of $T_c$ ($T_{cmax}$) in different materials. This model integrates three modules: nearest neighbor graph representation (NGR), communication message passing (CMP), and graph attention (GAT). It reveals a close relationship between $T_{cmax}$ and chemical bonding, showing that shorter bond lengths favor higher $T_{cmax}$, consistent with existing domain knowledge. However, this model can only predict $T_c$ and cannot determine whether a material is superconducting. Additionally, expert input is required for further screening to eliminate particularly unreasonable materials, such as insulators.

The above AI methods primarily focus on combining with theoretical calculations. In contrast, the work in~\cite{Pogue2023} introduces a closed-loop machine learning (Closed-loop ML) approach, which integrates machine learning with experimental feedback to accelerate the discovery of superconducting materials. Using an active learning strategy, this method iteratively selects and experimentally validates materials predicted by the machine learning model, then feeds the experimental results back into the model to continually improve its predictions. Through this approach, the research team discovered a previously unreported superconductor in the Zr-In-Ni system and rediscovered five known superconductors that were not part of the training dataset, significantly enhancing the success rate of superconductor discovery. This closed-loop method highlights the potential of combining machine learning with experimental feedback to accelerate new material discoveries and demonstrates the critical role of experimental validation in refining machine learning predictions.

\subsection{Magnetic Materials}
Magnetic materials are substances capable of generating magnetization when subjected to a magnetic field, exhibiting the property of attracting ferromagnetic elements such as iron. Dating back approximately 2,500 years, humanity had discovered that certain materials possess the ability to attract iron. For centuries, the creation of artificial magnets was confined to a singular method: friction with a natural magnet or an existing artificial magnet, the latter having been initially produced through contact with a natural source. A pivotal moment arrived in 1820 when Hans Christian Oersted uncovered the magnetic effects induced by electric currents. This discovery paved the way for the invention of the electromagnet in 1825, a breakthrough that introduced a novel means to generate significantly stronger artificial magnetic fields. Consequently, this period marked the commencement of extensive scientific inquiry into the properties and applications of magnetic materials~\cite{cullity2011introduction}.

Based on magnetic properties, magnetic materials can be categorized three types: ferromagnetic~\cite{goldman2012handbook}, antiferromagnetic~\cite{jungwirth2016antiferromagnetic}, and altermagnetic~\cite{PhysRevX.12.040501,PhysRevX.12.031042,PhysRevX.12.040002,doi:10.7566/JPSJ.88.123702,altermagnetism-2,altermagnetism-3,altermagnetism-4,jungwirth2024altermagnets,PhysRevB.103.245127}. 
The difference between ferromagnetic materials and antiferromagnetic materials lies in the arrangement of the magnetic moments within a material and the resulting magnetic behavior. Ferromagnetic materials exhibit a wide range of magnetic domains~\cite{hubert2008magnetic}, where the magnetic moments within each domain are aligned in the same direction. However, the magnetic moments in different domains may be oriented differently. Under the influence of an external magnetic field, ferromagnetic materials can be strongly magnetized, and they possess a high coercivity, meaning a strong reverse magnetic field is required to change the direction of magnetization. After the removal of the external magnetic field, ferromagnetic materials maintain a certain level of remanence and exhibit high permeability.
The altermagnetic materials, despite exhibiting antiferromagnetic compensated magnetic order, show macroscopic time-reversal symmetry-breaking phenomena and spin polarization similar to ferromagnet. This seemingly contradictory behavior has led scientists to reclassify these materials based on spin-symmetry principles, thereby defining altermagnetism as a new magnetic phase. As research on altermagnetic materials deepens, more material candidates have been discovered with these properties, including insulators, semiconductors, metals, and high-temperature superconductors~\cite{PhysRevX.12.031042}.

Moreover, altermagnetic materials have garnered significant attention due to their novel physical effects, such as giant magnetoresistance (GMR)~\cite{GMR-PRX2022}, unconventional superconductivity~\cite{SC-AM, MCM-liu2023}, tunneling magnetoresistance (TMR)~\cite{GMR-PRX2022}, piezomagnetic effects~\cite{piezomagnetism-NC}, spin-splitting torque\cite{SST-PRL2021,SST-PRL2022,SST-PRL2022-2},  time-reversal odd anomalous effects~\cite{AHE-Sinova2022,AHE-RuO2-NE2022,AHE-MnTe-PRL2023,AHE-hou2023,altermagnetism-2,MOE-Yao2021,CTHE-Yao2024}, quantum anomalous Hall effects~\cite{QAH-npj2023}, higher-order topological states~\cite{HighoT-liu2024}, altermagnetic ferroelectricity~\cite{LiFe2F6-guo2023}, and strong spin-orbit coupling effects in light element altermagnetic materials~\cite{NiF3-qu2024} . Recent studies~\cite{tan2024bipolarizedweylsemimetalsquantum} have demonstrated the theoretical feasibility of realizing bipolarized Weyl semimetals and the quantum crystal valley Hall effect in two-dimensional altermagnetic materials.

With the development of science and technology, it is very important to find new magnetic materials with a greater operating temperature range and better performance. Recently, a comprehensive, experiment-based database of magnetic materials based on large language model have been created, named the northeast materials database~(NEMAD)~\cite{itani2024northeast}. Itani~\emph{et al.} pointed out that there are several ML models that can classify magnetic materials and predict the magnetic properties of these materials, but the lack of a comprehensive database of magnetic materials leads to a lack of generalization of the models, that is, the accuracy of the models is reduced in predicting different classes of new magnetic materials. In view of this, they based an application, GPTArticleExtractor~\cite{zhang2024gptarticleextractor}, to automatically extract data from scientific articles to build a comprehensive database of material properties. Similarly, an AI search engine named MatALtMag~\cite{MatAltMag} was proposed to accelerate the discovery of altermagnetic materials. The engine first undergoes pre-training using crystal graph convolutional neural networks~(CGCNN)~\cite{xie2018crystal} and the Materials Project database~\cite{MPdata} to learn the intrinsic features of material crystal structures. It then fine-tunes the classifier using a limited number of positive samples to accurately predict intercalated magnetism. This engine has successfully discovered 50 new altermagnetic materials, including metals, semiconductors, and insulators, significantly expanding the known range of altermagnetic materials. The altermagnetism of these materials has been validated through DFT calculations, revealing novel physical effects such as the anomalous Hall effect, anomalous Kerr effect, and topological properties. Furthermore, the discovery of four $i$-wave altermagnetic materials marks a major breakthrough in the study of altermagnetic phases. The AI-driven method demonstrates great potential in accelerating the discovery of altermagnetic materials.

Due to the relatively recent emergence of altermagnetic materials, there has been limited work employing machine learning algorithms in this field, and the potential of AI has yet to be fully tapped. It is anticipated that future research will utilize generative models to explore a broader chemical space in the search for novel altermagnetic materials, while also leveraging AI algorithms to accelerate the DFT validation process.

\subsection{Thermoelectric Materials}
Thermoelectric materials represent a class of functional materials capable of directly converting heat into electricity, and vice versa, through the Seebeck effect, which generates an electric current in response to a temperature gradient~\cite{rowe2018thermoelectrics}. These materials hold significant promise for applications in waste heat recovery and solid-state cooling technologies. Currently, it is estimated that approximately two-thirds of global energy consumption is lost as waste heat, and thermoelectric materials provide a potential pathway to reclaim this wasted energy, thereby improving overall energy efficiency~\cite{tan2016rationally}. Furthermore, thermoelectric devices, which require neither mechanical components nor harmful working fluids, present an environmentally friendly and sustainable solution for various applications, including space power systems, automotive and industrial waste heat recovery, and thermal management in microelectronics~\cite{mamur2021thermoelectric}.

The performance of thermoelectric materials is generally evaluated by the dimensionless figure of merit, ZT, defined as ZT = $\dfrac{\alpha^2 \sigma T}{\kappa}$, where $\alpha$ is the Seebeck coefficient, $\sigma$ is the electrical conductivity, $T$ is the absolute temperature, and $\kappa$ is the thermal conductivity~\cite{vedernikov1998af}. The term $\alpha \sigma$, known as the power factor, is a function of both carrier concentration $(n)$ and carrier mobility $(\mu)$~\cite{zevalkink2018practical}. Achieving a high ZT value requires materials that exhibit both high electrical conductivity and a high Seebeck coefficient, while maintaining low thermal conductivity. However, optimizing ZT is challenging due to the interdependence of these parameters; for example, increasing electrical conductivity often leads to a concomitant increase in thermal conductivity, which can diminish the desired improvement in ZT~\cite{gorai2017computationally}. This trade-off remains one of the central challenges in thermoelectric material research.

Recent advancements in computational methods, particularly through the integration of machine learning and high-throughput screening techniques, have opened new avenues for addressing this challenge. By leveraging large datasets and advanced algorithms, researchers can now identify promising thermoelectric materials more efficiently and accurately than with traditional trial-and-error approaches or first-principles calculations alone. Studies have demonstrated the potential of machine learning models, such as the random forest (RF) model combined with Bayesian optimization, to predict new M2X3-type thermoelectric materials with rhombohedral structures from large datasets, successfully identifying candidates with high thermoelectric performance~\cite{chen2022machine}. Similarly, unsupervised learning techniques, including K-means and Gaussian Mixture models, have been applied to cluster half-Heusler compounds, effectively pinpointing materials with favorable thermoelectric properties~\cite{jia2022unsupervised}. These approaches underscore the utility of machine learning in uncovering complex structure-property relationships relevant to thermoelectric performance.

For doped materials, the DopNet neural network architecture has been introduced to capture the effects of dopants accurately, allowing for the prediction of doped materials’ thermoelectric properties based on extensive datasets~\cite{na2021predicting}. This model has successfully predicted the impact of various dopants on crucial properties such as the Seebeck coefficient and electrical conductivity. In addition, artificial neural networks combined with techniques like combinatorial gradient thermal annealing have optimized the internal strain of bismuth telluride films, resulting in significant improvements to their Seebeck coefficient~\cite{sasaki2020identifying}. Such studies illustrate the importance of doping and strain engineering in enhancing thermoelectric performance and show how machine learning can assist in identifying critical relationships.

In the realm of high-throughput computing and screening, frameworks have been established to analyze compounds like diamond-like ABX2, identifying materials with high ZT values and highlighting key conductive mechanisms contributing to thermoelectric efficiency~\cite{li2019high}. Similarly, approaches that combine high-throughput ab initio calculations with deep neural networks have predicted the thermoelectric properties of IV-V-VI layered semiconductors, revealing high-performance candidates such as n-type Pb$_2$Sb$_2$S$_5$, with ZT values surpassing 1.0~\cite{gan2021prediction}. These integrated approaches expedite the discovery of new thermoelectric materials by leveraging machine learning and high-throughput screening. Additionally, the ToBaCCo 3.0~\footnote{\url{https://github.com/tobacco-mofs/tobacco_3.0}} code has been employed for large-scale screening of metal-organic frameworks (MOFs). Molecular dynamics simulations on 10,194 hypothetical MOFs were conducted to investigate the structural characteristics affecting thermal conductivity, revealing that parameters such as density, pore size, porosity, and surface area significantly influence thermal transport properties~\cite{islamov2023high}. Although machine learning models were not utilized in this study, the computational screening provided valuable insights into the structure-performance relationships governing MOF thermal conductivity.

High-throughput screening techniques have also been instrumental in identifying materials with excellent Peltier cooling performance among Heusler compounds, showcasing their potential for sustainable cooling applications~\cite{luo2022high}. The identification of compounds with desirable electronic structures and thermoelectric transport properties demonstrates the utility of high-throughput approaches for applications in efficient cooling technologies.

Finally, the integration of machine learning with experimental feedback has been proposed as a promising approach for accelerating material discovery. By iteratively refining machine learning models with experimental data, methods such as error correction learning (ECL) have successfully predicted high-performance thermoelectric materials with improved power factors, minimizing experimental trials and enhancing predictive accuracy~\cite{choubisa2023closed}.

Looking ahead, the convergence of machine learning and high-throughput computational methods is expected to play a pivotal role in advancing the inverse design of thermoelectric materials. As computational capabilities expand and larger, more diverse datasets become available, these technologies will continue to drive breakthroughs in addressing the challenges associated with optimizing thermoelectric properties. Machine learning models—such as Random Forests, neural networks, and clustering algorithms—have already demonstrated their ability to uncover intricate structure-property relationships across various material classes, including M2X3 compounds, half-Heuslers, and doped systems. Coupled with high-throughput screening, which facilitates the exploration of extensive material libraries, such as MOFs and Heusler compounds, researchers can swiftly identify high-ZT candidates and deepen their understanding of key factors affecting thermoelectric performance, such as thermal conductivity and electronic transport. Moreover, the integration of experimental feedback with machine learning models, as exemplified by the Error Correction Learning (ECL) approach, holds great promise for accelerating material discovery by continuously improving predictive accuracy. As these techniques become more refined and collaborations between computational and experimental domains intensify, they are expected to revolutionize the design and discovery of next-generation thermoelectric materials. Ultimately, these advancements will contribute to the development of highly efficient, environmentally sustainable energy systems for applications such as waste heat recovery, energy harvesting, and sustainable cooling technologies.

\subsection{Carbon-based Nanomaterials}
Carbon-based nanomaterials encompass structures composed of carbon atoms arranged in a hexagonal honeycomb lattice, including carbon nanotubes (CNTs) and graphene. Due to their distinct physical and chemical properties, these materials have garnered significant attention in materials science since their discovery~\cite{iijima1991helical,geim2007rise}. CNTs are classified into single-walled carbon nanotubes and multi-walled carbon nanotubes. The former consists of a single layer of rolled graphene, while the latter comprises multiple concentric graphene layers~\cite{mendoza2018functionalization}. These two forms exhibit notable differences in mechanical, electrical, and thermal properties. Graphene, a two-dimensional carbon-based nanomaterials, is formed by a single layer of carbon atoms in a hexagonal lattice, offering exceptional carrier mobility and mechanical strength~\cite{novoselov2004electric}.

Owing to their outstanding electrical conductivity, mechanical strength, and thermal conductivity, carbon-based nanomaterials are widely applied in various advanced technology sectors. For example, CNTs and graphene are used in the electronics industry for manufacturing miniaturized circuits and efficient semiconductor components, potentially surpassing the limitations of Moore’s Law~\cite{javey2003ballistic}. In the mechanical field, carbon-based nanomaterials are ideal for reinforcing composites due to their lightweight and high-strength characteristics, making them integral in high-performance structural materials for aerospace, automotive, and other industries~\cite{yu2000strength}. In the energy sector, the superior conductivity and energy storage capabilities of CNTs and graphene show great promise for improving the energy density of batteries and supercapacitors~\cite{an2001supercapacitors}. However, the synthesis and design of carbon-based nanomaterials involve complex multi-variable optimization challenges. Traditional experimental approaches struggle to address these challenges effectively, prompting the integration of AI technologies as a promising solution for inverse design and performance optimization of carbon-based nanomaterials.

Some researchers proposed an active learning model to elucidate the growth mechanisms of carbon-based nanomaterials on metallic substrates~\cite{zhang2024active}. This study combines molecular dynamics (MD) with the time-stamped force-bias Monte Carlo method, enhanced by the Gaussian approximation potential model for sampling. By simulating graphene growth on a copper substrate (Cu(111)), the researchers analyzed carbon monomer and dimer diffusion, the formation of carbon chains and rings, and the copper-assisted edge growth mechanism. The model dynamically generates machine learning potentials during the simulations, offering key insights into the growth processes of carbon materials on various metal substrates. Similarly, ML-enhanced molecular simulations investigate the high-temperature growth dynamics of CNTs, using the DeepCNT-22 machine learning force field to model interfacial behaviors, defect formation, and defect healing mechanisms during CNT growth~\cite{hedman2024dynamics}. By utilizing MLFFs trained on first-principles calculations, this study extends simulation time scales while maintaining computational accuracy, revealing atomic-level processes from CNT nucleation to growth. Additionally, combining machine learning with automation enhances the synthesis efficiency of carbon-based nanomaterials, as demonstrated by the AI-driven platform carbon copilot. This platform integrates Transformer-based language models (Carbon-GPT and Carbon-BERT), a robotic chemical vapor deposition (CVD) system, and data-driven ML models to optimize synthesis processes, significantly improving controllability and yield in the growth of CNTs and graphene, especially for horizontally aligned CNT arrays~\cite{li2024transforming}.

Beyond synthesis optimization, machine learning plays a pivotal role in predicting the physical properties of carbon-based nanomaterials, especially for key attributes like thermal and electrical conductivity. Compared to traditional methods, ML techniques provide more efficient and accurate solutions. For example, ML-based multi-scale modeling predicts the thermal conductivity of CNT-reinforced polymer composites, effectively capturing the impact of structural uncertainties on thermal conductivity by combining MD simulations with finite element analysis, regression tree models (random forests and gradient boosting machines), and deep neural networks~\cite{liu2022stochastic}. Similarly, an interpretable ML framework predicts the electrical conductivity of CNT/polymer composites, using stochastic multi-scale numerical models. This approach employs AI technique, random forests, and XGBoost models along with SHAP analysis to elucidate the influence of CNT structural parameters on conductivity, offering a robust theoretical framework for material design optimization\cite{elaskalany2024stochastic}. Another study applies ANN models to accurately predict the electrical conductivity of CNT-reinforced polymer composites, reducing computational costs while maintaining predictive accuracy across various conductive network structures~\cite{matos2019predictions}.

Modeling the structure-property relationship is critical for understanding the behavior of carbon-based nanomaterials. Machine learning is particularly adept at capturing the nonlinear relationships between complex structures and their properties, facilitating the design and optimization of new materials. For instance, a deep learning model called CNTNeXt, based on multi-layer synthesized images, predicts the mechanical properties of vertically aligned CNT (VACNT) forests. By utilizing a ResNeXt feature extractor paired with a random forest regression model, the model improves prediction accuracy by generating 2.5D morphological images that simulate real VACNT structures, enhancing understanding of CNT self-assembly processes~\cite{safavigerdini2023predicting}. Additionally, a neural network-based surrogate model (NN-EBE) efficiently predicts higher-order phenomena like buckling and nonlinear deformation for CNTs, providing computational advantages over traditional micromechanical models, particularly for large-scale simulations in composite material design~\cite{papadopoulos2018neural}. For electronic property predictions, machine learning predicts the electronic properties of graphene nanosheets based on geometric features. Using the density functional tight binding (DFTB) method to generate data, this approach applies linear regression, multilayer perceptrons, and support vector machines (SVM) to efficiently identify graphene structures with desirable electronic properties, aiding high-throughput material screening in electronic applications~\cite{fernandez2016geometrical}.

In terms of interface design and load transfer in composites, ML models have proven effective in optimizing chemical linkages and the mechanical behavior of CNTs within composites. Alred \emph{et al.} employed ML models to predict the electron density and mechanical properties of sulfur cross-linked CNTs~\cite{alred2018machine}. By generating training data from density functional theory and MD simulations, the study utilized multilayer perceptron neural networks and RF models for accurate predictions, significantly reducing computational costs. This research highlights the importance of ML in interface design for CNTs composites.

Machine learning techniques also apply to optimizing the biocompatibility of carbon-based nanomaterials, particularly in predicting their biocompatibility and toxicity. Research explores the use of AI and ML to address the limitations of traditional experimental methods, which are often time-consuming and costly. By combining extensive experimental validation data with CNT physical properties (\emph{e.g.}, length, diameter, surface functionalization) and the chemical composition of polymers, predictive models for biocompatibility and toxicity are developed. Using ANN, RF, and XGBoost models, with ANN excelling in capturing the complex interactions between CNTs and biological systems, the study utilizes K-fold cross-validation and regularization techniques to improve generalization, while SHAP analysis further enhances model interpretability~\cite{singh2020artificial}. Results demonstrate the efficiency of ANN in predicting toxicity responses, significantly reducing experimental time and cost while revealing critical relationships between CNT structure and biocompatibility. This research marks the first application of AI and ML in predicting nanomaterial biocompatibility, offering a novel tool for efficient toxicity assessment in future material design.

In the future, the continued advancement of ML and deep learning technologies will further automate and enhance the inverse design of carbon-based nanomaterials. Automated experimental platforms, combined with intelligent experimental planning, will accelerate the synthesis and screening of materials, particularly in complex multi-variable systems. Future active learning algorithms will dynamically update models with real-time feedback, improving design efficiency. In multi-scale modeling, future approaches will better integrate micro- and macro-scale features through the combination of first-principles calculations, MD simulations, and experimental data, establishing comprehensive multi-scale prediction frameworks. Additionally, the interpretability of ML models will become increasingly important, particularly in high-risk fields such as biomedicine and environmental safety, ensuring transparency and accountability in decision-making processes. Interdisciplinary collaboration will play a pivotal role in advancing carbon-based nanomaterials design, with the convergence of computational science, materials science, and biomedicine driving further innovations.In conclusion, future ML technologies will significantly improve the efficiency of carbon-based nanomaterials design, accelerating the transition from laboratory research to industrial applications and opening new avenues for the discovery and development of novel materials.

\subsection{Two-dimensional Materials}
\label{Two-dimensional Materials}
Two-dimensional  materials are those that consist of only one to a few atomic layers in the direction of thickness, while they have the potential to extend indefinitely in the plane. These materials are characterized by their unique physical and chemical properties, which differ significantly from their bulk counterparts. Different layers of materials can be held together by van der Waals forces. Materials that rely solely on these interlayer van der Waals interactions, without any covalent bonds between the layers, are referred to as van der Waals heterostructures. These heterostructures combine the unique properties of individual 2D layers to create materials with tailored functionalities, making them promising for a wide range of applications in electronics, photonics, and nanotechnology~\cite{novoselov20162d}.

Due to the extremely thin thickness, the movement of electrons in 2D materials is highly constrained, leading to the quantum confinement effect. This effect often results in novel physical properties, making 2D materials of great interest in fields such as electronics~\cite{Slager2013,PhysRevX.7.041069,po2017symmetry}, spintronics~\cite{bradlyn2017topological}, valleytronics~\cite{vergniory2019complete}, optoelectronics~\cite{zhang2019catalogue}, twistronics~\cite{chen2019high}, and slidetronics~\cite{horton2019high}. Common examples of 2D materials include graphene, which consists of a single layer of carbon atoms arranged in a honeycomb lattice~\cite{geim2009graphene}. Another important class of 2D materials is transition metal dichalcogenides, typically composed of one layer of transition metal atoms~(such as Mo or W) sandwiched between two layers of chalcogenide atoms~(such as S, Se or Te)~\cite{manzeli20172d}. For van der Waals heterostructures, interlayer coupling can be performed by regulating the interaction between different two-dimensional material layers~\cite{pei2022interlayer}.By changing the number of layers and the stacking order, the properties of the van der Waals heterostructures can be fine-tuned, and the 2D materials can be turned into topological insulators through specific layering structures~\cite{das2019topological}.

Currently, experimental methods for obtaining single or multi-layer 2D materials include exfoliating them from 3D bulk materials or directly growing them on substrates using chemical vapor deposition~(CVD). Exfoliation techniques involve peeling off individual layers from bulk materials, while CVD allows for the controlled growth of 2D materials on various substrates~\cite{yi2015review,zhang2013review}. However, discovering new 2D materials through experimental methods can be resource-intensive and time-consuming, requiring significant investment in both materials and research efforts. ML can learn to capture material feature from a large number of crystals, guiding the synthesis of new 2D materials~\cite{lu2022machine,ryu2022understanding}.

Topological insulators are a typical category of 2D materials, that behave as insulators in the body, but exhibit electrical conductivity at the surface or edge. This unique property stems from the topological properties of the material, \emph{i.e.}, the electron band structure of the material has a special topological invariant mathematically. In 2D topological insulators, the spin and momentum of electrons are locked together to form edge states. This phenomenon is called the quantum spin Hall effect~\cite{hasan2010colloquium,ando2013topological}. Topological insulators have a wide range of applications, including field-effect transistors in electronics, efficient optoelectronic devices in optics, topological qubits in quantum computing to enhance the stability of quantum systems, and magnetic topological insulators in magnetism. However, due to the lack of sufficient candidate materials, the research on topological insulators is limited. In traditional methods, materials are evaluated one at a time through a trial-and-error process, making the search of the material space resource-intensive and time-consuming. Thus, a pipeline for finding new 2D topological insulators is developed~\cite{schleder2021machine}. By using this method, 56 topological materials are identified , of which 17 are quantum spin Hall insulators, and 9 are not reported in the literature, with 3 showing large energy gaps suitable for room-temperature applications.

Van der Waals heterostructures is another category of materials that have garnered attention, consisting of 2D materials that are combined together. However, according to computational screenings it is suggested that the number of possible 2D materials could be in the thousands, leading to potentially millions of distinct heterointerface combinations~\cite{choudhary2017high,mounet2018two}. Then, a computational database, web applications, and machine learning models are developed to accelerate the design and discovery of 2D heterostructures~\cite{choudharyefficient}. Based on 674 non-metallic 2D materials, they generate 226,779 heterostructures and classify as 3 types, of which type-II is found to be the most common and type-III the least common. Other users can also use their network applications and machine learning model to generate 2D heterostructures and predict physical properties of the new 2D heterostructures.

\vpara{AI-driven experimental.}
By using AI-assisted experimental procedures, it is possible to synthesize 2D thin film materials through the way of automatic control~\cite{harris2024autonomous}. Specifically, the growth conditions are chosen by the experimenter to initialize the algorithm, which then autonomously grows 125 samples, the process is about ten times faster than the traditional process. In addition to supervised learning styles, with the existing positive samples and a large number of unlabeled samples, a ML model can be trained using semi-supervised or unsupervised learning~\cite{wang2021unsupervised}. A large number of potential 2D materials with known properties enable to be screened through high-throughput clustering. This approach can aid in the design of new 2D materials~\cite{frey2019prediction}. It is also possible to optimize the query strategy of AI based on reinforcement learning, and improve the classification performance of the model through rewards. At the same time, through the framework of active learning, only a limited number of labels can be iteratively obtained and updated more positive sample labels~\cite{li2024reinforced}.

The work described above uses ML model to directly design crystal structures, helping researchers identify 2D materials of interest. Additionally, it is possible to indirectly assist experimenters in synthesizing new 2D and quasi-1D materials by training a ML model to control the parameters of CVD through the integration of AI in the experimental process~\cite{rajak2020quantum,xu2021machine}. It can also predict the probability of synthesizing new materials based on the parameters of a given CVD and recommend the most favorable parameters~\cite{tang2020machine}. 

\vpara{2D materials datasets.}
At present, the number of known 2D materials is still relatively limited, far fewer than the number of 3D bulk materials. Therefore, the construction of a comprehensive 2D material database is a crucial task in condensed matter physics. Computational 2D Materials Database~(C2DB) is a database for 2D material discovery that is aided by ML~\cite{haastrup2018computational,gjerding2021recent}, C2DB first generates new crystals by replacing experimentally known crystal structures with reasonable atoms. Then, stable materials are selected through a series of steps, including structural relaxation, deduplication, and stability calculation by DFT. Finally, stable materials are screened to build a 2D material database. 2DMatPedia selects 3D materials with layered structures from the Materials Project database through high-throughput topological analysis. It generates new materials by performing atomic replacements after stripping the layers. Stable materials are then identified through DFT calculations and added to its database~\cite{zhou20192dmatpedia}. 

Simultaneously, AI can also be used to replace the DFT calculation part in the construction of 2D material databases. By using discriminative AI models, the properties of 2D materials can be directly predicted and classified~\cite{talirz2020materials,campi2023expansion}. A 2D perovskite database can predict band gaps and atomic charges using machine learning models~\cite{marchenko2020database}. This demonstrates that machine learning can assist in the design of new 2D mixed perovskite materials through material classification. 

Furthermore, new 2D materials can be generated and their properties can be predicted directly by AI, without the need for traditional computational methods such as DFT. The fabrication of solid-state devices with tailored optoelectronic, quantum emission, and resistive properties requires the design of 2D materials with point defects. Due to the strong correlation effect of electrons and the exponential growth of the defect site search space, traditional methods are unable to find these 2D materials on a large scale. Through deep transfer learning, a ML model is first pre-trained on 3D materials and then used to predict 2D materials that are likely to produce effective point defects. Based on these predicted 2D materials, another ML model is trained to map these materials from the initial graph structure to calculated defect properties~\cite{frey2020machine}. In addition, manual data extraction from publications is the mainstream method for collecting 2D materials. However, through natural language processing and text mining technologies in the field of AI, crystals with interesting properties can be automatically extracted from publications~\cite{xu2023small,brito2023network}.

\subsection{Photovoltaic  Materials}
Photovoltaic  conversion is one of the most important energy conversion methods mastered by humanity, enabling us to ``store light''. The search for high-performance photovoltaic materials has long been a crucial pursuit. Not only do they provide support for lighting and display, but they also contribute to reducing dependence on fossil fuels, thereby having a profound impact on addressing global climate change.
The essence of ``storing light'' lies in capturing excitons generated by photons. Enhancing photovoltaic conversion efficiency (PCE) generally involves two main strategies: improving photon-to-exciton conversion efficiency and designing effective junctions to capture excitons. Photovoltaic  conversion works in both directions, and these improvements have the potential to inspire new applications in fields such as solar cells, light-emitting diodes~(LEDs), laser diodes, photodetectors, and photocatalysis.

In the present, monocrystalline silicon is the most mainstream photovoltaic material. It is widely used in ground-based photovoltaic power generation due to its relatively high power generation efficiency (27.3\%~\cite{green2024solar}) and long service life. However, it suffers from several shortages. In classical studies, the ideal bandgap for single-junction solar cells is around 1.34 eV, which can theoretically achieve a maximum efficiency of 33.7\%~\cite{polman2016photovoltaic}. However, the bandgap of monocrystalline silicon is around 1.1 eV and cannot be adjusted~\cite{nelson2003physics}. The second is indirect bandgap. It requires phonons as a medium for light absorption, leading to lower absorption efficiency and making it difficult to reduce the material thickness~\cite{nelson2003physics}. The last is the high sensitivity to impurities. Although the silicon industry is mature, producing high-purity monocrystalline silicon remains costly~\cite{philipps2019photovoltaics}.

III-V compounds represent another important class of photovoltaic materials, offering significant advantages over silicon in certain applications. These compounds possess direct bandgaps that can be precisely tuned through doping, enabling their use across a broad spectrum of optoelectronic devices~\cite{streetman2000solid}. For instance, gallium nitride (GaN), with a bandgap of approximately 3.4 eV, is widely utilized in blue and ultraviolet LEDs~\cite{nakamura2013blue,levinshtein2001properties}, whereas gallium arsenide (GaAs), with a bandgap of about 1.42 eV, is commonly applied in red LEDs~\cite{levinshtein2001properties}. III-V compounds exhibit superior thermal stability, high defect tolerance, and exceptional efficiency, making them particularly suitable for applications in extreme environments, such as space-based photovoltaics~\cite{li2021brief}. Despite these advantages, III-V compounds also pose challenges, including high production costs, the inclusion of toxic elements, and moderate chemical stability, which may limit their widespread adoption~\cite{dimroth2007high,streetman2000solid}.
In the exploration of next-generation photovoltaic materials, perovskites and organic solar cells~(OSC) stand out as two prominent directions. 

\vpara{Perovskite materirals.}
Perovskite materials have been one of the most important subjects of study in materials science due to their outstanding optoelectronic properties and complex crystal structures~\cite{kojima2009organometal,snaith2013perovskites}. The term ``perovskite'' refers to materials that share the crystal structure of calcium titanate~(CaTiO$_3$), characterized by a general chemical formula of ABX$_3$, where A and B are cations, and X is an anion that bonds with both cations~\cite{tilley2016perovskites}. These materials typically adopt a cubic structure, but various distortions can lead to alternative configurations~\cite{tilley2016perovskites}. Such structural flexibility allows for the incorporation of a wide range of elements, making the perovskite family remarkably diverse, with potentially tens of thousands of possible compositions. Moreover, when considering element substitution and doping, the number of potential perovskite variants could exceed millions~\cite{tao2021machine}. 

The surge of interest in perovskites began with early demonstrations of their potential in solar cells, and the power conversion efficiency~(PCE) of perovskite solar cells~(PSCs) has risen dramatically over the past decade, reaching certified values above 25\%~\cite{royperovskite2022}. This exceptional performance is largely due to the unique properties of perovskites, such as their high light absorption coefficients, direct and tunable bandgaps, high charge-carrier mobilities, and long carrier diffusion lengths. However, the prototypical organic–inorganic hybrid lead halide perovskites are highly sensitive to environmental factors, including moisture, oxygen, and temperature, and their performance depends on the presence of toxic lead Pb.  As a result, current research is increasingly focused on identifying stable, non-toxic alternatives to traditional lead halide perovskites. Addressing these challenges will be crucial for transitioning perovskites from the laboratory to practical optoelectronic applications.

Machine learning has demonstrated significant potential in accelerating the discovery of high-performance perovskite materials, especially in terms of predicting photovoltaic efficiency and identifying stable, synthesizable structures. Various machine learning models have been successfully deployed to predict key properties of PSCs, including PCE, short-circuit current density, open-circuit voltage, fill factor, and external quantum efficiency~\cite{liumachine2024}. In a recent work, several models are compared for PCE prediction and XGBoost is identified as the most accurate~\cite{mohantycomprehensive2023}. Furthermore, XGBoost has also been utilized to predict recombination losses in PSCs, revealing insights into dominant recombination mechanisms~\cite{akbarunveiling2024}. 

Beyond performance prediction, machine learning plays an essential role in assessing structural stability and guiding the synthesis of promising candidates. Machine learning interatomic potentials models have been developed to predict the crystal structures of hybrid organic-inorganic perovskites~(HOIPs), enabling researchers to evaluate the stability of these compounds accurately~\cite{karimitariaccurate2024}. Additionally, MLIP has been applied to explore the dynamic behaviors of metal halide perovskites, shedding light on their structural characteristics under different conditions~\cite{ranhalide2023}. Through a combination of machine learning and molecular dynamics simulations, researchers have also investigated the stability of HOIPs~\cite{biinfluence2024}. Moreover, machine learning methods are actively used to identify synthesizable PSCs, which is crucial for material validation in practical applications~\cite{wuuniversal2024}. Moreover, machine learning aids in the search for non-toxic perovskite materials as well. Active learning has been applied to identify lead-free white-light LEDs by training models on oxide perovskites and datasets selected from six halide perovskites using active learning methods~\cite{maidata2024}. This study successfully predicted photoluminescence quantum yield, finding a significant correlation between ionic radii and PLQY. A simpler machine learning model has also been applied to identify non-toxic PSCs~\cite{lirational2023}, focusing on the importance of the $d_{10}$ orbital in double perovskite doping. Improving interpretability and identifying chemical patterns is another key area of development in perovskite-related machine learning research. Genetic Algorithms (GA) have been used as a pre-screening tool to aid graph convolutional networks~(GCN) in bandgap prediction, facilitating the identification of relevant chemical trends~\cite{choubisainterpretable2023}. Moreover, decision tree models have been enhanced to improve interpretability in applications to solid-state chemistry, enabling researchers to identify significant descriptors across diverse materials, including perovskites, spinels, and rare-earth intermetallics~\cite{selvaratnaminterpretable2023}.

\vpara{Organic solar cells.}
Organic solar cells~(OSCs) represent another prominent class of next-generation photovoltaic materials. While inorganic materials are known for their efficiency and durability, their production can be costly. In contrast, OSCs offer distinct advantages, such as lightweight and flexible structures that can be easily fabricated into various shapes and sizes, along with the potential for low-cost and scalable production. However, OSCs generally have lower efficiency and shorter lifetimes, presenting a significant challenge for widespread adoption~\cite{solak2023advances}. The primary device architecture is bulk-heterojunction, featuring an active layer composed of interpenetrating networks of donor and acceptor materials, which facilitates efficient charge separation and transport~\cite{brabec2010polymer}. The initial breakthrough in acceptor materials came with the use of fullerene derivatives, such as C$_{60}$~\cite{sariciftciphotoinduced1992}, which exhibited promising efficiency and laid the groundwork for further advancements in OSCs. Soluble fullerene-based acceptors like PC$_{61}$BM and PC$_{71}$BM have since been developed and attracted significant attention. However, the inherent limitations of fullerene acceptors, such as weak absorption in the solar spectrum, difficulty in tuning energy levels, and the propensity for aggregation and crystallization—have hindered their further development and highlighted the need for alternative materials~\cite{li2012polymer}.

In recent years, the focus has shifted toward non-fullerene acceptors~(NFAs), which have shown great promise due to their tunable energy levels, improved absorption properties, straightforward synthesis, and better morphological stability. A major milestone was reached in 2019 with the introduction of Y6~\cite{yuansinglejunction2019}, a new benchmark NFA that achieved a power conversion efficiency~(PCE) of 18\%~\cite{solakadvances2023}, garnering widespread attention for its high performance. Since its introduction, Y6 and related NFAs have become a central topic in the field, with research efforts dedicated to further improving performance, understanding the underlying mechanisms responsible for the high efficiency, and exploring new applications. The next phase of development in OSCs will likely involve optimizing Y6 derivatives to further enhance performance and stability, as well as the continued search for novel NFAs with unique properties. These efforts aim to push the boundaries of efficiency while addressing the remaining challenges related to the lifetime of organic solar cells, paving the way for more practical and commercially viable OSC technologies. In materials science, decision tree algorithms have been widely utilized to accelerate the discovery of new materials, particularly for predicting optical and electronic properties of molecules and materials. The general workflow typically includes collecting molecular datasets, mapping molecular structures to descriptor spaces, and then employing machine learning models to learn and predict key properties like PCE. This approach has seen extensive application across various studies. Random forest~(RF) is used to predict PCE for non-fullerene acceptors~\cite{zhanghighefficiency2022}, aiding in the development of new acceptor materials for MP6 donors. In~\cite{morishitamachine2024}, RF was applied to predict JSC, assisting in the creation of new donor molecules compatible with non-fullerene acceptors~(NAs). Additionally, XGBoost is leveraged for high-throughput screening to predict PCE for DA pairs~\cite{liuaccelerating2022}. 

Comparative studies have also been conducted on various decision tree models, with RF often emerging as advantageous in PCE prediction. RF is used to screen and predict performance for non-fullerene acceptors~(NFA) and broader DA pairs~\cite{sutharmachine2023}~\cite{limachine2024}. Meanwhile, other studies have observed that gradient boosting models exhibit superiority in predicting JSC and VOC, as noted in~\cite{sutharmachinelearningguided2023}. To further enhance model performance, some studies explore algorithmic improvements. In a recent work~\cite{miyakeimproved2022}, by introducing artificial failure data, RF’s PCE predictions for NFAs are observed to be enhanced. Another approach integrates geometric graph neural networks~(GNN), automatically extracting features from molecular structures, and then uses decision tree models~(such as LightGBM) as a backend for analysis, as described in~\cite{wangefficient2023}. Additionally, due to the lag in available datasets, leveraging large language models to retrieve literature can expand material libraries and accelerate the discovery of novel materials, which is poined in~\cite{shettyaccelerating2024}.

\subsection{Catalyst Materials}
Catalyst materials are defined as a type of substance that has the capacity to alter the rate of a chemical reaction. It is important to note that they are not themselves involved in the final product of the reaction; however, they serve to accelerate or decelerate the rate of the reaction by modifying the path of the reaction and reducing the activation energy required for the reaction to occur. A catalyst retains its chemical nature and quality both before and after a reaction, and therefore, it can be reused on numerous occasions.

The history of the application of catalysts can be traced back several hundred years. Over time, catalyst technology has developed in a direction that is increasingly efficient, environmentally friendly and sustainable. It has made significant contributions to a number of fields, including the chemical industry, energy, environmental protection, life science and medicine. In recent years, the field of material synthesis has witnessed a period of accelerated development, largely due to the advent of advanced AI technologies. The synthesis of catalyst materials has been no exception to this trend. Given the extensive scope of catalyst materials and the diverse applications of corresponding ML technology, it was deemed prudent to select a representative sample of work for presentation, taking into account the limitations of author's knowledge in this field.

Descriptors play an important role in improving the accuracy of ML models. While ML techniques can improve the accuracy of predictions, it is often the case that descriptors determine the upper limit of the prediction~\cite{mou2023machine}. 
Some researchers concentrate their efforts on the process of searching for additives in the electrochemical deposition of copper catalysts for the reduction of CO$_{2}$~(CO2RR)~\cite{guo2021machine}. The selection of an appropriate combination of additives for the preparation of the catalyst is a challenging process. To solve this problem, a strategy comprising three rounds of learning, integrating experimental outcomes and ML, is devised and subsequently applied to an additive library. The process enabled researchers to identify crucial chemical components and to synthesise the requisite molecules. This work represents a significant contribution to the field of experiment-based descriptors.

Furthermore, the analysis of descriptors for product selectivity in the oxidative coupling of methane (OCM) reaction has been conducted using ML and physical quantities derived from the periodic table~\cite{ishioka2022designing}. One of the principal objectives of this study is to investigate the selectivity of C$_{2}$H$_{4}$/C$_{2}$H$_{6}$ (C$_{2}$s). The process employs the use of hierarchical clustering, random forest classifier, and support vector classifier. Eventually, three previously unreported catalysts with high C$_{2}$s are identified as potential candidates, \emph{i.e.} Ti-V-Ce-BaO, Y-Y-Eu-TiO$_{2}$ and La-Pr-Hf-BaO, and their performance is subsequently verified through experimental analysis. The authors additionally examine five additional related descriptors. The authors additionally examine five related descriptors and conclude that high C$_{2}$s values are associated with low first ionization energies, electron affinities and electronegativities, as well as high second ionization energies and densities.

In addition to experiment-based descriptors, there has been some work utilizing theory-guided descriptors.
The SISSO~(sure independence screening and sparsifying operator) descriptors are expressed as nonlinear functions of intrinsic properties of the clean catalyst surface~\cite{andersen2019beyond}. The authors demonstrated that their method was more general and accurate in predicting the adsorption energy of alloys on mixed metal surfaces, even when based on training data that included only pure metals. In the event that a considerable number of features are available, the compressive sensing method SISSO represents an appropriate solution~\cite{ouyang2018sisso}. This approach has also been applied to the study of other materials, including perovskite oxides and halides~\cite{bartel2019new} and doped transition metal oxides~\cite{xu2020data}. However, this method still has some limitations, such as the presence of nearly degenerate models, stability issues when dealing with data perturbations, and unclear physical interpretations~\cite{mou2023bridging}. There is also a lot of work on theory-guided descriptors from other perspectives, such as intrinsic atomic property~\cite{ren2022universal}, electronic and structural property~\cite{li2017high,li2020adaptive,lu2020neural} and others~\cite{wang2020electric,zhong2021electronic,gu2022nitrogen}. Similarly, efforts have been made to identify descriptors that integrate theoretical and experimental data~\cite{williams2019enabling,karim2020coupling,artrith2020predicting,zhu2022all}.

Additionally, research has been conducted from alternative perspectives on the subject of catalysts. For example, the utilization of active learning in conjunction with DFT calculations to develop an efficient copper-aluminium electrocatalyst for the conversion of CO$_{2}$ to ethylene is worthy of considerable praise~\cite{zhong2020accelerated}. This approach has resulted in the highest reported Faradaic efficiency to date. In their study, the researchers employ probabilistic models with Gaussian processes, trained with $\textit{ab initio}$ data and a set of multifidelity features, to identify high-performance ABO$_{3}$-type cubic perovskites capable of catalyzing the oxygen evolution reaction~(OER)~\cite{li2020adaptive}. The method is successful in identifying several known perovskites, which demonstrate superior performance compared to the benchmark LaCoO$_{3}$. A series of perovskites with favourable properties are also obtained, including KRbCo$_{2}$O$_{6}$, BaSrCo$_{2}$O$_{6}$, KBaCo$_{2}$O$_{6}$, and others.

The discovery of single-atom alloy catalysts~(SAACs) represents a significant area of focus within the scientific community~\cite{han2021single}. In this particular study, they combine first-principles calculations with compressed-sensing data-analytics methodology. As a result of this approach, they are able to finalize more than 200 unreported candidate SAACs, some of which exhibit greater stability than previously observed. Concurrently, their investigation highlights the significance of data analysis in avoiding bias in catalytic design.

A recent publication presents an active learning workflow for the generation of fuel cell catalysts~\cite{yin2024machine}. The authors concentrate their research on ternary alloys in the form of Pt$_{2}$CoM~(in the practical application of proton exchange membrane fuel cells, PtCo is the most promising alloy catalyst, while M represents another base metal element). Guided by theoretical calculations, the researchers ultimately prepared the Pt$_{2}$CoCu and Pt$_{2}$CoNi compounds through experimental synthesis. These materials exhibited a large electrochemically active surface area of approximately 90 m$^{2}$/g$_{Pt}$ and a high specific activity of approximately 3.5 mA/cm$^{2}$.

Researchers leverage local machine learning capabilities to rapidly and accurately identify structure descriptors by integrating fundamental physical attributes with graph convolutional neural networks~\cite{chen2022universal}. They successfully identify 43 high-performance alloys as electrocatalysts for the hydrogen evolution reaction, with some candidates already validated. To further validate the method's precision, a comprehensive study of AgPd is conducted through ab initio calculations in a realistic electrocatalytic environment.

An intriguing prospect emerges from the utilization of language models in catalyst discovery~\cite{mok2024generativelanguagemodelcatalyst}. The discovery of catalyst materials is presented with the aid of a large language model, designated CatGPT. The focus is on the two-electron oxygen reduction reaction (2e-ORR), with the model being applied to identify the catalyst for this process. Ultimately, several materials that are not present in the database are identified, including RhSe, CdAg, SnAu, and others.

The field of technology catalysts is vast, encompassing a multitude of applications for ML technology and its associated techniques. The aforementioned examples represent a selection of the most influential or recently developed outcomes, and a comprehensive account of all relevant findings is beyond the scope of this discussion.

\subsection{High-Entropy Alloys}
High-Entropy Alloys~(HEAs) are a class of innovative materials made from five or more elements in nearly equal proportions. The traditional idea of making alloys is to add other reinforcing elements to a metal, which has been used extensively throughout human history. In 2004, the discovery of HEAs offers a distinctive alloying strategy~\cite{yeh2004nanostructured, CANTOR2004213}. These new materials are referred to as HEAs due to their increased configurational entropy, which was believed to be the key factor in their stabilization. Although it was later found that configurational entropy was not as important as first assumed~\cite{george2019high,OTTO20132628}.
HEAs have many excellent mechanical properties and a large component space, which provides a good platform for material research.

However, when faced with the design of HEAs, conventional methods hit a major roadblock. Considering only the usual elements of the periodic table, this spans a huge composition space, which cannot be managed by conventional material design approaches, such as the calculation of phase diagrams and density-functional theory~\cite{rao2022machine}.

The past decades have witnessed the rapid development of machine learning, which suggests that it is likely to make a difference in the area of material design for HEAs~\cite{liu2023machine, liu2024machine}. Some researchers utilize several ML models including some simple deep nueral networks and select a conditional random search to be the inverse predictor to design the new HEAs~\cite{wang2023neural}. And they successfully find two HEAs with better ultimate tensile strength and total elongation than the input datasets. Support vector machine is also utilized to tackle problems of HEAs~\cite{chau2023support}. In their approach, the emphasis is on the framework of hyperparameter tuning and the use of weighted values. Through experiments, they discovere that their architecture performs very well and even exceeds the results of artificial neural network~(ANN). There is also some work on comparing the effectiveness of different methods. Researchers test three different ML algorithms, namely K-nearest neighbours, SVM, and two ANNs, \emph{i.e.} the unsupervised self-organizing maps and the supervised multi-layer feed-forward neural network~(MLFFNN)~\cite{HUANG2019225}. They conclude that MLFFNN performs better. There is also some work that has tested Logistic Regression, Decision Tree, SVM, Random Forest, Gradient Boosting Classifier, and ANN. They claim ANN exhibits the highest accuracy~\cite{krishna2021machine}. 

All of the work described above is an attempt to employ ML techniques, but does not fully incorporate the characteristics of HEAs themselves. The dataset for HEAs is not large enough, so there is a need of the architectures that can be trained from small samples to produce sufficiently adaptive structures. Some recent work considers this issue to some degree.

The first work to mention is about active learning. A typical example illustrates the successful implementation of an iterative approach for the synthesis of new HEAs~\cite{rao2022machine}. Their workflow is a closed-loop where ML techniques, DFT, thermodynamic calculations and experiments are used in the process. Their research focuses on the design of high-entropy Invar alloys with low thermal expansion coefficients~(TEC), and they investigate FeNiCoCr HEAs and FeNiCoCrCu HEAs. From their data, it is known that the HEAs find have lower TEC and higher configurational entropy compared to past materials. In addition to the ability of discovering new HEAs, one of the greatest strengths of this workflow is its efficiency: the entire process of their work takes only a few months, whereas in the past the corresponding discoveries could have taken several years and more experiments. It is notable that other efforts have proposed frameworks with a similar structure~\cite{li2022towards}.

Some researchers employ a deep network architecture with residual connections to predict the phase formation in HEAs~\cite{zhu2022phase}. And compared to traditional NN, the overall accuracy of this framework could achieve 81.9\%. This work provides a new approach for realizing phase formation prediction of HEAs and is also of great significance for designing new HEAs. 

Some researchers have devised a novel type of NN, designated the elemental convolution neural network~(ECNet), which is capable of attaining global element-wise representations~\cite{wang2022element}. The authors investigate FeNiCoCrMn/Pd systems using ECNet and the data obtained from DFT. The authors also use transfer learning to make enhancements to the performance of the network. By employing this framework, the concentration-dependent formation energies, magnetic moments, and local displacements in a number of sub-ternary and binary systems can be obtained.

The application of ML to the analysis of materials microstructure provides a basis for the reverse engineering of alloys, which is then integrated with the accumulated knowledge of human experience. As a result of employing their own methodologies, the team makes a discovery that leads to the identification of a novel alloy, provisionally named as 9\% Cr steel~\cite{pei2021machine}. The authors posit that the success of the approach can be attributed to the optimization of NN structures and the fine-tuning of associated parameters.

Some of the work described above does effectively mitigate the problem of small size of datasets. However, the quality of the datasets and the selection of appropriate ML techniques remain problematic.

\subsection{Porous Materials}
Porous materials are a class of materials with a pore structure filled with holes or voids of varying sizes. The size, shape and distribution of the pores determine the properties of the material. Porous materials usually have a high specific surface area, low density and good adsorption capacity, which give them a wide range of applications in a variety of fields, such as the chemical industry, energy, environment and biomedicine. Zeolites are a representative class of porous materials. Newly emerging porous materials, including metal-organic framework materials~(MOFs), covalent organic framework materials~(COFs), and carbon-based porous materials, have expanded the range of applications of porous materials~\cite{slater2015function, thomas2020much}.

In recent years, there have been significant developments in the field of AI, with notable advances in the area of porous material synthesis. Given that zeolites were synthesised earlier and are more widely used than other porous materials, it is unsurprising that research has been conducted into the AI-assisted synthesis of zeolites. In their study, the researchers employ a generative adversarial network, designated as ZeoGAN, to generate 121 crystalline porous materials~\cite{kim2020inverse}. The neural network receives inputs in the form of energy and material dimensions, and the results demonstrate that zeolites with a desired range of 4 kJ/mol methane heat of adsorption can be produced with reliability. Four years later, work employing diffusion modelling in the context of porous materials is published~\cite{park2024inverse}, and the corresponding architecture is known as ZeoDiff. The authors assert that the diffusion model outperforms the ZeoGAN with regard to structural validity, exhibiting an improvement in performance of over 2000-fold. Furthermore, the authors implement conditional generation, namely the generation of structures with user-desired properties, with ZeoDiff. In order to achieve conditional generation, it is necessary to make adjustments to the network, such as integrating an additional channel that is closely relevant to the property of interest.

The synthesis of MOFs is a more complex and challenging process than that of zeolites. There are over 100 known species of atoms that can form MOFs, with an average number of atoms per unit cell that is significantly higher than that of zeolites. The variation autoencoder~(VAE) has already successfully applied generative modelling to MOFs~\cite{yao2021inverse}. A VAE framework, designated SMVAE, is employed for the generation of MOFs. The structural validity of SMVAE is demonstrated to be 61.5\%. By focusing on the adsorption capacity of CO$_{2}$ in their model, the authors demonstrate the ability to modulate the CO$_{2}$ adsorption capacity and selectivity of the material. The top-performing MOF they have discovered has a CO$_{2}$ capacity of 7.55mol/kg and a selectivity over CH$_{4}$ of 16. A generative AI framework called GHP-MOFassemble is proposed in the study.~\cite{park2024generative}. The GHP-MOFassemble method is employed to synthesise MOF linkers, which are then utilized with one of three pre-selected metal nodes~(Cu paddlewheel, Zn paddlewheel, Zn tetramer) to form MOFs with a primitive cubic topology. The generation of linker molecules is achieved through the utilization of DiffLinker~\cite{igashov2024equivariant}. Molecular fragments with high expression levels are extracted from the existing database. The aforementioned molecular fragments are employed to obtain linkers, which in turn facilitate the generation of MOFs. Subsequently, the MOFs were evaluated using a predictive model, resulting in the identification of six MOFs with a CO$_{2}$ capacity exceeding 2 mmol/g, indicative of high performance.
Researchers put forth a coarse-grained~(CG) diffusion model, designated MOFDiff, which is capable of generating CG MOF structures through a denoising diffusion process~\cite{fu2023mofdiff}. The authors put forward the proposition that template-based methodologies can serve to constrain the search space and preclude the inclusion of viable materials. Consequently, they have devised a process for the generation of MOFs which entails the direct identification of coarse-grained building blocks within three-dimensional coordinates. As a subsequent processing stage, an additional MOF assembly procedure is necessary to orientate the components and ascertain the interconnectivity between them. The utilization of MOFDiff has led to the identification of potential candidates for high CO$_{2}$ working capacity, with nine of the current top ten being generated by this framework. The rencent work makes use of signed distance functions~(SDFs)~\cite{park2024multi}. They introduce a latent diffusion model, MOFFUSION, which utilizes SDFs as the input representation of MOFs. The model demonstrates a high structural validity of 81.7\%. Additionally, the authors illustrate its capacity for conditional generation across a range of data modalities, including numeric, categorical, text data, and their combinations.

In addition to zeolites and MOFs, research has been conducted into the use of AI in the inverse design of other porous materials. To illustrate, a methodology employing ML approaches for the identification of highly porous carbon materials has been devised~\cite{wang2023machine}. A methodology employing quantitative structure–property relationships in conjunction with machine learning techniques has been developed for the purpose of forecasting the properties of COFs. This methodology utilizes the structural characteristics of the solvents and COF building blocks~\cite{kumar2021synthesis}. Furthermore, the integration of machine learning (ML) with porous media has facilitated a multitude of advancements across diverse research domains~\cite{delpisheh2024leveraging,d2022machine}. The application of ML technology in the synthesis of porous materials is now widely acknowledged. Further attempts are currently underway.

\section{The Development of AI Methods in Materials Science}

\begin{figure*}[t]
		\centering  
		\includegraphics[width=1.0\linewidth]{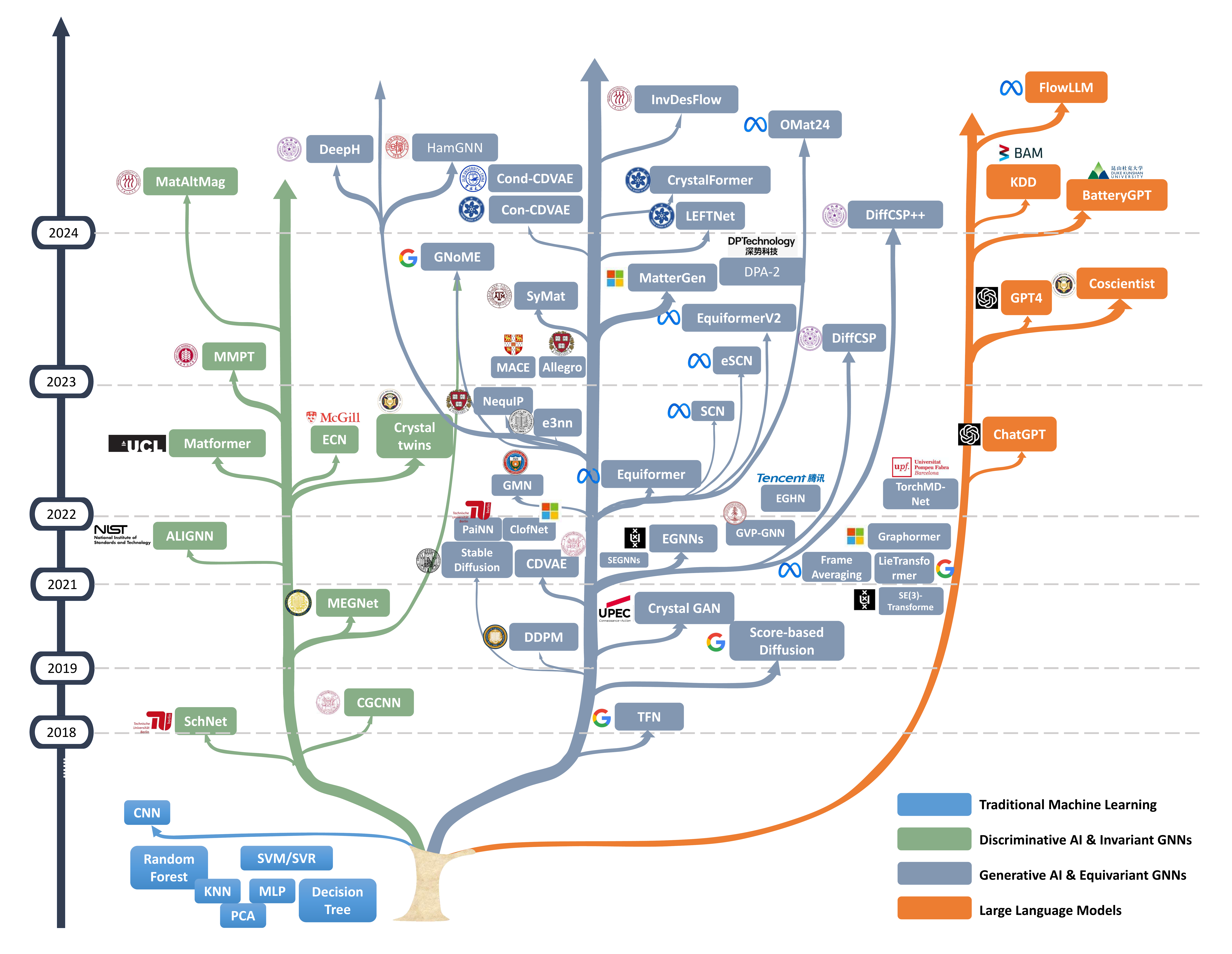}
		\caption{The rapid advancement of AI technologies accelerates materials discovery in various ways. Blue indicates traditional machine learning; green represents the development of invariant GNNs and their application in discriminative AI for predicting material properties; deep blue denotes the progress of equivariant GNNs, generative AI techniques, and their integration for structural predictions. Orange signifies the role of LLMs in expediting materials discovery.}
		\label{fig:aitec} 
\end{figure*}

\label{sec-3}
In this chapter, we present a comprehensive review of the advancements in AI technologies within the field of materials science. Drawing upon recent trends and the growing significance of AI, we explore traditional machine learning methods, geometric graph neural networks (GNNs), discriminative AI, generative AI, and large language models (LLMs). As illustrated in Figure~\ref{fig:aitec}, GNNs are classified into invariant and equivariant models. Invariant GNNs are extensively applied in discriminative AI for predicting material properties, thereby enabling high-throughput screening. In contrast, equivariant GNNs are predominantly employed in generative AI to facilitate structural predictions of materials. Next, we provide a detailed overview of the development of these algorithms and their applications in materials discovery.

\subsection{Traditional Machine Learning Methods}
We categorize techniques such as random forests, convolutional neural networks (CNN), support vector machines (SVM), and multilayer perceptrons (MLP) as traditional machine learning methods. These algorithms are among the most commonly applied in materials discovery, with each offering distinct advantages in addressing different design challenges. Random Forest, as an ensemble method, combines the predictions of multiple decision trees to reduce the risk of overfitting from individual trees. Its formula is given by
\begin{equation}
    \hat{y} = \frac{1}{N}\sum_{i=1}^{N}h_i(x),
\end{equation} 
where $\hat{y}$ represents the final prediction, $N$ denotes the number of trees in the ensemble, and $h_i(x)$ refers to the prediction made by the $i$-th tree~\cite{breiman2001random}. The decision tree, renowned for its interpretability, performs classification by recursively splitting the data. While decision trees are prone to overfitting, this risk can be mitigated by controlling the depth of the tree. Common splitting criteria include the Gini coefficient and entropy, with entropy defined as:
\begin{equation}
    \text{Entropy}(D) = - \sum_{i=1}^{n} p_i \log(p_i),
\end{equation} 
where $D$ is the dataset, and $p_i$ is the probability of class $i$ in the dataset~\cite{quinlan1986induction}. The SVM finds an optimal hyperplane that maximizes the margin between classes; for nonlinear data, it uses kernel functions to map data to a higher-dimensional space, optimizing $\min_{w, b} \frac{1}{2} |w|^2$ to determine the hyperplane, where $w$ is the weight vector and $b$ is the bias term~\cite{cortes1995support}. The MLP is a neural network with at least one hidden layer, which minimizes mean squared error (MSE) via backpropagation to learn from complex data. Its MSE formula is 
\begin{equation}
    \text{MSE} = \frac{1}{n} \sum_{i=1}^{n} (y_i - \hat{y}_i)^2,
\end{equation} 
where $n$ denotes the number of samples, $y_i$ is the true value, and $\hat{y}_i$ is the predicted value~\cite{rumelhart1986learning,goodfellow2016deep}. Furthermore, MLP utilizes nonlinear activation functions, such as ReLU, to capture complex data features. Together, these algorithms exhibit strong capabilities in addressing the high-dimensional, nonlinear challenges inherent in materials science and design.

In the design of nonlinear responses for mechanical metamaterials, neural networks, when combined with evolutionary strategies within an optimization framework, enable the precise engineering of metamaterials with specified stress-strain behaviors~\cite{deng2022inverse}. PCA generates a stress-strain dataset consisting of 7,500 elements, effectively reducing data dimensionality to facilitate neural network training. With precise stress-strain behavior design, the relative error in the test set is as low as 4.8$\%$, supporting personalized material design for applications such as soft robotics and energy absorption systems. This method introduces innovation by controlling the nonlinear response of metamaterials through geometric parameter optimization, thereby enhancing the efficiency of nonlinear mechanical metamaterial design and expanding their potential in smart device applications.

In alloy optimization, multiple regression algorithms are employed to predict the hardening curves of boron steels, thereby improving alloy design. Using experimental hardening curve data from 62 boron steels, with input variables including chemical composition, Jominy bar distance, and material hardness as the output, a 10-fold cross-validation is applied to evaluate various regression algorithms~\cite{geng2022data}. The random forest model demonstrates the highest correlation coefficient and lowest error, facilitating the design of boron steel alloys with enhanced hardening performance. This model shows considerable potential for commercial application in software such as JMatPro~\footnote{\url{https://jmatpro.cn/}}, underscoring the significant role of machine learning in optimizing material performance.

In the field of optical metasurface solar absorber design, decision tree and random forest regressors are employed to optimize geometric parameters for spectral absorption~\cite{ijaz2023machine}. Using data from full-wave electromagnetic simulations, PCA reduces dimensionality, enhancing computational efficiency. The random forest model achieves high prediction precision for spectral absorption, with an $R^2$ value of 0.99, improving the design efficiency of solar absorbers for green energy applications and paving new pathways in optical material design. To accelerate the development of polymeric biomaterials, various machine learning models, including deep learning, random forest, and Gaussian process models, process the physical properties of polymers encoded by SMILES notation and molecular descriptors~\cite{mcdonald2023applied}. These machine learning models significantly enhance the design of novel polymer materials, optimizing physical, electrical, and rheological properties under data-scarce conditions, thereby accelerating progress in polymer biomaterials within the medical field. 

In real-time topology optimization, support vector regression (SVR) and k-nearest neighbors (KNN) models generate optimized material distributions~\cite{lei2019machine}. Through feature extraction and PCA for dimensionality reduction, simplified design variable vectors from direct optimization serve as training inputs. The model adapts structures rapidly to different external loads, supporting applications in short beam structure optimization and illustrating the potential of machine learning in real-time structural optimization for complex industrial structures.

CNN~\cite{lecun2015deep} represent a powerful class of deep learning algorithms, particularly well-suited for processing images and sequential data due to their capability to efficiently extract complex features. Widely applied in computer vision and related scientific fields, CNN perform a series of convolution operations across multiple layers, progressively abstracting high-dimensional input data into lower-dimensional representations. This hierarchical feature extraction enhances their effectiveness in tasks such as pattern recognition and predictive modeling. Mathematically, the convolution operation in each layer $l$ with filter $k$ is expressed as:
\begin{equation}
    \mathbf{f}_{ij}^{k} = \sum_{m=0}^{M-1} \sum_{n=0}^{N-1} \mathbf{W}_{mn}^{k} \cdot \mathbf{X}_{(i+m)(j+n)} + \mathbf{b}^{k},
\end{equation} 
where $\mathbf{f}_{ij}^{k}$ denotes the output feature map for filter $k$ at spatial position $\mathbf{W}_{mn}^{k}$ represents the weights in the convolutional kernel $k$ of size $\mathbf{X}_{(i+m)(j+n)}$ is the input feature map from the previous layer, and $\mathbf{b}^{k}$ is the bias term associated with filter $k$. Unlike traditional neural networks, CNN use a combination of convolutional layers, pooling layers, and fully connected layers, allowing them to detect spatial and positional relationships within structured data effectively. In materials science, CNN have become increasingly instrumental in optimizing material properties and structural design, yielding notable advancements in the topology optimization and inverse design of complex materials.

A CNN architecture based on ResNet~\cite{he2016deep} is proposed for the topology optimization of nonlinear structures, specifically targeting large-deformation hyperelastic materials such as neo-Hookean materials~\cite{abueidda2020topology}. The study generates an extensive dataset, consisting of 15,000 data pairs for linear elastic materials, 18,000 pairs for hyperelastic materials, and 20,000 pairs for linear stress materials, all aimed at optimizing material distribution to achieve maximal structural performance. By integrating the residual learning capabilities of ResNet with the U-net architecture, the model attains a Dice similarity coefficient of 0.964 in nonlinear material topology optimization, significantly improving both predictive accuracy and optimization efficiency. This method offers a robust tool for structural optimization of hyperelastic materials and highlights the potential of machine learning in addressing complex nonlinear topology optimization challenges in materials science.

A transfer learning framework integrating CNN with simplified machine learning (SML) techniques is applied to design a novel steel alloy with superior rotating bending fatigue resistance~\cite{wei2022use}. Utilizing a database of 411 samples, the study extracts static mechanical performance features and employs a transfer learning model to predict fatigue strength, leading to the design of Alloy R—an innovative alloy with fatigue strength that significantly surpasses that of existing alloys. This novel application of transfer learning effectively reduces data requirements, providing a practical solution for high-cost property predictions such as fatigue strength. Moreover, it establishes a theoretical foundation for exploring the relationship between fatigue performance and alloy design, offering valuable insights for material engineers. Additionally, a CNN-based approach is developed to predict the toughness and strength of composite materials under crack conditions, with finite element models generating data that treat composite materials as 8x8-pixel images with binary values representing different material properties (\emph{e.g.}, soft or hard)~\cite{gu2018novo}. The CNN model achieves prediction accuracy exceeding 98$\%$, enabling high-performance composite design even under data-scarce conditions. This research highlights the potential of machine learning to facilitate accurate design with minimal data, providing a foundational method for reverse engineering in composite materials, and demonstrating the effectiveness of ML in enhancing material design efficiency and predictive accuracy. 

In this section, we review the development and application of traditional machine learning algorithms in inverse design of materials. These algorithms have provided significant impetus for material discovery. Although they emerged early, this does not imply that they are outdated; different problems and scenarios require the selection of appropriate algorithms.

\subsection{Geometric Graph Neural Networks}
\label{GNNs}
The prediction of properties and structures of crystalline materials has long been a crucial task in materials science. Due to the need to account for complex symmetry constraints, crystal data are challenging to model using conventional networks like CNN. Geometric GNNs are models designed to process graph data with geometric information (\emph{e.g.}, spatial coordinates and angles) and are well-suited for studying spatial structures like molecules, proteins, and materials. Based on scientific requirements, geometric GNNs are categorized into two types. The first type, Invariant GNNs, maintains invariant outputs under euclidean transformations, making them ideal for predicting properties like band gaps, formation energies, and $T_c$, which are independent of the absolute position and orientation of the material structure. The second type, Equivariant GNNs, updates both invariant and equivariant features such that outputs change in the same way as the inputs under geometric transformations, making them particularly effective for capturing directional relationships in applications like material structure prediction and generation.

\begin{table*}[t]
\caption{Basic notations. Following Jiaqi Han~\cite{han2024surveygeometricgraphneural} notation conventions, we present the commonly used symbols in this table.}
\label{tab:notations}
\renewcommand\arraystretch{1.2}
\begin{tabular}{c p{0.8\linewidth}} % Use p{width} for the second column
    \toprule
    \textbf{Notation} & \textbf{Description} \\
    \midrule
    
    $\gG \coloneqq (\mA, \mH)$ & Represents a graph with $N$ nodes, characterized by its adjacency matrix $\mA$ and node feature matrix $\mH$. \\
    
    $\vec{\gG} \coloneqq (\mA, \mH, \vec{\mX})$ & A geometric graph incorporating a 3D coordinate matrix $\vec{\mX}$ in addition to $\mA$ and $\mH$. \\
    
    $\gN_i$ & The set of nodes neighboring node $i$. \\
    
    $\vh_i \in \mathbb{R}^{C_h}$ & The feature of node $i$, containing $C_h$ attributes. \\
    $\vu$ & The global state vector \\
    $\bigoplus$ &  This denotes an aggregation of neighboring node features, like sum, mean, or max.\\
    
    $\vec{\vx}_i \in \mathbb{R}^3$ & The 3D spatial coordinates of node $i$. \\
    
    $\vec{\mV}_i \in \mathbb{R}^{3 \times C}$ & The multi-channel 3D vector representing node $i$. \\
    
    $\vec{\mV}_i^{(l)} \in \mathbb{R}^{(2l+1) \times C_l}$ & The type-$l$ irreducible vector associated with node $i$. \\
    
    $\ve_{ij} \in \mathbb{R}^{C_e}$ & The edge feature vector from node $j$ to node $i$. \\
    
    $G, g$ & The group $G$ and its element $g$, relevant to transformations in the graph. \\

    $\times, \otimes$ & Operators for vector operations: cross product $\times$ and Kronecker product $\otimes$. \\
    
    $\otimes_{\text{cg}}$ & Clebsch-Gordan tensor products. \\
    
    $Y^{(l)}(\vec{\vx}) \in \mathbb{R}^{2l+1}$ & The type-$l$ spherical harmonic vector evaluated at point $\vec{\vx}$. \\
    
    $\mD^{(l)}(g)$ & The $l$-th degree Wigner-D matrix for rotation $g \in \mathrm{SO}(3)$. \\
    
    $\phi, \psi, \varphi, \sigma$ & Functions realized through MLP. \\   

    $\gL$ & Loss function.\\

    $\mW$ & Weight matrix.\\
    
    \bottomrule
\end{tabular}
\end{table*}

\vpara{Graph Neural Networks.}
GNNs differ from traditional neural networks in that they operate on graph-structured data rather than tensor-structured data. A graph $\gG \coloneqq (\mA, \mH)$ represents a graph with $N$ nodes, characterized by its adjacency matrix $\mA$ and node feature matrix $\mH$. The core idea of GNNs is message passing, where, as the graph propagates forward, the features stored at node \( i \), denoted as \( \vh_i \), are updated based on information from its neighborhood
\begin{align}
    \vh_i' = 
    \sigma_{upd}
    \left( 
        \vh_i,~
        \bigoplus_{j\in\gN_i} \vh_j
    \right).
\end{align}
The equation illustrates how the feature \(\vh_i'\) of node \(i\) is updated through a message passing mechanism. In this framework, the function \(\sigma_{upd}\), which is typically a nonlinear activation function, takes as inputs both the original feature \(\vh_i\) and an aggregated summary of the features \(\vh_j\) from neighboring nodes \(j\) within the set \(\gN_i\). The aggregation operator \(\bigoplus\) can encompass operations such as summation, averaging, or taking the maximum. This methodology enables node \(i\) to effectively incorporate information from its local neighborhood, thereby enriching its updated feature representation.
This mirrors the behavior of physical systems, where long-range correlations are governed by local interactions. Specifically, when features include spatial information—typically represented as three-dimensional vectors—the networks are categorized as geometric GNNs, denoted as \(\vec{\gG} \coloneqq (\mA, \mH, \vec{\mX})\), where a geometric graph incorporates a 3D coordinate matrix \(\vec{\mX}\) in addition to \(\mA\) and \(\mH\).

To establish a framework for categorizing various types of geometric GNNs, we first present the message passing neural network (MPNN)~\cite{gilmerneural2017}. In this framework, features $h_i$ are associated with nodes and are iteratively updated at each hidden layer
\begin{align}
    \vh_i' = 
    \sigma_{upd}
    \left( 
        \vh_i,~
        \bigoplus_{j\in\gN_i} \vm_{ij}
    \right),
\end{align}
while messages $m_{ij}$ are stored on the edges, serving as update buffers for the connected nodes.
\begin{align}
    \vm_{ij}' = 
    \sigma_{msg}
    \left( 
        \vh_i,~\vh_j,~
        \bigoplus_{(kl)\in\gN_{(ij)}} \vm_{kl},~
    \right).
\end{align}
Messages guide the feature updates and propagate along the edges, thereby reinforcing the locality principle in GNNs. 

\vpara{Invariant Geometric GNNs.}
Invariant geometric GNNs can be classified according to the scalar geometric quantities they utilize for message passing, such as pairwise distances, triplet-wise angles, and quadruplet-wise torsion angles. In the following, we will introduce one model from each category as an example.

SchNet~\cite{schuttschnet2018} is one of the earliest invariant GNNs models, developed as a variant of the deep tensor neural network~\cite{schuttquantumchemical2017} for modeling atomic structures. In SchNet, continuous filters are employed to capture interaction terms based on the pairwise distances between atoms
\begin{align}
    \vh_{i}' = \vh_{i} + \sum_{j \in \gN_i} \vm_{ij}' ~ \vh_{j},
\end{align}
\begin{align}
    \vm_{ij}' = \mW(\| \vec{\vx}_i-\vec{\vx}_j \|),
\end{align} 
where \( \mW \) denotes the learnable filter. The filter employs Gaussians as radial basis functions in its first layer to encode pairwise distances.
\begin{align}
    e_{k}(\| \vec{\vx}_i-\vec{\vx}_j \|) = 
    \text{\rm{exp}}(-\gamma (\| \vec{\vx}_i-\vec{\vx}_j \| - \mu_k)^2).
\end{align}
Notably, later studies~\cite{huforcenet2021} have shown that spherical Bessel functions serve as more effective basis.
Other distance-based models include CGCNN~\cite{xiecrystal2018} and PhysNet~\cite{unkephysnet2019}. While simple and effective, these models are limited in that they cannot distinguish between molecular structures that have identical pairwise distances but differ in bond angles.

DimeNet~\cite{gasteigerdirectional2020} goes to triplet information by introducing directional message passing 
\begin{align}
    \vm_{ji}' = \sigma_{msg}
    \left(
        \vm_{ji},~
        \bigoplus_{j \in \gN_j \backslash \{i\}} \vm_{kj}
    \right),
\end{align}
where $\vm_{ji}$ denotes a directional message passed from node $j$ to node $i$, which is updated by iterating over all messages directed toward node $j$, excluding those from node $i$. The angle information $\angle ijk$ can be computed, enabling the differentiation of structures with varying bond angles. While DimeNet represents an advancement over SchNet by capturing angular information, it still cannot resolve torsions or distinguish chiral structures.

SphereNet~\cite{liuspherical2022} also applies a directional message passing approach, but with further refinement. It introduces ``spherical message passing'' using a local spherical coordinate system. The messages encode not only polar angles, such as $\angle ijk$ and $\angle ijk'$, but also the azimuthal angle between $\vec{\vx}_k-\vec{\vx}_j$ and $\vec{\vx}_{k'}-\vec{\vx}_j$. This allows SphereNet to capture complete geometric information around node \(j\), including quadruplet torsional angles. Consequently, SphereNet can distinguish chiral structures, achieving SE(3)-invariant, whereas prior models were limited to E(3)-invariance. Although there still are failure cases where SphereNet cannot distinguish certain configurations, such scenarios are theoretically possible but highly improbable in natural settings. Later research~\cite{yueplugandplay2024} proposed quaternion message passing, which encodes these geometric features more efficiently. Other models in this category include GemNet~\cite{klicperagemnet2024} and ComENet~\cite{wangcomenet2022}.

\vpara{Equivariant Geometric GNNs.}
Geometric GNNs achieves equivariance in two main ways: one approach achieves scalarization-based models through inner product operators, while the other utilizes group representation theory, spherical harmonics, and tensor product methods. Finally, we introduce several works that integrate attention mechanisms into Geometric GNNs models, significantly enhancing their expressive power.

To achieve equivariance, there is still much work~\cite{egnn,schutt2017schnet,kohler2019equivariant} implemented a scalarization-based models approach, with equivariant graph neural networks (EGNNs)~\cite{egnn} being the most prominent. This model does not require costly higher-order representations; instead, it first converts the 3D coordinates into invariant scalars, specifically squared distances $\|\vec{\vx}_i-\vec{\vx}_j  \|^2$, and then uses this invariant information, node representations, and edge attributes to perform edge message passing:
\begin{align}
\label{eq:egnn_scalar_message}
    \vm_{ij} &= \sigma_1\left(\vh_i,\vh_j, \|\vec{\vx}_i-\vec{\vx}_j  \|^2, \ve_{ij}\right),
\end{align}
where $\ve_{ij}$ represents the edge attributes between the $i$-th and $j$-th nodes, and $\vh_i$ denotes the node representations of the $i$-th node. Geometric messages can be expressed as:
\begin{align}
    \label{eq:egnn_vector_message}
    \vec{\vm}_{ij} &= (\vec{\vx}_i-\vec{\vx}_j)\sigma_2\left(\vm_{ij}\right),
\end{align}
where $\vec{\vx}_i$ and $\vec{\vx}_j$ represent the coordinates of the $i$-th and $j$-th nodes, respectively. Node features are updated by passing invariant messages:
\begin{align}
\label{eq:egnn_scalar}
    \vh'_i &= \sigma_3\left(\vh_i,  \sum\nolimits_{j\in\gN_i}\vm_{ij} \right),
\end{align}
the position of each atom is then updated using a vector field along the radial direction; in other words, each atom’s position is updated by a weighted sum of all relative vector differences:
\begin{align}
    \label{eq:egnn_vector_update}
    \vec{\vx}'_i &= \vec{\vx}_i + \gamma\sum\nolimits_{j\in\gN_i}\vec{\vm}_{ij},
\end{align}
where $\sigma_1, \sigma_2, \sigma_3$ are multi layer perceptrons, and $\gamma$ is a predefined constant equal to 1/(M-1), where M is the number of atoms.

In addition to coordinates, a node's vector information can include attributes like velocity and acceleration. GMN~\cite{huang2022equivariant} introduced a multi-channel vector representation $\vec{\mV}_i \in \mathbb{R}^{3 \times C}$ for each node to capture these attributes. Prior to message passing, these vectors undergo a scalarization process: $ \frac{\vec{\mV}_{ij}^\top \vec{\mV}_{ij}}{\|\vec{\mV}_{ij}^\top \vec{\mV}_{ij}\|_F} $, where $\vec{\mV}_{ij} = \vec{\mV}_i - \vec{\mV}_j$. This subtraction ensures that $\vec{\mV}_{ij}$ remains translation-invariant. Many studies, such as ClofNet~\cite{du2022se}, achieve scalarization and preserve equivariance by constructing local frames. Other works, like those in~\cite{kofinas2021rototranslated, kofinas2023latent}, follow similar approaches. Specifically, the message passing is conducted as follows:
\begin{align}
    \vm_{ij}&=\sigma_1\left(\vh_i,\vh_j, \vec\mV_{ij}^\top\vec\mF_{ij}\right),
\end{align}
where $\vec\mF_{ij}$ is translation-invariant~\cite{han2024surveygeometricgraphneural}. Many other GNNs achieve scalarization through inner product operators while maintaining equivariance, such as GVP-GNN~\cite{gvpgnn}, LEFTNet~\cite{du2024new}, Frame Averaging~\cite{puny2021frame}, and EGHN~\cite{han2022equivariant}.

By combining the Wigner-D matrix $\mD^{(l)}(g)$~\cite{TFN2018}, spherical harmonics $Y^{(l)}(\vec{\vx}) \in \mathbb{R}^{2l+1}$, and the clebsch-gordan (CG) tensor product $\otimes_{\text{cg}}$, e3nn~\cite{e3nn} and TFN~\cite{TFN2018} offer a neural network approach capable of handling 3D data while preserving equivariance to rotations, translations, and inversions. This integration enables the network to learn richer and more robust feature representations. The Wigner-D matrix represents the irreducible representations of the SO(3) group (the 3D rotation group) and describes the rotational behavior of angular momentum in quantum mechanics. Wigner-D matrices ensure that network layers maintain equivariance under rotation. Spherical harmonics, functions defined on the unit sphere that are equivariant under SO(3), can be extended to the entire $\mathbb{R}^3$ space and are used to construct equivariant polynomials. In e3nn, spherical harmonics are applied to build equivariant convolutional layers that can process point cloud data in 3D space. Their equivariance enables the network to learn features that are invariant to rotation. The CG tensor product combines two irreducible representations of angular momentum into a new irreducible representation that retains equivariance. The e3nn framework leverages the CG tensor product to combine outputs from different layers and construct complex equivariant operations, such as convolutions and attention mechanisms. The CG tensor product ensures that these operations remain equivariant under rotation, allowing learned patterns to be invariant to transformations in 3D space. By employing higher-order representations ($l > 1$), these networks can capture complex 3D spatial relationships, which traditional convolutional neural networks struggle to achieve. 

Steerable E(3) equivariant graph neural networks (SEGNNs)~\cite{segnns} can handle geometric and physical information contained in node and edge attributes, such as position, force, velocity, or spin. SEGNNs integrate this information into the message passing and node updating functions and introduce a new class of equivariant activation functions, based on manipulable node attributes and manipulable multi-layer perceptrons , allowing for the injection of geometric and physical information into the node updates. Neural equivariant interatomic potentials (NequIP)~\cite{NequIP} use a similar approach to learn interatomic potentials from ab initio calculations. NequIP provides significant improvements in the accuracy and efficiency of molecular dynamics simulations, facilitating the application of this method across a broader range of research fields. In summary, there is still much work to be done on neural networks based on the integration of Wigner-D matrices, spherical harmonics, and CG Tensor Products, such as DimeNet~\cite{DimeNet}, SCN~\cite{scn2022}, eSCN~\cite{eSCN}, MACE~\cite{MACE2023}, PaiNN~\cite{PaiNN}, and Allegro~\cite{Allegro}.

Transformers have demonstrated remarkable capabilities in the field of large language models, primarily due to their attention mechanism. A natural consideration is to incorporate this attention mechanism into graph neural networks. Metaproposed the equivariant graph attention transformer for 3D atomistic graphs (Equiformer)~\cite{equiformerV1}, which combines the Transformer architecture with SE(3)/E(3) equivariant features based on irreducible representations, introducing an equivariant graph attention mechanism. By replacing traditional dot product attention with multi-layer perceptron attention and integrating nonlinear message passing, this approach enhances the attention expressiveness of transformer. The network employs equivariant operations, such as independent linear transformations, layer normalization, and depth-wise tensor products, to handle different types of vectors while processing 3D graph structures through embedding layers and Transformer blocks. Meta further introduced EquiformerV2~\cite{equiformerV2}, an advancement over Equiformer, with key architectural improvements to enhance stability, computational efficiency, and model expressivity. EquiformerV2 incorporates an additional normalization layer preceding attention weight computation to stabilize the training process, and it refines the nonlinear activation function to more effectively handle representations across different orders. This approach uses specialized activation functions for each order and independently normalizes vectors to preserve their relative importance. Leveraging eSCN convolutions, EquiformerV2 simplifies the computation of SO(3) tensor products by transforming them into SO(2) linear operations, significantly reducing computational complexity. These innovations enable EquiformerV2 to achieve improved accuracy and efficiency on large-scale and complex 3D atomic graph datasets, particularly through the effective utilization of higher-order irreducible representations.

A prominent method integrating attention mechanisms with GNNs is Graphormer~\cite{graphoformer}, a GNNs built on the standard Transformer architecture. Graphormer excels in a range of graph representation learning tasks by proficiently encoding the structural information of graphs. It introduces several straightforward yet effective structural encodings—namely, centrality encoding, spatial encoding, and edge encoding—which enhance its ability to model graph-structured data. Theoretical analysis further underscores Graphormer’s expressive power, proving that numerous popular GNNs variants can be viewed as special cases within its overarching framework. There are many similar approaches that combine attention mechanisms with GNNs, such as SE(3)-Transformer~\cite{fuchs2020se}, TorchMD-Net~\cite{TorchMD-NET}, and LieTransformer~\cite{hutchinson2021lietransformer}.

\subsection{Discriminative AI}
\label{ALIGNN}
\vpara{Introduce to Discriminative AI.}
A common task in inverse design of materials involves leveraging AI to predict the physico-chemical properties of materials, enabling high-throughput screening based on these properties. Discriminative models are primarily employed to directly predict the category (\emph{e.g.}, superconducting material, magnetic material) or label (\emph{e.g.}, formation energy, \(T_c\)) associated with a given input (\emph{e.g.}, crystal structure, chemical composition). These models focus on learning how to map input data to the corresponding output labels. The objective of discriminative models is to infer the conditional probability distribution \(P(y \mid x)\) of the target variable \(y\), given the input features \(x\).

\vpara{CGCNN.} 
The crystal graph convolutional neural network (CGCNN)~\cite{xie2018crystal} is one of the earliest tools to use graph neural networks for predicting material properties. Compared to DFT calculations, CGCNN achieves comparable or slightly better accuracy in predicting properties such as formation energy, band gap, Fermi level, bulk modulus, shear modulus, and Poisson's ratio, with significantly faster computational speed. By analyzing the energy of each site in perovskite structures, CGCNN aids in the discovery of new stable perovskite materials, significantly reducing the search space for high-throughput screening.

Specifically, CGCNN uses an undirected multilateral graph to represent atoms and chemical bonds as nodes and edges, respectively. In this graph, the \emph{i}-th node is represented by the feature vector $\vh_i$, which corresponds to the properties of the \emph{i}-th atom. Each edge~(\emph{i},\emph{j})$_k$ is represented by the feature vector $\vm_{(i,j)_k}$, corresponding to the \emph{k}-th bond between atoms \emph{i} and \emph{j}. Atoms are considered to be connected if the distance between them does not exceed 6Å. For discrete values, such as atomic numbers, a unique one-hot vector is used for direct embedding. For continuous values, such as bond lengths, the value range is divided into 10 intervals, and the value is encoded using a one-hot vector.

In the model, two types of neural networks are employed: graph convolutional networks (GCN) and fully connected networks (FCN). Convolutional layers are used to aggregate information from the neighboring atoms around each atom,
\begin{align}
    \vh_i^{(t+1)f} = \text{Conv}\left(\vh_i^{(t)},\vh_j^{(t)},\vm_{(i,j)_k}\right),(i,j)_k\in\gG,
\end{align}
where \(\vh_i^{(T)}\) represents the feature vector of node \(i\) after \(T\) convolutions. The pooling layer uses normalized summation as the pooling function to generate a feature vector representing the entire graph. Finally, the target property \(\hat{y}\) is predicted through two fully connected layers. A loss function \(J(y, \hat{y})\) is defined, and the model parameters are updated by minimizing this loss function,
\begin{align}
    \min_{\mW} J(y,f(\gC;\mW)),
\end{align}
where the $\mW$ represent weights of the CGCNN, $f$ is a function means that maps the crystal $\gC$ to the target property $\hat{y}$.For more specific details, please refer to the original paper~\cite{xie2018crystal}.

\vpara{MEGNet.}
The materials graph network (MEGNet)~\cite{chen2019graph} is a general material graph network model capable of accurately predicting the properties of both molecules and crystals. The model outperforms previous machine learning models in predicting various properties. Due to the lack of embedding the entire crystal structure in earlier models, MEGNet introduces global state attributes to the graph representation, including temperature, pressure, and entropy. MEGNet is trained on the QM9~\cite{ramakrishnan2014quantum} molecular dataset and the Materials Project (MP) crystal dataset. On the 13 properties in QM9, MEGNet outperforms previous models in predicting 11 of them. On the MP dataset, which includes 60,000 crystals, MEGNet not only outperforms earlier ML models but also achieves higher accuracy than DFT on a larger dataset.

In the model, the feature vector of the edge is updated by applying a specific update rule that incorporates information from both the node features and the edge's previous state. This process ensures that the edge representation evolves through successive layers, capturing the interactions between atoms in the crystal structure
\begin{align}
    \vm_k' = \phi_e\left(\vh_{i}\oplus\vh_{j}\oplus\vm_{ij}\oplus\vu\right),
\end{align}
where $\phi_e$ is the bond update function, $\oplus$ is the concatenation operator, $\vh_i$ and $\vh_j$ are the \emph{i}-th atom and \emph{j}-th atom corresponding feature vector respectively, and $\vu$ is the global state vector. Next, the feature vector of each atom is updated by
\begin{align}
    \overline{\vh}_i^e = \frac{1}{N_i^e}\sum_{k=1}^{N_i^e}\{\vm_{ij}'\}, \\
    \vh_i' = \phi_v\left(\overline{\vh}_i^e\oplus\vh_i\oplus\vu\right),
\end{align}
where $N_i^e$ is the degree of atom i, and $\phi_v$ is the atom update function. Finally, the global state attribute is updated,
\begin{align}
    \overline{\vu}^e = \frac{1}{N^e}\sum_{k=1}^{N^e}\{\vm_{ij}'\},\\
    \overline{\vu}^v = \frac{1}{N^v}\sum_{i=1}^{N^v}\{\vh_i'\},\\
    \vu' = \phi_u\left(\overline{\vu}^e\oplus\overline{u}^v\oplus\vu\right),
\end{align}
where $\phi_u$ is the global state update function. 
The model combines two layers of FCN and MEGNet modules into a large MEGNet block. Multiple MEGNet blocks are then connected through a residual network, enabling the model to have deep hidden layers.

\vpara{ALIGNN.}
The atomistic line graph neural network (ALIGNN)~\cite{choudhary2021atomistic} introduces an innovative GNN architecture that alternates message passing between atomic bond graphs and line graphs (capturing bond angles) to explicitly incorporate critical geometric information often overlooked by traditional GNN models. Compared to models solely based on interatomic distances, ALIGNN significantly improves the accuracy of material property predictions. Its dual-graph message-passing mechanism, combined with edge-gated convolution, efficiently updates both node and edge representations. Tests on datasets like JARVIS-DFT, Materials Project, and QM9 demonstrate that ALIGNN not only outperforms existing GNN models (such as CGCNN, MEGNet, and SchNet) in predicting properties like formation energy and band gaps but also maintains high computational efficiency and robust generalization. By incorporating bond angle information into the GNN architecture, this work provides a powerful tool for accelerating materials design and discovery, with its open-source code and datasets further promoting advancements in materials science research.

In ALIGNN models, the node feature embedding and convolution design is similar to CGCNN, but the feature vector is updated using the edge-gate graph convolution. 
\begin{align}
    \hat{\vm}_{ij}=\frac{\sigma(\vm_{ij})}{\sum_{k\in \gN_i}\sigma(\vm_{ik})+\epsilon},
\end{align}
\begin{align}
    \vh_i'=\vh_i+\text{SiLU}\left(\text{Norm}\left(\mW_s\vh_i+\sum_{j\in \gN_i}\hat{\vm}_{i,j}\mW_d\vh_j\right)\right),
\end{align}
\begin{align}
\vm_{ij}'=\vm_{ij}+\text{SiLU}\left(\text{Norm}\left(\mA\vh_i+\mB\vh_j+\mC\vm_{ij}\right)\right),
\label{gatedGN}
\end{align}
where $\mW_s$, $\mW_d$, $\mA$, $\mB$ and $\mC$ are learnable parameters, $\sigma$ is sigmoid function, SiLU is sigmoid linear unit, and Eq.~\ref{gatedGN} is equivalent to gating term in CGCNN. 

An ALIGNN layer consists of edge-gated graph convolutions on both the bond graph (\(g\)) and its line graph (\(L(g)\)). To avoid confusion between the node and edge features in the atomistic graph and its line graph, we denote atoms, bonds (nodes in the line graph), and triplets (angles between bonds) as \(h\), \(m\), and \(t\), respectively. The line graph convolution generates bond messages \(m\), which are propagated to the atomistic graph and combined with atomic features \(h\) to update bond representations,
\begin{align}
    H',\vt' = \text{EdgeGatedGraphConv}\left(L(g),\vm,\vt\right),\\
    \vh',\vm' = \text{EdgeGatedGraphConv}\left(g,\vh,H'\right).
\end{align}

\vpara{OGCNN.}
The orbital graph convolutional neural network (OGCNN) is a crystal graph convolutional network that incorporates atomic orbital interaction features~\cite{karamad2020orbital}. Beyond accounting for orbital contributions, the authors introduce a technique known as the orbital field matrix (OFM)~\cite{lam2017machine}, which captures orbital interactions by leveraging the electron configurations of both the central atom and its neighboring atoms. The study evaluates various properties, including formation energy, band gap, and Fermi energy. Results indicate that the model achieves improved predictive accuracy over CGCNN, underscoring the critical role of orbital-orbital interactions. The OFM is defined as
\begin{equation}
    \mX^{c}={\bm{O}^{c}}^{T}+\sum_{n=1}^{M}{\bm{O}^{c}}^{T}{\bm{O}^{n}}\theta_{cn}\zeta(r_{cn})
\end{equation}
where $\mX^{c}$ represents the OFM for the central atom. ${\bm{O}^{c}}$ and ${\bm{O}^{n}}$ represent the 1D binary vectors for the central atom and neighboring atoms. $\mM$ denotes the number of neighboring atoms. $\theta_{cn}$ denotes the solid angle between center-neighbor pairs in the Voronoi cell and $\zeta(r_{cn})$ denotes the function related to the distance between central and neighboring atoms, which can be selected as required.
The total framework contains four modules: input module, encoder-decoder modules, graph convolution module and output module. The input module processes the atom configuration and OFM features, and output vectors containing information about basic atoms and OFM. The purpose of the second module is to learn significant information among atoms by a MLP. The operation in the graph convolution module can be expressed as 
\begin{equation}
    \begin{aligned}
        \bm{V}_{i}'=\bm{V}_{i}+&\sum\sigma(\bm{z}_{(i,j)_{k}}\bm{W}_{f}+\bm{b}_{f})\\&\odot g(\bm{z}_{(i,j)_{k}}\bm{W}_{s}+\bm{b}_{s})
    \end{aligned}
\end{equation}
where $\bm{z}_{(i,j)}=\bm{V}_{i}\oplus \bm{V}_{j}\oplus \bm{u}_{(i,j)_{k}}$, $\bm{u}_{(i,j)_{k}}$ contains the information of kernelized distance features, $\sigma$ denotes a sigmoid function and $g$ denotes the softplus function. The OGCNN architecture uses a total of three convolutional operations and what follows is a summation operation. finally, in the output module, a pooling layer and a fully connected network are implemented to map to the desired result.

\vpara{ECN.}
The equivariant crystal network (ECN)~\cite{kaba2022equivariant} incorporates spatial group symmetries into Graph Neural Networks (GNNs). ECN emphasizes that symmetry invariance is a necessary condition for practical applications. The authors experiment using the Materials Project dataset, filtering it to retain only data relevant to three-dimensional materials. They then predict physical properties such as formation energy, Fermi energy, band gap, and magnetic moment per atom. The authors frame their work as both regression and classification tasks. Experimental results show improved performance compared to previous models, supporting the authors' claim that symmetry provides a beneficial inductive bias.

In practice, using models that are equivariant to the actual symmetry of the data can be more beneficial in terms of the model's expressiveness. This is because the corresponding group is significantly smaller than the symmetric group, and thus, equivariance to a smaller group may reduce parameter sharing, making the model more expressive. Moreover, the author emphasizes the importance of equivariance, noting that invariant functions can be constructed by combining equivariant layers with an output pooling layer. Additionally, equivariant functions can be used to predict local properties, such as charge distribution and magnetization.

The construction of an ECN is based on two key principles. The first challenge is the difficulty of supervised learning when a dataset encompasses a diverse range of crystal structures for a specific species. To address the discrepancies inherent in the unit cell structures, the researchers introduce the direct product of groups. The second principle in constructing the ECN is the equivariant message-passing framework. The authors define the layer based on the update equation within this message-passing framework.

The aforementioned section outlines the fundamental structure of the ECN. The model takes as input a graph representing the crystal structure and generates a single encoded feature vector for each atom. This is followed by the integration of hidden layers and average pooling layers, culminating in a two-layer MLP for output prediction.

\vpara{Matformer.}
Matformer~\cite{yan2022periodic} is specifically designed for periodic graph representation learning to predict the properties of crystalline materials. It computes various physical quantities, such as formation energy, band gap, and total energy, across multiple benchmark datasets, surpassing existing baseline methods such as ALIGNN~\cite{choudhary2021atomistic}, SchNet~\cite{schutt2017schnet}, and CGCNN~\cite{xie2018crystal}, demonstrating notable performance improvements. The model achieves periodic invariance through a specialized graph construction method, ensuring that the learned representations remain invariant to translations of the unit cell boundaries. Matformer effectively encodes periodic patterns, capturing the lattice size and orientation of the crystal. Additionally, Matformer employs an attention-based architecture, integrating edge attention and geometric information encoding, thereby enhancing its capability to process multi-atom crystal graphs. 
Experimental results indicate that Matformer outperforms baseline methods on the Materials Project and JARVIS datasets, offering advantages in both training and inference speed. By incorporating periodic orbital interaction features alongside fundamental atomic features, Matformer significantly enhances the accuracy of material property predictions, showing promising potential for applications, particularly in the field of materials discovery.

Before introducing periodic invariance, it is advisable to take a look at unit cell E(3) invariance. Unit cell E(3) invariance can be seen as a function $f:(\mA,\mX,\mL)\to\chi$, where the atom feature matrix~$\mA$, and the position matrix~$\mX$, are to describe unit cell itself, and $\mL$ is to describe how a unit cell repeat itself in different directions. Thus it is straightforward to obtain from the definition that no matter if you rotate, reflect, or translate the cell, the structure of the cell itself remains unchanged. After taking into account the periodic invariance, the authors define that unit cell E(3) function $f$ is periodic invariance if $f(\mA,\mX,\mL)=f(\phi(\hat{\mA},\hat{\mX},\alpha\mL,p), \alpha\mL)$ holds for all $p\in\mathbb{R}^{3}$ and $\alpha\in\mathbb{N}_{+}^{3}$, where $\phi$ is a function that $\phi:(\hat{\mA},\hat{\mX},\alpha\mL,p)\to(\mA,\mX)$. The authors also emphasize the importance of disrupting period invariance, namely for the same crystal this would lead to different crystal graphs. And periodic pattern encoding is expressed by $\mL$ to better represent infinite structures of crystals. 

Some researchers use both multi-edge graph construction and fully-connected graph construction, which both satisfy the periodic invariance to build their Matformer~\cite{xie2018crystal}. And the radius-based method is used for better performance in experiments. Then they add self-connecting edges to encode periodic pattern, namely $\mL$. For a direction vector~$l_{i}$ of~$\mL=[l_{1},l_{2},l_{3}]^{T}\in \mathbb{R}^{3\times3}$, it obviously contains two parts, which is the length $\Vert l_{i}\Vert_{2}$ and the orientation. Thus it is necessary to encode the orientation. The authors utilize additional distances to encode angles between two direction vectors. For example, the angle between $l_{1}$ and $l_{2}$ can be derived from $\Vert l_{1}\Vert_{2}$, $\Vert l_{2}\Vert_{2}$ and an additional distances $\Vert l_{1}+l_{2}\Vert_{2}$.  

In addition, the authors discuss the impact of the introduction of angular information. The utilization of angular information in practical tests does not seem to give much hints about accuracy, but it triples the time cost. This result may be due to that periodic invariant graph construction and periodic
patterns encoding in Matformer already give enough information for the tests.

\vpara{CT.}
Crystal twins (CT) is a self-supervised learning framework capable of learning from large unlabeled datasets~\cite{magar2022crystal}. This work represents one of the first applications of self-supervised learning methods to the prediction of crystalline properties, whereas prior to this, self-supervised learning was primarily applied to molecular systems. The authors conduct experiments on 14 datasets to evaluate the performance of the CT model and calculate various properties, such as exfoliation energy, band gap, and formation energy. The CT model demonstrates strong performance in most cases, although its results may not always be the best. Additionally, the authors assess the effectiveness of augmentation methods and argue that the use of all three techniques enhances the overall effectiveness of the experiments.

The authors propose a set of main processes. Firstly a CGCNN is used in pre-training to be an encoder to learn the effective representation of the crystal system, here the authors propose two different schemes based on Barlow Twins~\cite{zbontar2021barlow} and SimSiamese~\cite{chen2020simple} loss functions. After this, the weights are shared to initialize the encoder and fine-tuning for downstream tasks will be done with the help of labeled data. In this work, the authors also introduce three augmentation methods, namely random perturbations, atom masking and edge masking.

We focus on the description of two pre-training frameworks. In CT$_{Barlow}$, the CGCNN encoder would generate representations of augmented instances from same crystal systems. And the goal of pre-training here is to align the cross-correlation matrix of the two embeddings as closely as possible with the identity matrix. The Barlow Twins loss function can be defined as 
\begin{equation}
    \gL_{BT}\triangleq\sum_{i}(1-C_{ii})^{2}+\lambda\sum_{i}\sum_{j\neq i}C_{ij}^{2},
\end{equation}
where $\textbf{\textit{C}}$ is the cross-correlation matrix, as
\begin{equation}
    C_{ij}\triangleq\frac{\sum_{b}\mZ_{b,i}^{A}\mZ_{b,j}^{B}}{\sqrt{(\mZ_{b,i}^{A})^2}\sqrt{(\mZ_{b,j}^{B})^2}}.  \label{LossBarlowMatrix}
\end{equation}
In equation~(\ref{LossBarlowMatrix}), $b$ represents the index of the data within the batch and $A$ and $B$ both denote the augmented instances, while $i$ and $j$ represents the index of the vector dimensions of the projector output, namely $\mZ^{A}$ and $\mZ^{B}$.
Another model is pretrained using the SimSiamese loss function with the goal of maximizing the cosine similarity between the embeddings generated by the encoder for two augmented instances. Furthermore, in CT$_{SimSiam}$, one branch has a stop-gradient operation while the other branch has a prediction header after the graph encoder. The loss function in this case can be defined as 
\begin{equation}
    \begin{aligned}
        \gL_{SimSiam}&\triangleq\frac{1}{2}\sum_{b}(\mathcal{D}(\mP_{b}^{A},stopgrad(\mZ_{b}^{B}))\\
        &+\mathcal{D}(\mZ_{b}^{B},stopgrad(\mZ_{b}^{A}))),
    \end{aligned}
\end{equation}
where $stoopgrad$ represents disabling back propagation of gradients. And the distance of two vectors can be defined as 
\begin{equation}
    \mathcal{D}(\mP_{b}^{A},\mP_{b}^{B})=-\frac{\mP_{b}^{A}}{\Vert\mP_{b}^{A}\Vert_{2}}\cdot\frac{\mZ_{b}^{B}}{\Vert\mZ_{b}^{B}\Vert_{2}}.
\end{equation}

\vpara{MMPT.}
Mutex masked pre-training (MMPT) is a self-supervised pre-training framework designed to address challenges that hinder the prediction of crystal properties, particularly the limited availability of labeled crystal data and the constraints of quantum chemistry~\cite{yu2023crystal}. The masking techniques employed in MMPT are inspired by the BERT framework~\cite{devlin2018bert}. The authors conduct their experiments using a subset of the Open Quantum Materials Database (OQMD)~\cite{kirklin2015open}. Compared to other supervised and self-supervised methods, MMPT demonstrates strong performance. The experimental results indicate that MMPT alleviates, to some extent, the issue of limited labeled data. Furthermore, the authors assert that their results outperform those of the Crystal Twins method, attributing this improvement to the effective utilization of E(3) invariance and the periodic invariance of crystals.

It is helpful to introduce the main framework of MMPT. A crystal is described using three vectors~$\gM=(\mA,\mX,\mL)$. First of all there are two encoders here, one is lattice encoder and the other is structure encoder. At the beginning, the lattice encoder encodes the lattice matrix~$\mL$ into a lattice representation~$\vh_{L}$ which contains the information of periodicity. The structure encoder first encodes $\gM$ into a structure representation~$\vh_{s}$ with the help of a periodic invariance multi-graph~(PIMG) module, and then two mutex representations $\vg_{s}$ and $\bar{\vg}_{s}$ are formed from $\vh_{S}$ by mutex masking. The next step is to decode. The lattice decoder decodes $\vh_{L}$ into $\vp_{L}$, and the coordinate decoder and the atom decoder decodes $\vg_{s}$,  $\bar{\vg}_{s}$ into $\vp_{C}$, $\bar{\vp}_{C}$ and $\vp_{A}$, $\bar{\vp}_{A}$, respectively. Finally, crystal reconstruction is done by $(\vh_{L},\vp_{C},\vp_{A})$ and $(\vh_{L},\bar{\vp}_{C},\bar{\vp}_{A})$. And $\vp_{A}$ and $\bar{\vp}_{A}$ are used for atom-type contrastive learning, which emphasizes the role of the species of the atoms, and the periodic attribute learning~(PAL), which can guarantee the introduction of the periodicity property.

Metux masking is a spectial point of this work. In a nutshell, $\vg_{s}$ can be defined as 
\begin{equation}
     \vg_{s}^{i}=\left\{
    \begin{aligned}
        &\vh_{s}^{i} & &{i\notin \mathcal{M}} \\
        &[MASK] & &  {i\in \mathcal{M}}
    \end{aligned}
    \right.,
\end{equation}
where $\mathcal{M}$ is a uniform distribution. And $\bar{\textbf{g}}_{s}^{i}$ can be derived from the mutex mask of $\mathcal{M}$, namely $\bar{\mathcal{M}}$. This mask process allows the model to learn through structural relationships between two complementary sets of atoms. And based on the mutex masks, the authors design crystal reconstruction and atom-type contrastive learning. The authors implement crystal reconstruction by optimizing the loss function $\mathcal{L}_{REC}$. 

Since the crystal is described by three vectors, it is natural that $\mathcal{L}_{REC}$ is also divided into three parts, namely $\mathcal{L}_{A}$, $\mathcal{L}_{X}$ and $\mathcal{L}_{L}$. $\mathcal{L}_{A}$ corresponds to $\mA$, which is the cross entropy between real atom types and predicted atom types. $\mathcal{L}_{X}$ corresponds to $\mX$, which is based on the distance from the coordinate of each atom to the center coordinate. $\mathcal{L}_{L}$ corresponds to $\mL$.

Due to the influence of atom types for chemical properties, the learning of atom types is also referenced in this work. The authors utilize the Barlow Twins loss to consider the significance of atom types. They align cross-correlation matrix of $\vp_{A}^{i}$ and $\bar{\vp}_{A}^{i}$ as closely as possible with the identity matrix to do the optimization.

The last point relates to periodicity-related arrangements. The authors introduce PIMG and PAL to ensure the validity of periodicity. Crystal can be seen as multi-graph~$\gG \coloneqq (\mA, \mH)$, where $\mA$ represents the set of atom nodes and $\mH$ represents the set of edges which are relevant atom pairs. The authors ensure periodic invariance by constructing edges using Euclidean
distance. Then the authors design a multi-graph attention mechanism to capture structural patterns. First a linear transformation is applied to the initial feature vectors of the nodes, which is parameterized by the weight~$\mW$. Then the new atom node can be derived by 
\begin{equation}
    \begin{aligned}
        &r_{ij}=f_{a}(\mW\va_{i},\mW\va_{j}),\\
        &\varepsilon_{ij}=\frac{exp(LeakyReLU(r_{ij}))}{\sum_{k\in\mathcal{N}_i}exp(LeakyReLU(r_{ik}))},\\
        &\hat{\va}_{i}=\phi_{FC}(\sum_{j\in\mathcal{N}_{i}}\varepsilon_{ij}\cdot\va_{i}),
    \end{aligned}
\end{equation}
where $f_{a}$ denotes a single-layer feed-forward neural network and $\varepsilon$ is a correlation coefficient. $LeakyReLU$ is the activation function and $\phi_{FC}$ is a fully-connected layer. The final step is to send the multi-graph and new atom nodes into DimeNet++~\cite{gasteiger2020fast}, which is an E(3) invariance graph neural network, to learn the crystal structure representations.
The periodic attribute learning module focuses on learning about the three components associated with periodicity: discrete direction, unit cell position and distance between nodes. They utilize three attribute learners to study these three sections and optimize the prediction by computing $\gL_{CAA}$, which contains three losses corresponding to three parts mentioned above.

\subsection{Generative AI Accelerates Materials Discovery.}\label{ActivatingLearning}
In recent years, diffusion generative models have gained significant attention for their realistic effects in image generation, such as Stable Diffusion~\cite{rombach2021highresolution}. Methods based on diffusion generative models, such as denoising diffusion probabilistic models (DDPM)~\cite{ho2020denoising} and Score-based Diffusion~\cite{song2021Score-based}, have been widely applied in molecular docking (AlphaFold3~\cite{abramson2024accurate}, DiffDock~\cite{ gabriele2023diffdock}) and molecular generation (Torsional Diffusion~\cite{jing2022torsional}). 
In the context of materials discovery, there are three categories of materials: known knowns, known unknowns, and unknown unknowns. While the largest existing explorations of previously uncharted crystalline materials are on the order of $10^6$ to $10^7$ materials, the potential space of stable inorganic compounds, even within the realm of quaternary compounds and fixed stoichiometry, reaches an astounding $10^{10}$ possibilities. The third category, the ``unknown unknowns,'' represents the most challenging yet vast frontier. Generative models present one of the most effective means to glimpse into this largely uncharted space.
In the field of material structure prediction, notable generative AI methods, including CDVAE~\cite{CDVAE2022}, CrystalGAN~\cite{CrystalGAN}, FlowMM~\cite{FlowMM}, DiffCSP~\cite{jiao2024crystal} and DiffCSP++~\cite{jiao2024space}, have also emerged. This section will focus on these generative AI approaches.

\vpara{CDVAE.}
The crystal diffusion variational autoencoder (CDVAE)~\cite{xie2021crystal} stands as a prominent early algorithm developed to generate stable periodic structures of materials. It employs a combination of joint equivariant diffusion and SE(3)-equivariant graph neural networks, explicitly encoding interactions across periodic boundaries while preserving permutation, translation, rotation, and periodic invariance. Furthermore, CDVAE introduces a noise conditional score network~(NCSN)~\cite{song2021Score-based} as the VAE decoder to generate more realistic material structures, utilizing Langevin dynamics to refine the generation process. The algorithm also establishes three benchmark datasets specifically for material generation, along with a set of physically meaningful evaluation tasks and metrics. Extensive experiments validate superior performance of CDVAE in tasks such as material reconstruction, generation, and property optimization, underscoring its significant potential in advancing the field of materials discovery.
The workflow of CDVAE consists of several key steps. First, a periodic material encoder is employed to map the material $\mM$ into a latent representation. Then, three independent MLPs predict the composition $\vc$, lattice $\mL$, and number of atoms $N$, respectively. A conditional score matching decoder with equivariance is used to denoise the atomic coordinates and the probability distribution of atom types. To account for the periodicity of atomic coordinates, a constraint is incorporated into the loss function. Finally, during each denoising step, atomic types and coordinates are updated using Langevin dynamics.
Con-CDVAE~\cite{YE2024100003} is a modified version of CDVAE designed to directly generate crystal structures based on crystal properties. Incorporating the conditional generation method of DALL-E2~\cite{DALL-E2}, Con-CDVAE achieves this goal by introducing new modules specifically designed for crystal latent representations in CDVAE. In the first training phase, Con-CDVAE trains the core framework of CDVAE and applies MLPs to the crystal latent representations to predict crystal properties, aligning crystal structures with their corresponding properties. In the second training phase, Con-CDVAE introduces a prior module inspired by DALL-E2, which takes crystal properties as input to generate crystal latent representations. With these new modules, Con-CDVAE enables the direct generation of crystal structures from crystal properties. The model has been applied to conditional generation tasks based on formation energy and bandgap, showing that the results for conditional generation of formation energy outperform those for bandgap. While achieving conditional generation, the model retains the crystal latent variables, potentially enabling manipulable structure generation such as latent variable interpolation. However, a key challenge lies in the design and training of the prior module to generate diverse and accurate crystal latent variables from the given crystal properties.
Cond-CDVAE~\cite{Luo2024} is a deep learning-based generative model, the conditional crystal diffusion variational autoencoder, designed for crystal structure prediction (CSP). This model generates physically realistic crystal structures based on user-defined material compositions and external conditions, such as pressure. By leveraging Cond-CDVAE, the research team successfully generates high-fidelity crystal structures under various pressure conditions, achieving an accuracy of 59.3$\%$, with an impressive 83.2$\%$ accuracy for structures containing fewer than 20 atoms—surpassing traditional global optimization-based CSP methods. Furthermore, a comprehensive dataset (MP60-CALYPSO) comprising over 670,000 local minimum structures, including both ambient and high-pressure crystal configurations, is established. The generated structures demonstrate superior convergence rates and fewer ionic steps during DFT local optimization, indicating that they are closer to local energy minima. Cond-CDVAE integrates discrete chemical compositions and continuous pressure attributes, utilizing SE(3)-equivariant graph neural networks to encode crystal structures, thereby ensuring invariance under permutations, translations, rotations, and periodicity. Evaluation results underscore the model's high accuracy and reliability in crystal structure prediction, illustrating the substantial potential of deep learning generative models to accelerate the discovery and design of novel materials.

\vpara{MatterGen.}
MatterGen~\cite{zeni2023mattergen} operates through a two-step process. In the first step, a base model is pre-trained to generate stable and diverse materials. The model is trained on data from the Materials Project and Alexandria databases, where stability is defined by DFT relaxation to a local energy minimum, with the energy per atom deviating no more than 0.1 eV from the reference database values. Novelty is ensured by verifying that the generated structures do not exist in the Alexandria dataset. In benchmark comparisons using the same Materials Project dataset, MatterGen outperforms CDVAE by achieving a higher proportion of stable, unique, and novel (S.U.N.) materials, while also showing lower RMSD values, which indicates better alignment with DFT-relaxed structures. Furthermore, when trained on a larger dataset, MatterGen demonstrates even stronger performance.
In the second step, MatterGen fine-tunes the base model using adapter modules, which allow for customization based on specific chemical compositions, symmetry requirements, and desired electronic, magnetic, and mechanical properties. This adaptability makes MatterGen a versatile and powerful tool for targeted material discovery across a wide range of application domains. A similar generative algorithm is SyMat~\cite{joshi2023expressive}.

\vpara{DiffCSP.}
Crystal structure prediction by joint equivariant diffusion (DiffCSP)~\cite{jiao2024crystal} is a specialized algorithm for crystal structure prediction. It outperforms existing baseline methods, such as CDVAE, across multiple benchmark datasets, while offering reduced computational cost compared to DFT-based approaches. DiffCSP generates crystal structures that satisfy periodic E(3) invariance, ensuring stability in translation, rotation, and periodicity. Furthermore, DiffCSP demonstrates strong scalability, making it applicable not only to fixed-composition crystal structure prediction but also to tasks involving atomic type generation and property optimization. By employing a joint equivariant diffusion process, DiffCSP simultaneously operates on lattice vectors and atomic fractional coordinates to preserve periodic invariance, utilizing denoising models and message-passing mechanisms to improve prediction accuracy. The model demonstrates exceptional performance in crystal structure and property prediction experiments, underscoring its considerable potential to enhance both the accuracy and efficiency of crystal structure prediction.
In crystal structures, atoms are periodically arranged within the unit cell, denoted as $\gM=(\mA, \mX, \mL)$, where $\mA \in \sR^{h \times N}$ indicates atom types, $\mX \in \sR^{3 \times N}$ represents the Cartesian coordinates of each atom, and $\mL \in \sR^{3 \times 3}$ is the lattice matrix defining crystal periodicity. Any atom's Cartesian coordinates and type within the crystal are expressed as $\{(\va_i',\vx_i')|\va_i'=\va_i,\vx_i'=\vx_i + \mL\vk, \forall \vk\in\sZ^{3 \times 1}\}$. A relationship between Cartesian and fractional coordinates is given by $\vx = \sum_{i=1}^3 f_i \vl_i$. For the generation process, DiffCSP adopt a fractional coordinate system $\gM=(\mA, \mF, \mL)$.
With $\mL$ as a continuous variable and the one-hot encoding of $\mA$ treated similarly, a standard Denoising Diffusion Probabilistic Model~\cite{DDPM2020} can be used to generate $\mL$ and $\mA$, with the loss function:
\begin{equation}
    \gL_{\mL / \mA}=\E_{\vepsilon\sim\gN(0,\mI)}[\|\vepsilon - \hat{\vepsilon}_{\mL / \mA}(\gM_t,t)\|_2^2],
\end{equation}
the equivariant denoising model $\phi$ predicts the denoising terms $\hat{\vepsilon}_\mL(\gM_t, t)$ and $\hat{\vepsilon}_\mA(\gM_t, t)$. For $\mF$, periodicity is handled using a Score-Matching (SM) based framework~\cite{song2021Score-based}.

\vpara{DiffCSP++.} 
DiffCSP++~\cite{jiao2024space} is an advanced crystal structure generation algorithm specifically optimized to account for space group constraints. The algorithm demonstrates exceptional performance in generating crystal structures that adhere to specific space group symmetries. Compared to both learning-based and DFT-based methods, DiffCSP++ not only exhibits enhanced performance but also achieves notable success in ab initio crystal generation tasks, producing crystals with valid compositions and stable structures. Experimental results consistently show that DiffCSP++ surpasses existing baseline methods in both crystal structure generation and property statistics, underscoring its substantial potential in materials design.

DiffCSP++ considers the constraints of space groups to generate crystal structures. Space groups are collections of symmetry operations of crystals, constrained by the o(3) invariance of the lattice matrix and the Wyckoff positions of fractional coordinates. According to the Polar Decomposition ~\cite{hall2013lie}, the lattice matrix $\mL \in \sR^{3 \times 3}$ can be decomposed as $\mL = \mQ \exp(\mS)$, where $\mQ$ is an orthogonal matrix and $\mS$ is a symmetric matrix. Any symmetric matrix can be expressed using six symmetric basis functions, with coefficients that are o(3) invariant:
\begin{equation}
    \mS = \sum_{i=1}^{6} k_i \mB_i
\end{equation}
\begin{align*}
\mB_1 &= \begin{pmatrix} 0 & 1 & 0 \\ 1 & 0 & 0 \\ 0 & 0 & 0 \end{pmatrix}, & 
\mB_2 &= \begin{pmatrix} 0 & 0 & 1 \\ 0 & 0 & 0 \\ 1 & 0 & 0 \end{pmatrix}, \\
\mB_3 &= \begin{pmatrix} 0 & 0 & 0 \\ 0 & 0 & 1 \\ 0 & 1 & 0 \end{pmatrix}, & 
\mB_4 &= \begin{pmatrix} 1 & 0 & 0 \\ 0 & -1 & 0 \\ 0 & 0 & 0 \end{pmatrix}, \\
\mB_5 &= \begin{pmatrix} 1 & 0 & 0 \\ 0 & 1 & 0 \\ 0 & 0 & -2 \end{pmatrix}, & 
\mB_6 &= \begin{pmatrix} 1 & 0 & 0 \\ 0 & 1 & 0 \\ 0 & 0 & 1 \end{pmatrix},
\end{align*}
by determining the coefficients $\vk$ of six symmetric basis functions $\mB$, DiffCSP++ ascertains the lattice matrix $\mL$. During noise addition and denoising, these coefficients can be transformed. The 230 space groups are categorized into six crystal families $G_{family}$~(Triclinic, Monoclinic, Orthorhombic, Tetragonal, Hexagonal, Cubic), each imposing distinct constraints on the $k_i$. These families define six projection spaces, ensuring consistent projections to the prior crystal family space throughout the crystal generation process. The wyckoff positions ($W_{positions}$) represent the symmetry of equivalent atoms in the unit cell, with $N$ fractional coordinates $\mF \in \R^{3 \times N}$ derived from $N'$ basic fractional coordinates $\mF' \in \R^{3 \times N'}$. Noise addition and denoising can be performed on these basic coordinates $\mF'$, allowing for the determination of all atoms' fractional coordinates within the unit cell. The coefficients $\vk$ are treated as continuous variables, and the one-hot encoded atomic types of Wyckoff positions are similarly considered, both generated using standard DDPM. Due to the periodic nature of the Wyckoff positions' fractional coordinates, DiffCSP++ utilizes a SM framework for generation.
The CrystalFormer~\cite{cao2024CrystalFormer} also takes into account the constraints of space groups to generate crystals.

\vpara{CrystalFormer.}
CrystalFormer~\cite{cao2024CrystalFormer} is a transformer-based autoregressive model specifically designed to generate crystal materials that respect space group symmetries. The model learns the discrete and sequential nature of Wyckoff positions to directly predict the type, position, and lattice parameters of symmetry-inequivalent atoms within the lattice, thereby generating crystal structures. Compared to traditional crystal structure prediction methods, CrystalFormer shows significant advantages in tasks such as symmetry structure initialization and element substitution. Additionally, CrystalFormer facilitates property-guided materials design by integrating solid-state chemistry knowledge and heuristic rules, enabling systematic exploration of crystal materials and the discovery of novel superconductors with low Ehull (<0.1 eV/atom). To achieve these capabilities, CrystalFormer incorporates algorithms such as markov chain monte carlo~\cite{brooks1998markov} sampling, gaussian mixture models~\cite{reynolds2009gaussian}, and von Mises distributions, effectively incorporating space symmetries and other physical constraints during the generation process. This work offers new insights for the discovery of high-temperature superconductors and advances the field of materials modeling and discovery.

\vpara{GNoME.}
Graph networks for materials exploration (GNoME)~\cite{merchant2023scaling} marks a significant advancement in the generation and discovery of novel materials through an innovative active learning algorithm. Departing from conventional approaches that rely on sampling from existing datasets, GNoME challenges the prevailing assumption that newly generated materials must adhere to the same data distribution as the training set. By employing an active learning framework, it efficiently generates millions of novel crystal structures. This methodology uncovers 2.2 million potentially stable materials, many of which transcend traditional chemical intuition. Of these, 381,000 materials are integrated into the materials database, with 736 structures experimentally validated for stability, thereby enriching the field with valuable resources for further research and practical applications in materials science.

The GNoME framework operates through two key modules: symmetry-aware partial substitutions combined with random structure search, and GNN-based modeling of material properties. These components drive two independent material discovery pipelines. The Structural Pipeline focuses on evaluating the stability of crystal frameworks without considering specific atomic types, filtering randomly generated structures using GNoME to retain potentially stable frameworks. The compositional pipeline, on the other hand, takes chemical formulas as input to GNoME, identifying stable chemical combinations to explore novel material compositions. After structures are selected through these pipelines, DFT calculations are performed to further validate their structural stability. Stable materials are then added to the training set for subsequent iterations, creating an iterative active learning loop. This process facilitates the discovery of new materials that extend beyond existing data distributions.

\begin{figure*}[t]
		\centering  
		\includegraphics[width=0.9\linewidth]{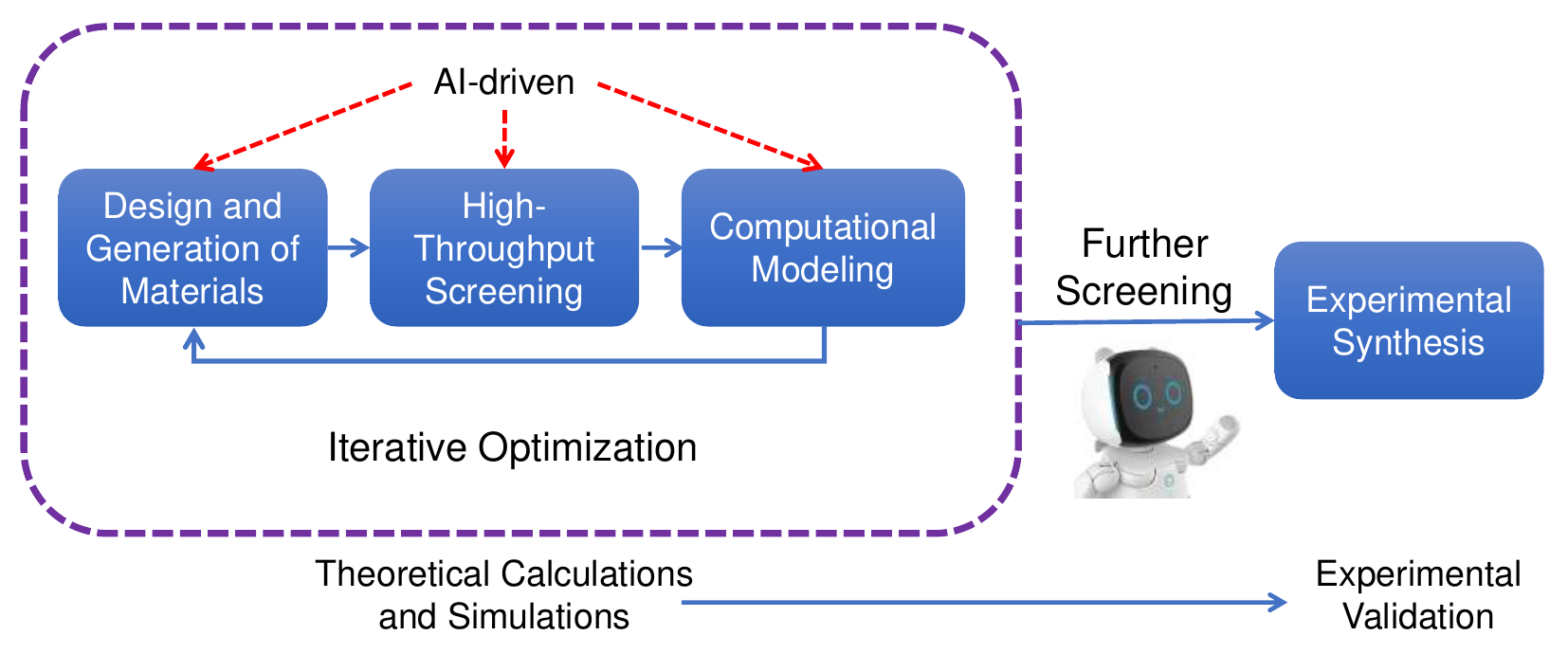}
		\caption{The general process of materials inverse design, which is divided into two main parts: the AI-driven theoretical calculation part and the experimental validation part. The theoretical calculation part is further subdivided into material design and generation, high-throughput screening, and computational modeling.}
		\label{fig:InvDesFlow} 
\end{figure*}

\vpara{InvDesFlow.}
The InvDesFlow~\cite{xiaoqiAIsuper} develops an AI-driven workflow for discovering high-$T_c$ superconductors that are not present in any existing database. In this survey, we summarized the general process for inverse design of materials, as shown in Figure~\ref{fig:InvDesFlow}. The process is divided into two main parts: theoretical calculations and experimental validation. The theoretical calculations are further subdivided into material design and generation, high-throughput screening, and computational models. Unlike traditional inverse design, the entire theoretical part is fully AI-driven, with AI implementing the entire theoretical simulation process.

\vpara{OMat24.}
Meta has introduced Open Materials 2024 (OMat24)~\cite{OMat24}, a comprehensive dataset featuring over 110 million density functional theory based calculations, emphasizing structural and compositional diversity. This dataset spans non-equilibrium atomic crystal structures and varied elemental compositions, offering a rich resource for materials discovery. Utilizing a pre-trained EquiformerV2 model, Meta achieved state-of-the-art results on the Matbench Discovery~\footnote{\url{https://matbench-discovery.materialsproject.org/}} leaderboard. Through extensive experimentation, they explored the impact of different training strategies on model performance, providing valuable insights for future advancements. By openly sharing both the dataset and model, Meta empowers the research community to build upon and refine these foundational resources, fostering progress in AI-driven materials science.

\vpara{FlowLLM.}
Meta has introduced FlowLLM~\cite{FlowLLM2024}, an advanced crystal generation model that integrates LLMs with Riemannian flow matching (RFM) to enable the design of novel crystalline materials. FlowLLM first fine-tunes an LLMs to capture an effective foundational distribution of metastable crystals within textual representations. Following the conversion of text to graph representations, the RFM model further refines LLMs-generated samples through iterative optimization of atomic coordinates and lattice parameters. FlowLLM surpasses existing state-of-the-art methods by over threefold in stable material generation rates. The structures produced by FlowLLM are notably closer to relaxed states, substantially reducing post-processing costs while improving both the efficiency and precision of material generation, thereby marking a significant advancement in the field of materials science.

\subsection{Large Language Models}
In recent years, LLMs, such as GPT-3.5 Turbo~\cite{brown2020language}, GPT-4~\footnote{\url{https://openai.com/gpt-4}}, and BERT~\cite{devlin2018bert}, have been widely adopted across various scientific fields, becoming essential tools for natural language processing and knowledge generation. OpenAI's GPT-3.5 Turbo and GPT-4 are particularly popular due to their ability to generate human-like text, answer complex questions, and facilitate knowledge synthesis. These models are powered by transformer-based architectures~\cite{vaswani2017attention}, where self-attention mechanisms enable the efficient processing of large-scale textual data across diverse applications. Similarly, Google's BERT, distinguished by its bidirectional training approach, captures context from both directions in a text, making it especially effective for tasks that require nuanced understanding, such as question-answering.  
In addition to these general-purpose models, domain-specific LLMs like Coscientist~\cite{Boiko2023},SciBERT~\cite{beltagy2019scibert} and MatSciBERT~\cite{gupta2021matscibert} have been developed to improve performance in particular scientific domains. SciBERT, based on the BERT architecture, is fine-tuned on scientific literature, enabling more precise information retrieval and classification within scientific texts. Similarly, MatSciBERT, tailored for materials science, incorporates domain-specific language and terminology to enhance its effectiveness in analyzing material-related data and generating specialized tags.  
Building upon these foundational and domain-specific models, researchers are increasingly employing LLMs for inverse material design, using these models to effectively predict and optimize material properties, as discussed in the following sections.

In sustainable concrete design, a methodology using LLMs to develop alkali-activated concrete formulations aims at reducing the carbon footprint and environmental impact of concrete production~\cite{volker2024LLMs}. Researchers utilize GPT-3.5 Turbo and GPT-4 models to create a ``Knowledge-Driven Design'' (KDD) system, which integrates domain-specific knowledge with experimental feedback to enhance the accuracy and efficiency of material design. Through the automated generation of low-calcium concrete formulations, property optimization, and performance prediction, this study demonstrates a marked improvement in predictive accuracy for complex composite mixtures. This innovative approach introduces a sustainable paradigm in concrete design, supporting the development of an environmentally friendly concrete industry. In the field of battery technology, particularly for rapid charging applications, the BatteryGPT system leverages LLMs to accelerate information extraction and innovation~\cite{zhao2024potential}. Utilizing retrieval-augmented generation technology, BatteryGPT aggregates data from over 2,200 publications and incorporates advanced technical methodologies to facilitate knowledge generation and optimization in fast-charging battery technology. BatteryGPT provides domain-specific literature reviews and recommends the latest solutions for fast-charging materials, such as Li$_3$P-enhanced solid electrolyte interfaces and laser patterning. This system significantly enhances the efficiency of information integration and processing in material innovation, supporting advancements in battery technology aimed at extending electric vehicle range and improving energy storage. In the study of microstructure evolution within materials, LLMs generate code and simulate complex multiphysics-coupled models~\cite{satpute2024exploring}. Researchers employ ChatGPT to produce code based on phase-field models to simulate the evolution of material microstructures over time. Although the model-generated code requires further validation and adjustments for complex equations, this research systematically explores the potential of LLMs in microstructure modeling. This novel approach highlights the utility of LLMs in materials education and research, offering new perspectives for integrated computational materials engineering (ICME)~\footnote{\url{https://link.springer.com/book/10.1007/978-3-030-40562-5}}. From a broader perspective, automation and knowledge integration benefits of LLMs within materials science are further investigated~\cite{lei2024materials}. The study shows that LLMs not only retrieve knowledge and generate code through natural language instructions but also conduct multi-level literature analysis and generate tags, such as for 3D microstructure analysis and micrograph labeling. By leveraging fine-tuned, domain-specific models like SciBERT~\cite{beltagy2019scibert} and MatSciBERT~\cite{gupta2022matscibert}, this research demonstrates that LLMs effectively process complex materials science data and produce high-quality structured information. This advancement supports the automation of materials science workflows with significant applications in rapid charging and renewable energy integration.

In summary, LLMs have shown remarkable potential in inverse materials design, transforming traditional materials innovation processes. Leveraging advanced language processing and knowledge synthesis capabilities, LLMs have greatly enhanced the accuracy of property prediction and optimization, while also providing sustainable solutions to pressing challenges in materials science. For instance, in sustainable concrete design, LLMs enable researchers to develop eco-friendly materials that lower carbon emissions. In battery technology, LLMs facilitate the integration of vast information, accelerating innovation in fast-charging batteries that support extended electric vehicle ranges and improved energy storage. Additionally, LLMs contribute to the simulation of material microstructures and multiphysics modeling, offering new possibilities for ICME and materials education. As applications of LLMs in inverse materials design continue to expand, these models are anticipated to play a critical role in property prediction, novel material discovery, and the automation of knowledge integration. Coupled with domain-specific models, LLMs promise breakthroughs in processing complex materials data and generating high-quality information, supporting the development of automated scientific workflows. Future research may further explore robust algorithm designs and the integration of real-time data analysis with experimental feedback, thereby enhancing the breadth and depth of LLMs applications in materials science and providing technical support for sustainable and intelligent materials innovation.

\subsection{Dataset}

\begin{table}[t]
\caption{Some typical databases. Listed in the table are their name, amount of data, description, and reference.}
\label{tab:datasets}
    \begin{tabular}{llll}
    \hline
     \textbf{Datasets}&\textbf{Size}  &\textbf{Description}  &\textbf{Reference}  \\ \hline
     QM7&> 7,000  &DFT calculations  &\cite{blum2009970}  \\
     QM9&> 134,000  &DFT calculations  &\cite{ramakrishnan2014quantum}  \\
     MD17&>100,000  &MD simulation data  &\cite{chmiela2017machine,chmiela2018towards} \\
     MP &>150,000 & Materials Project &\cite{jain2013commentary} \\
     OQMD &>1,226,000 &Quantum Materials &\cite{kirklin2015open} \\
     OC20 &>133,940,000 &Open Catalyst 2020 &\cite{chanussot2021open} \\
     JARVIS-DFT &>80,000 &DFT calculations &\cite{choudhary2020joint} \\ 
     PubChemQC &>85,000,000  &DFT calculations &\cite{Nakata2023,ullah2024molecularquantumchemicaldata} \\ 
     \hline
    \end{tabular}
\end{table}

As we can see from the previous section, there are a number of datasets that are typically used when testing the performance of a model. In Table~\ref{tab:datasets} we briefly introduce a few common material datasets.

Different researchers have created different datasets for different kinds of materials. In the specific research, we will choose different datasets according to the needs and select the appropriate data from the dataset to conduct the experiments. At the same time, different researchers choose different prior work with some impact as benchmark, such as CGCNN~\cite{xie2018crystal} and ALIGNN~\cite{choudhary2021atomistic}. Some researchers use different frameworks as different benchmarks for different parts in their work.

\section{Future Directions}
\label{sec-future}
The previous chapters have provided a comprehensive overview of the way that AI has accelerated advancements in the inverse design of functional materials, alongside a rapid evolution of AI technologies within the materials science domain. In this chapter, we will delve into potential future directions for AI across each critical stage of the reverse design workflow for functional materials. Key stages in this workflow encompass the design and generation of novel materials, high-throughput screening for target functional properties, theoretical validation of candidate materials through computational modeling, and the experimental synthesis and testing necessary to confirm material performance. By advancing and designing AI algorithms that deeply integrate with each of these stages, we aim to accelerate every step of the process, allowing AI to fully realize its transformative potential in the inverse design of materials.
\begin{figure*}[t]
		\centering  
		\includegraphics[width=1.0\linewidth]{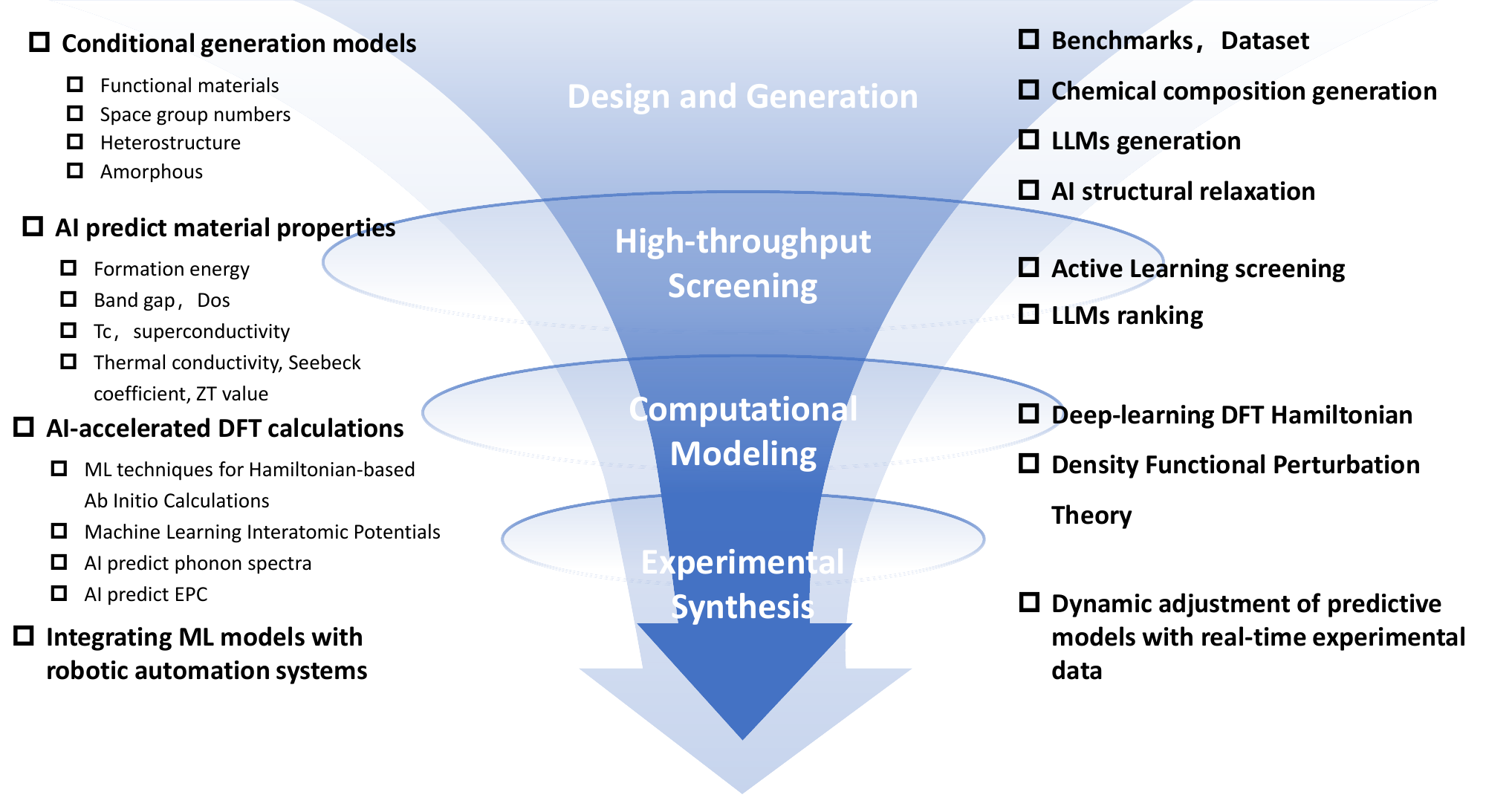}
		\caption{The future of AI-driven inverse design of materials. This figure outlines the potential development directions of AI in key stages of the functional material reverse design workflow, including the design and generation of new materials, high-throughput screening, computational modeling for candidate validation, and experimental synthesis and testing. By optimizing the integration of AI algorithms, these AI technique can accelerate efficiency and innovation in materials discovery.}
		\label{fig:pubcit} 
\end{figure*}

\subsection{Design and Generation of New Materials}
For the inverse design and generation of new functional materials, current generative models remain constrained in their capabilities and fail to meet the complex requirements of material design. The following challenges represent areas where we anticipate AI advancements may bring impactful solutions in the future. The first one is that the AI-generated material structures are frequently not in a stable ground state, necessitating additional assessment and structural relaxation. Enhancing generative models to address this limitation remains a key objective for future development. The second one is the generative models based on chemical composition are currently scarce, despite being frequently needed in experimental science. Large language models may provide such capabilities in the future. The third one is that we need conditional generation methods where a partially known structure or chemical formula serves as input, producing a complete structure or formula. For example, given part of a heterostructure, we aim to generate the remaining portion. The forth one is that the space group family-based generation algorithms, such as DiffCSP++, have been realized, but algorithms allowing direct control over space group numbers for new structure generation have yet to be developed. Moreover, the generative AI algorithms specifically designed for diverse functional materials have yet to be developed. And a significant number of materials exist in amorphous rather than crystalline forms, highlighting the need for future development of generative algorithms specifically for amorphous materials. Furthermore, the generative algorithms discussed above necessitate the development of more equitable benchmarks for assessing model performance. Consequently, the release of suitable datasets and corresponding benchmarks represents a crucial task.

\emph{Creation and updating of datasets.}
With the assistance of AI, the development of materials science has indeed gained a lot of progress, but at the same time, there are some urgent problems that need to be solved. Firstly, there are already quite large databases for some of the materials such as MP~\cite{jain2013commentary}, OQMD~\cite{kirklin2015open}, OC20~\cite{chanussot2021open} and so on, however there is still a lack of sufficient data for some complex materials such as HEAs. Some methods are designed to deal with these materials~\cite{rao2022machine}. But the creation of a database of complex materials is also required. Second, With the assistance of AI, it is true that many new materials have been discovered that exist in theory, but these new materials may not actually be synthesized in actual experiments. And these materials are in dire need of experimental validation to determine whether they should be added to the database. However, the progress of experiments is often not enough. Thus, we could create a candidate database that records AI-generated materials. The third point is that many AI-generated materials do not have a standardized nomenclature. Many researchers use simple numbers to represent generated materials right in their own papers, which is not conducive to the documentation of new materials as well as the retrieval of new materials. All of the above are problems that need to be solved or optimized with respect to the database. Also we have noticed some databases related to AI have been established, which could be seen as a future direction.

\emph{Large language models.}
The future prospects of LLMs in inverse material design are promising. As the technology matures, LLMs are expected to bring profound changes in autonomous material discovery and efficient data integration. Firstly, a significant potential of LLMs in inverse material design lies in achieving autonomous material discovery and design systems. These systems can rapidly adjust design strategies while interpreting human instructions and dynamically adapt to different tasks and material properties, ultimately realizing an automated material research and development (R\&D) process. For instance, the LLMatDesign~\cite{jia2024llmatdesign} project demonstrated the potential of LLMs to achieve autonomous material design through instruction fine-tuning in low-sample environments. Its framework allows the model to perform material screening and optimization based on design instructions and chemical knowledge. With future enhancements in model optimization on datasets and improved multimodal integration capabilities, LLMs are expected to be more flexible in complex material tasks, even forming ``autonomous laboratories'' to accelerate the material R\&D process. The HoneyComb system~\cite{zhang2024honeycomb} builds upon this by using its high-quality knowledge base, MatSciKB~\cite{zhang2024honeycomb}, and computational tool center, ToolHub, to enable real-time data updates and cleaning support, providing a stronger data foundation and scientific computing capabilities for autonomous material design. This indicates that future LLMs will not only generate and evaluate material design schemes but also continuously learn and improve through knowledge bases, laying the groundwork for fully automated material discovery.
Secondly, future LLMs will make significant strides in the efficient integration and extraction of material data. The key to achieving this goal lies in building high-quality multimodal datasets and enhancing the model’s ability for complex reasoning. Several studies have emphasized the necessity of dataset expansion and multimodal information integration. For example, Miret \emph{et al.}~\cite{miret2024llms} proposed a six-step interactive roadmap that progressively improves the model's reasoning performance in complex material tasks by integrating text, images, and experimental data. Similarly, there is another reference~\cite{lei2024materials} highlighted the importance of high-quality multimodal data for enhancing the model’s understanding of the complexity of the materials science field. The study pointed out that by dynamically integrating experimental data, LLMs can significantly enhance their ability to parse information related to material properties and structures, thereby advancing the automation of the material discovery process. These studies indicate that through the construction and integration of multimodal information, future LLMs will exhibit stronger reasoning capabilities in complex materials science tasks, especially under conditions of limited data, providing reliable support for the design and validation of new materials.

In conclusion, the future development prospects of LLMs in inverse material design are extensive. As the capabilities of models in data processing, knowledge integration, and complex reasoning continue to improve, LLMs are anticipated to significantly accelerate the material R\&D process, promoting autonomous and intelligent material design. By dynamically acquiring and parsing multimodal data and continuously learning from the latest research developments in materials science, LLMs can effectively support all stages of material discovery, from screening to structural optimization and experimental plan generation, gradually achieving a more comprehensive and efficient material innovation process. This technological advancement will not only help scientists address current research challenges more rapidly but also open new research avenues in the field of materials science, laying a solid foundation for future technological innovation.

\subsection{High-Throughput Screening of Functional Materials}
In recent years, high-throughput screening of target functional materials has predominantly utilized invariant graph neural networks to classify and predict material properties, subsequently establishing specific thresholds for material selection. This approach has led to numerous significant contributions, enabling the accurate prediction of several physical properties, including formation energy, band gap, Debye temperature, and density of states. However, certain physical properties remain challenging to predict with high precision due to their inherent complexity. The following are potential issues that require further investigation.

The rapid development of AI technology in recent years has provided new momentum for the inverse design of superconducting materials, leading to the establishment of a comprehensive workflow for discovering new high-$T_c$ superconductors. However, there remains considerable potential for enhancement across various dimensions. For instance, the application of AI to accurately assess whether a material exhibits superconductivity, predict the $T_c$ of any superconductor—whether characterized by conventional or unconventional properties and defined by its chemical formula or crystal structure—expedite the DFT verification process for electron-phonon coupling through AI integration, and provide more precise guidance for experimental directions using AI. Ultimately, we anticipate that advancements in AI will illuminate new physics related to superconducting mechanisms in the future.

In the case of complex materials, such as high-entropy alloys, the primary challenge associated with AI-assisted computation is the limited size of the available dataset. As previously stated, a number of workflows have been proposed in recent years to address this class of materials~\cite{rao2022machine,li2022towards,nie2024active}. However, the corresponding methods require further promotion and application to assess their practicality and reasonableness. Concurrently, the necessity for high-quality datasets for complex materials persists. Moreover, there is also work using activating learning to discover new and efficient catalysts. This idea is very similar to the ``exploration'' and ``exploitation'' concepts in reinforcement learning. In the context of expanding the existing dataset, exploitation allows AI to generate more data of the same distribution, while exploration enables to realize the step of activating learning. However, current activating learning still requires DFT or other methods for validation, and the verified samples must be manually added to the dataset for the next iteration. Therefore, if the steps of activating learning can be automated, future exploration of crystals through AI could significantly improve efficiency.

Note that in AI-driven high-throughput screening of functional materials, significant gaps remain in material classification and property prediction, particularly in terms of accuracy and handling complex properties. Future LLMs hold promise for advancing material recommendation, ranking, and structural design through frameworks like natural language embeddings, which generate compositional and structural feature vectors of materials~\cite{qu2023leveraging}. Such models aim to better represent materials, identify under-explored spaces, and streamline material optimization. Despite these advancements, current machine learning methods often struggle with reliably predicting key properties—such as thermal conductivity, Seebeck coefficient, and ZT value—due to the challenges in modeling nonlinear relationships within complex material systems, including thermoelectric materials~\cite{gorai2017computationally}. Additionally, the accurate prediction of dopant and structural effects on material performance remains limited, with models like DopNet capturing only partial insights into these intricate dependencies~\cite{na2021predicting}. Addressing these challenges requires AI models that can integrate multi-scale data and adapt to experimental feedback, ultimately improving robustness and reliability in predictions. Furthermore, systems like the HoneyComb illustrate the need for interpretable AI models, which are essential for achieving precise material recommendation and structural optimization, especially for high-stakes applications requiring practical and reliable adoption.

\subsection{Computational Modeling for the Validation of Candidate Materials}
The majority of conventional materials calculation methods are based on the DFT. Nevertheless, the DFT-based approaches are constrained in their ability to accommodate a comprehensive range of materials, largely due to the influence of intricate terms such as the exchange-correlation functional. It is therefore unsurprising that there is a tendency to develop new calculation methods with a view to improving the quality of the calculations. Furthermore, the combination of \textit{ab initio} calculation with AI is also popular and promising avenue of research.

In some related work, the authors focus on predicting complex terms with the assistance of AI. For example, some researchers employ ML techniques to predict EPC~\cite{zhong2023accelerating,gibson2024accelerating,haldar2024machine} and other researchers have employed AI to predict phonon spectra~\cite{okabe2024virtual,fang2024phonon}. Furthermore, there are promising avenues for researchers to utilize AI in optimizing alternative versions of DFT, beyond the prevalent Kohn-Sham DFT~\cite{zhang2024overcoming}.
Attention should be paid to recent work about DeepH~\cite{li2022deep,gong2023general,li2024deep,li2023deep,tang2024deep,gong2024generalizing,tang2024improving,wang2024deeph,li2024neural,wang2024universal}, which propose a framework about using ML techniques to study the Hamiltonian of materials to conduct \textit{Ab initio} calculation. This method is highly efficient and achieves high accuracy on many materials. And this approach is now being generalized to a wider range of systems, with a number of similar works having been done, such as HamGNN~\cite{zhong2023transferable, zhong2024universal, su2023efficient}. More research and applications along these lines are bound to emerge in the future.

To summarize, we can actually see the following points worth trying for the combination of AI and material calculation methods. The most important one is that the lack of a generalized network structure or a generalized approach to network selection, and the answer of how ML technology can be integrated into material computation methods is not conclusive and there are still many perspectives worth testing. Moreover, there is still a certain threshold for the various methods, which implies the necessity of writing relevant software packages or developing relevant software.

\subsection{Experimental Synthesis and Testing for Validating Material Performance}
As evidenced by recent research, a workflow that integrates computational prediction with experimental synthesis can be an invaluable tool for studying complex materials with limited datasets~\cite{rao2022machine,li2022towards,nie2024active}. The proposed workflow allows researchers to obtain actual materials directly, which is not always possible in other fields where materials prediction involves a disconnect between computational prediction and experimental synthesis. It should be noted that the time cost of this workflow is not insignificant and requires continuous improvement.

AI-guided experimentation holds significant promise for advancing functional material development, particularly in thermoelectric materials, where experimental synthesis and optimization remain complex and resource-intensive. In the field of thermoelectric materials, AI-driven approaches, such as error correction learning models, refine predictive accuracy through iterative experimental feedback, enabling faster identification of high-performance materials~\cite{choubisa2023closed}. Future research in AI-guided experiments focuses on enhancing adaptability and scalability for large-scale synthesis processes, incorporating real-time data from experimental setups to dynamically adjust predictive models. Furthermore, AI-guided experimental planning proves invaluable for efficiently testing the effects of doping, strain, and nano-engineered structures, which are critical for improving ZT values in thermoelectric materials~\cite{sasaki2020identifying}. By integrating ML models with automation platforms, such as robotic experimental systems, researchers explore complex parameter spaces more efficiently, facilitating the development of sustainable and high-performance materials for applications like waste heat recovery and thermal management in microelectronics.

\section{Conclusion}
\label{sec-conclusion}
In this paper, we delve into the history of inverse design of materials to pinpoint the essence of AI-driven material design and to underscore the pivotal role of AI technology in the process of inverse design of materials. Furthermore, we conduct a comprehensive review of the latest endeavors aiming at enhancing AI-driven inverse design processes, encompassing recent AI-based discoveries on typical materials and the research progress and technological trajectories of AI in the field of materials science. These collective efforts have significantly contributed to the recent wave of advancements in AI-driven inverse design of materials. While existing AI-driven inverse design of materials has yielded promising results, particularly in the inverse design of functional materials for few-shot learning scenarios, the future development of AI technology in materials science and its application in key areas remain open questions. Especially with the burgeoning growth of large language model technology, it is crucial to explore which aspects of materials science can benefit from improvements and optimizations offered by large language models. We hope that our insights will inspire further efforts in this field and propel the development of AI-driven inverse design of materials forward.

\section*{Note and Contribution}
\textbf{Author contributions:}
The contributions of all authors are listed as follows:
Ze-Feng Gao and Zhong-Yi Lu lead this survey program; Ze-Feng Gao and Xiao-Qi Han designed the structure of this paper; Ze-Feng Gao drafted the abstract and Section 1; 
Xiao-Qi Han, Xin-De Wang, Meng-Yuan Xu, Zhen Feng and Bo-Wen Yao drafted Section 2, Section 3. All faculty authors drafted various topics in Section 4. Peng-Jie Guo provided comments to the manuscript, and Xiao-Qi Han and Ze-Feng Gao proofread the whole paper. The authors would like to thank Li-Jun Chen and Xi-Wen Liu for useful discussion.

\textbf{Survey Writing.} This survey was planned during a discussion meeting held by our research team, with the objective of summarizing the latest advancements in AI-driven inverse design of materials into a highly readable report. Then, we have extensively revised the writing and contents in several passes. Due to the space limit, we can only include a fraction of existing AI methods in Figure~\ref{fig:aitec} by setting the selection criterion. We release the initial version on November 14, 2024, and this latest version on November 26, 2024.

\textbf{Seeking for Advise.} Despite our best efforts, this survey may still have many shortcomings, such as the potential omission of important references, models, methods, or topics, as well as the possibility of imprecise expressions and discussions. We will continue to update this survey and strive to enhance its quality as much as possible. For our research team, the process of writing this survey is also a journey of learning about AI-driven inverse design of materials research. Readers who have constructive suggestions to improve this survey are welcome to send emails to our authors. We will revise future versions based on the comments and suggestions received and acknowledge those readers who have contributed constructive suggestions in our survey.

\textbf{Update log.}
In this part, we regularly maintain an update log for the submissions of this survey to arXiv:

$\bullet$  First release on November 14, 2024: the initial version.

$\bullet$  Update on November 26, 2024: add several related work in Section~\ref{sec-2} and Section~\ref{sec-3}, revise Figure~\ref{fig:aitec} and Table~\ref{tab:datasets}, add the Figure~\ref{fig:InvDesFlow}, improve the writing, and correct some minor errors.

\bibliographystyle{IEEEtran}
\bibliography{ref_article}

\begin{thebibliography}{100}
\providecommand{\url}[1]{#1}
\csname url@samestyle\endcsname
\providecommand{\newblock}{\relax}
\providecommand{\bibinfo}[2]{#2}
\providecommand{\BIBentrySTDinterwordspacing}{\spaceskip=0pt\relax}
\providecommand{\BIBentryALTinterwordstretchfactor}{4}
\providecommand{\BIBentryALTinterwordspacing}{\spaceskip=\fontdimen2\font plus
\BIBentryALTinterwordstretchfactor\fontdimen3\font minus \fontdimen4\font\relax}
\providecommand{\BIBforeignlanguage}[2]{{%
\expandafter\ifx\csname l@#1\endcsname\relax
\typeout{** WARNING: IEEEtran.bst: No hyphenation pattern has been}%
\typeout{** loaded for the language `#1'. Using the pattern for}%
\typeout{** the default language instead.}%
\else
\language=\csname l@#1\endcsname
\fi
#2}}
\providecommand{\BIBdecl}{\relax}
\BIBdecl

\bibitem{zunger2018inverse}
A.~Zunger, ``Inverse design in search of materials with target functionalities,'' \emph{Nature Reviews Chemistry}, vol.~2, no.~4, p. 0121, 2018.

\bibitem{lee2023machine}
J.~Lee, D.~Park, M.~Lee, H.~Lee, K.~Park, I.~Lee, and S.~Ryu, ``Machine learning-based inverse design methods considering data characteristics and design space size in materials design and manufacturing: a review,'' \emph{Materials Horizons}, 2023.

\bibitem{wang2022inverse}
J.~Wang, Y.~Wang, and Y.~Chen, ``Inverse design of materials by machine learning,'' \emph{Materials}, vol.~15, no.~5, p. 1811, 2022.

\bibitem{long2024generative}
T.~Long, Y.~Zhang, and H.~Zhang, ``Generative deep learning for the inverse design of materials,'' \emph{arXiv preprint arXiv:2409.19124}, 2024.

\bibitem{wu2024physics}
Y.~Wu, ``Physics-informed machine learning methods for inverse design of multi-phase materials with targeted mechanical properties,'' 2024.

\bibitem{chen2021generative}
L.~Chen, W.~Zhang, Z.~Nie, S.~Li, and F.~Pan, ``Generative models for inverse design of inorganic solid materials,'' \emph{J. Mater. Inform}, vol.~1, no.~4, 2021.

\bibitem{abu2023inverse}
M.~Abu-Mualla and J.~Huang, ``Inverse design of 3d cellular materials with physics-guided machine learning,'' \emph{Materials \& Design}, vol. 232, p. 112103, 2023.

\bibitem{onnes1911further}
H.~K. Onnes, ``Further experiments with liquid helium,'' in \emph{Proceedings of the KNAW}, vol.~13.\hskip 1em plus 0.5em minus 0.4em\relax sn, 1911, pp. 1910--1911.

\bibitem{nagamatsu2001superconductivity}
J.~Nagamatsu, N.~Nakagawa, T.~Muranaka, Y.~Zenitani, and J.~Akimitsu, ``Superconductivity at 39 k in magnesium diboride,'' \emph{nature}, vol. 410, no. 6824, pp. 63--64, 2001.

\bibitem{kim2020inverse}
B.~Kim, S.~Lee, and J.~Kim, ``Inverse design of porous materials using artificial neural networks,'' \emph{Science advances}, vol.~6, no.~1, p. eaax9324, 2020.

\bibitem{vidmar2011dirac}
D.~Vidmar, ``The dirac equation and the prediction of antimatter,'' \emph{PDF document provided on the internet by the Universidade Federal do Rio Grande do Sul}, 2011.

\bibitem{anderson1932apparent}
C.~D. Anderson, ``The apparent existence of easily deflectable positives,'' \emph{Science}, vol.~76, no. 1967, pp. 238--239, 1932.

\bibitem{bardeen1957theory}
J.~Bardeen, L.~N. Cooper, and J.~R. Schrieffer, ``Theory of superconductivity,'' \emph{Physical review}, vol. 108, no.~5, p. 1175, 1957.

\bibitem{bardeen1955theory}
J.~Bardeen, ``Theory of the meissner effect in superconductors,'' \emph{Physical Review}, vol.~97, no.~6, p. 1724, 1955.

\bibitem{cooper1956bound}
L.~N. Cooper, ``Bound electron pairs in a degenerate fermi gas,'' \emph{Physical Review}, vol. 104, no.~4, p. 1189, 1956.

\bibitem{DFT-1}
\BIBentryALTinterwordspacing
G.~Kresse and J.~Furthm\"uller, ``Efficient iterative schemes for ab initio total-energy calculations using a plane-wave basis set,'' \emph{Phys. Rev. B}, vol.~54, pp. 11\,169--11\,186, Oct 1996. [Online]. Available: \url{https://link.aps.org/doi/10.1103/PhysRevB.54.11169}
\BIBentrySTDinterwordspacing

\bibitem{DFT-2}
\BIBentryALTinterwordspacing
J.~P. Perdew, K.~Burke, and M.~Ernzerhof, ``Generalized gradient approximation made simple,'' \emph{Phys. Rev. Lett.}, vol.~77, pp. 3865--3868, Oct 1996. [Online]. Available: \url{https://link.aps.org/doi/10.1103/PhysRevLett.77.3865}
\BIBentrySTDinterwordspacing

\bibitem{DFT-3}
\BIBentryALTinterwordspacing
P.~E. Bl\"ochl, ``Projector augmented-wave method,'' \emph{Phys. Rev. B}, vol.~50, pp. 17\,953--17\,979, Dec 1994. [Online]. Available: \url{https://link.aps.org/doi/10.1103/PhysRevB.50.17953}
\BIBentrySTDinterwordspacing

\bibitem{novoselov2004electric}
K.~S. Novoselov, A.~K. Geim, S.~V. Morozov, D.-e. Jiang, Y.~Zhang, S.~V. Dubonos, I.~V. Grigorieva, and A.~A. Firsov, ``Electric field effect in atomically thin carbon films,'' \emph{science}, vol. 306, no. 5696, pp. 666--669, 2004.

\bibitem{pople1976theoretical}
J.~A. Pople, J.~S. Binkley, and R.~Seeger, ``Theoretical models incorporating electron correlation,'' \emph{International Journal of Quantum Chemistry}, vol.~10, no. S10, pp. 1--19, 1976.

\bibitem{park2024has}
H.~Park, Z.~Li, and A.~Walsh, ``Has generative artificial intelligence solved inverse materials design?'' \emph{Matter}, vol.~7, no.~7, pp. 2355--2367, 2024.

\bibitem{vaswani2017attention}
A.~Vaswani, ``Attention is all you need,'' \emph{Advances in Neural Information Processing Systems}, 2017.

\bibitem{CDVAE2022}
\BIBentryALTinterwordspacing
T.~Xie, X.~Fu, O.-E. Ganea, R.~Barzilay, and T.~Jaakkola, ``Crystal diffusion variational autoencoder for periodic material generation,'' 2022. [Online]. Available: \url{https://arxiv.org/abs/2110.06197}
\BIBentrySTDinterwordspacing

\bibitem{jiao2024crystal}
\BIBentryALTinterwordspacing
R.~Jiao, W.~Huang, P.~Lin, J.~Han, P.~Chen, Y.~Lu, and Y.~Liu, ``Crystal structure prediction by joint equivariant diffusion,'' \emph{NeurIPS}, vol.~36, pp. 17\,464--17\,497, 2023. [Online]. Available: \url{https://proceedings.neurips.cc/paper_files/paper/2023/file/38b787fc530d0b31825827e2cc306656-Paper-Conference.pdf}
\BIBentrySTDinterwordspacing

\bibitem{chen2024mattergpt}
Y.~Chen, X.~Wang, X.~Deng, Y.~Liu, X.~Chen, Y.~Zhang, L.~Wang, and H.~Xiao, ``Mattergpt: A generative transformer for multi-property inverse design of solid-state materials,'' \emph{arXiv preprint arXiv:2408.07608}, 2024.

\bibitem{choudhary2024atomgpt}
K.~Choudhary, ``Atomgpt: Atomistic generative pretrained transformer for forward and inverse materials design,'' \emph{The Journal of Physical Chemistry Letters}, vol.~15, no.~27, pp. 6909--6917, 2024.

\bibitem{fernandez2024denoising}
P.~Fernandez-Zelaia, S.~Thapliyal, R.~Kannan, P.~Nandwana, Y.~Yamamoto, A.~Nycz, V.~Paquit, and M.~M. Kirka, ``Denoising diffusion probabilistic models for generative alloy design,'' \emph{Additive Manufacturing}, p. 104478, 2024.

\bibitem{lyu2024microstructure}
X.~Lyu and X.~Ren, ``Microstructure reconstruction of 2d/3d random materials via diffusion-based deep generative models,'' \emph{Scientific Reports}, vol.~14, no.~1, p. 5041, 2024.

\bibitem{gao2023ai}
Z.-F. Gao, S.~Qu, B.~Zeng, Y.~Liu, J.-R. Wen, H.~Sun, P.-J. Guo, and Z.-Y. Lu, ``Ai-accelerated discovery of altermagnetic materials,'' \emph{arXiv preprint arXiv:2311.04418}, 2023.

\bibitem{ansari2024dziner}
M.~Ansari, J.~Watchorn, C.~E. Brown, and J.~S. Brown, ``dziner: Rational inverse design of materials with ai agents,'' \emph{arXiv preprint arXiv:2410.03963}, 2024.

\bibitem{merchant2023scaling}
A.~Merchant, S.~Batzner, S.~S. Schoenholz, M.~Aykol, G.~Cheon, and E.~D. Cubuk, ``Scaling deep learning for materials discovery,'' \emph{Nature}, vol. 624, no. 7990, pp. 80--85, 2023.

\bibitem{OMat24}
L.~Barroso-Luque, M.~Shuaibi, X.~Fu, B.~M. Wood, M.~Dzamba, M.~Gao, A.~Rizvi, C.~L. Zitnick, and Z.~W. Ulissi, ``Open materials 2024 (omat24) inorganic materials dataset and models,'' \emph{arXiv preprint arXiv:2410.12771}, 2024.

\bibitem{chen2022inverse}
Y.~Chen, Z.~Lan, Z.~Su, and J.~Zhu, ``Inverse design of photonic and phononic topological insulators: a review,'' \emph{Nanophotonics}, vol.~11, no.~19, pp. 4347--4362, 2022.

\bibitem{liu2024machine}
S.~Liu and C.~Yang, ``Machine learning design for high-entropy alloys: models and algorithms,'' \emph{Metals}, vol.~14, no.~2, p. 235, 2024.

\bibitem{han2024surveygeometricgraphneural}
\BIBentryALTinterwordspacing
J.~Han, J.~Cen, L.~Wu, Z.~Li, X.~Kong, R.~Jiao, Z.~Yu, T.~Xu, F.~Wu, Z.~Wang, H.~Xu, Z.~Wei, Y.~Liu, Y.~Rong, and W.~Huang, ``A survey of geometric graph neural networks: Data structures, models and applications,'' 2024. [Online]. Available: \url{https://arxiv.org/abs/2403.00485}
\BIBentrySTDinterwordspacing

\bibitem{reiser2022graph}
\BIBentryALTinterwordspacing
P.~Reiser, M.~Neubert, A.~Eberhard, L.~Torresi, C.~Zhou, C.~Shao, H.~Metni, C.~van Hoesel, H.~Schopmans, T.~Sommer, and P.~Friederich, ``Graph neural networks for materials science and chemistry,'' \emph{Communications Materials}, vol.~3, no.~1, p.~93, 2022. [Online]. Available: \url{https://doi.org/10.1038/s43246-022-00315-6}
\BIBentrySTDinterwordspacing

\bibitem{lvovsky2013novel}
\BIBentryALTinterwordspacing
Y.~Lvovsky, E.~W. Stautner, and T.~Zhang, ``Novel technologies and configurations of superconducting magnets for mri,'' \emph{Supercond. Sci. Technol.}, vol.~26, p. 093001, 2013. [Online]. Available: \url{https://api.semanticscholar.org/CorpusID:121285609}
\BIBentrySTDinterwordspacing

\bibitem{bruzzone2018high}
P.~Bruzzone, W.~H. Fietz, J.~V. Minervini, M.~Novikov, N.~Yanagi, Y.~Zhai, and J.~Zheng, ``High temperature superconductors for fusion magnets,'' \emph{Nucl. Fusion}, vol.~58, p. 103001, 2018.

\bibitem{mirhosseini2020superconducting}
M.~Mirhosseini, A.~Sipahigil, M.~Kalaee, and O.~Painter, ``Superconducting qubit to optical photon transduction,'' \emph{Nature}, vol. 588, p. 599, 2020.

\bibitem{gambetta2017building}
J.~M. Gambetta, J.~M. Chow, and M.~Steffen, ``Building logical qubits in a superconducting quantum computing system,'' \emph{NPJ Quantum Inf.}, vol.~3, p.~2, 2017.

\bibitem{degen2017quantum}
\BIBentryALTinterwordspacing
C.~L. Degen, F.~Reinhard, and P.~Cappellaro, ``Quantum sensing,'' \emph{Rev. Mod. Phys.}, vol.~89, p. 035002, Jul 2017. [Online]. Available: \url{https://link.aps.org/doi/10.1103/RevModPhys.89.035002}
\BIBentrySTDinterwordspacing

\bibitem{onnes1911resistance}
\BIBentryALTinterwordspacing
H.~K. Onnes, ``The resistance of pure mercury at helium temperatures,'' \emph{Commun. Phys. Lab. Univ. Leiden, b}, vol. 120, 1911. [Online]. Available: \url{{https://cir.nii.ac.jp/crid/1572543024905854592}}
\BIBentrySTDinterwordspacing

\bibitem{Meissner1933}
\BIBentryALTinterwordspacing
W.~Meissner and R.~Ochsenfeld, ``Ein neuer effekt bei eintritt der supraleitfähigkeit,'' \emph{Naturwissenschaften}, vol.~21, no.~44, pp. 787--788, 1933. [Online]. Available: \url{https://doi.org/10.1007/BF01504252}
\BIBentrySTDinterwordspacing

\bibitem{gavaler1973superconductivity}
J.~Gavaler, ``Superconductivity in nb--ge films above 22 k,'' \emph{Applied Physics Letters}, vol.~23, no.~8, pp. 480--482, 1973.

\bibitem{bednorz1986possible}
J.~G. Bednorz and K.~A. M{\"u}ller, ``Possible high t c superconductivity in the ba- la- cu- o system,'' \emph{Zeitschrift f{\"u}r Physik B Condensed Matter}, vol.~64, no.~2, pp. 189--193, 1986.

\bibitem{PhysRevLett.58.908}
\BIBentryALTinterwordspacing
M.~K. Wu, J.~R. Ashburn, C.~J. Torng, P.~H. Hor, R.~L. Meng, L.~Gao, Z.~J. Huang, Y.~Q. Wang, and C.~W. Chu, ``Superconductivity at 93 k in a new mixed-phase y-ba-cu-o compound system at ambient pressure,'' \emph{Phys. Rev. Lett.}, vol.~58, pp. 908--910, Mar 1987. [Online]. Available: \url{https://link.aps.org/doi/10.1103/PhysRevLett.58.908}
\BIBentrySTDinterwordspacing

\bibitem{zhi2008superconductivity}
R.~Zhi-An, L.~Wei, Y.~Jie, Y.~Wei, S.~Xiao-Li, C.~Guang-Can, D.~Xiao-Li, S.~Li-Ling, Z.~Fang, and Z.~Zhong-Xian, ``Superconductivity at 55 k in iron-based f-doped layered quaternary compound sm[o$_{1-x}$f$_{x}$] feas,'' \emph{Chinese Physics Letters}, vol.~25, no.~6, p. 2215, 2008.

\bibitem{PhysRevLett.130.256002}
\BIBentryALTinterwordspacing
J.~Ying, S.~Liu, Q.~Lu, X.~Wen, Z.~Gui, Y.~Zhang, X.~Wang, J.~Sun, and X.~Chen, ``Record high 36 k transition temperature to the superconducting state of elemental scandium at a pressure of 260 gpa,'' \emph{Phys. Rev. Lett.}, vol. 130, p. 256002, Jun 2023. [Online]. Available: \url{https://link.aps.org/doi/10.1103/PhysRevLett.130.256002}
\BIBentrySTDinterwordspacing

\bibitem{Sun-nature}
H.~Sun, M.~Huo, X.~Hu, J.~Li, Z.~Liu, Y.~Han, L.~Tang, Z.~Mao, P.~Yang, B.~Wang, J.~Cheng, D.-X. Yao, G.-M. Zhang, and M.~Wang, ``Signatures of superconductivity near 80 $\rm{K}$ in a nickelate under high pressure,'' \emph{Nature}, vol. 621, no. 7979, pp. 493--498, 2023.

\bibitem{drozdov2015conventional}
A.~Drozdov, M.~Eremets, I.~Troyan, V.~Ksenofontov, and S.~I. Shylin, ``Conventional superconductivity at 203 kelvin at high pressures in the sulfur hydride system,'' \emph{Nature}, vol. 525, no. 7567, pp. 73--76, 2015.

\bibitem{Stanev2018}
\BIBentryALTinterwordspacing
V.~Stanev, C.~Oses, A.~G. Kusne, E.~Rodriguez, J.~Paglione, S.~Curtarolo, and I.~Takeuchi, ``Machine learning modeling of superconducting critical temperature,'' \emph{npj Computational Materials}, vol.~4, p.~29, Jun 2018. [Online]. Available: \url{https://doi.org/10.1038/s41524-018-0085-8}
\BIBentrySTDinterwordspacing

\bibitem{ward2016general}
L.~Ward, A.~Agrawal, A.~Choudhary, and C.~Wolverton, ``A general-purpose machine learning framework for predicting properties of inorganic materials,'' \emph{npj Computational Materials}, vol.~2, no.~1, pp. 1--7, 2016.

\bibitem{li2024deeplearningapproachsearch}
\BIBentryALTinterwordspacing
J.~Li, W.~Fang, S.~Jin, T.~Zhang, Y.~Wu, X.~Xu, Y.~Liu, and D.-X. Yao, ``A deep learning approach to search for superconductors from electronic bands,'' 2024. [Online]. Available: \url{https://arxiv.org/abs/2409.07721}
\BIBentrySTDinterwordspacing

\bibitem{zhang2024superbandelectronicbandfermisurface}
\BIBentryALTinterwordspacing
T.~Zhang, C.~Suo, Y.~Wu, X.~Xu, Y.~Liu, D.-X. Yao, and J.~Li, ``Superband: an electronic-band and fermi surface structure database of superconductors,'' 2024. [Online]. Available: \url{https://arxiv.org/abs/2409.09419}
\BIBentrySTDinterwordspacing

\bibitem{GNN2008}
F.~Scarselli, M.~Gori, A.~C. Tsoi, M.~Hagenbuchner, and G.~Monfardini, ``The graph neural network model,'' \emph{IEEE transactions on neural networks}, vol.~20, no.~1, pp. 61--80, 2008.

\bibitem{GNNs2020}
Z.~Wu, S.~Pan, F.~Chen, G.~Long, C.~Zhang, and S.~Y. Philip, ``A comprehensive survey on graph neural networks,'' \emph{IEEE transactions on neural networks and learning systems}, vol.~32, no.~1, pp. 4--24, 2020.

\bibitem{GCN2016}
T.~N. Kipf and M.~Welling, ``Semi-supervised classification with graph convolutional networks,'' \emph{arXiv preprint arXiv:1609.02907}, 2016.

\bibitem{GraphSAGE2017}
W.~Hamilton, Z.~Ying, and J.~Leskovec, ``Inductive representation learning on large graphs,'' \emph{Advances in neural information processing systems}, vol.~30, 2017.

\bibitem{GAT2017}
P.~Veli{\v{c}}kovi{\'c}, G.~Cucurull, A.~Casanova, A.~Romero, P.~Lio, and Y.~Bengio, ``Graph attention networks,'' \emph{arXiv preprint arXiv:1710.10903}, 2017.

\bibitem{CONVgraph2016}
M.~Defferrard, X.~Bresson, and P.~Vandergheynst, ``Convolutional neural networks on graphs with fast localized spectral filtering,'' \emph{Advances in neural information processing systems}, vol.~29, 2016.

\bibitem{choudhary2022designing}
K.~Choudhary and K.~Garrity, ``Designing high-tc superconductors with bcs-inspired screening, density functional theory, and deep-learning,'' \emph{NPJ Comput. Mater.}, vol.~8, no.~1, p. 244, 2022.

\bibitem{alignn2021}
K.~Choudhary and B.~DeCost, ``Atomistic line graph neural network for improved materials property predictions,'' \emph{NPJ Comput. Mater.}, vol.~7, no.~1, p. 185, 2021.

\bibitem{bcsHalignn}
D.~Wines and K.~Choudhary, ``Data-driven design of high pressure hydride superconductors using dft and deep learning,'' \emph{Materials futures}, vol.~3, no.~2, p. 025602, 2024.

\bibitem{zhao2024machine}
J.~ZHAO, J.~WANG, D.~HE, J.~LI, Y.~SUN, X.-Q. CHEN, and P.~LIU, ``Machine learning model for predicting the critical transition temperature of hydride superconductors,'' \emph{Acta Metall Sin}, vol.~60, no.~10, pp. 1418--1428, 2024.

\bibitem{li2024machine}
J.~Li, L.~Wei, X.~Shi, L.~Shi, J.~Si, P.-F. Liu, and B.-T. Wang, ``Machine learning accelerated discovery of superconducting two-dimensional janus transition metal sulfhydrates,'' \emph{Physical Review B}, vol. 109, no.~17, p. 174516, 2024.

\bibitem{cerqueira2024searching}
T.~F. Cerqueira, Y.-W. Fang, I.~Errea, A.~Sanna, and M.~A. Marques, ``Searching materials space for hydride superconductors at ambient pressure,'' \emph{Advanced Functional Materials}, p. 2404043, 2024.

\bibitem{2023_3dsc}
T.~Sommer, R.~Willa, J.~Schmalian, and P.~Friederich, ``3dsc-a dataset of superconductors including crystal structures,'' \emph{Scientific Data}, vol.~10, no.~1, p. 816, 2023.

\bibitem{stablediffusion2022}
R.~Rombach, A.~Blattmann, D.~Lorenz, P.~Esser, and B.~Ommer, ``High-resolution image synthesis with latent diffusion models,'' in \emph{Proceedings of the IEEE/CVF conference on computer vision and pattern recognition}, 2022, pp. 10\,684--10\,695.

\bibitem{DDPM2020}
J.~Ho, A.~Jain, and P.~Abbeel, ``Denoising diffusion probabilistic models,'' \emph{Advances in neural information processing systems}, vol.~33, pp. 6840--6851, 2020.

\bibitem{song2019generative}
Y.~Song and S.~Ermon, ``Generative modeling by estimating gradients of the data distribution,'' \emph{Advances in neural information processing systems}, vol.~32, 2019.

\bibitem{alphafold3}
J.~Abramson, J.~Adler, J.~Dunger, R.~Evans, T.~Green, A.~Pritzel, O.~Ronneberger, L.~Willmore, A.~J. Ballard, J.~Bambrick \emph{et~al.}, ``Accurate structure prediction of biomolecular interactions with alphafold 3,'' \emph{Nature}, pp. 1--3, 2024.

\bibitem{xiaoqiAIsuper}
\BIBentryALTinterwordspacing
X.-Q. Han, Z.~Ouyang, P.-J. Guo, H.~Sun, Z.-F. Gao, and Z.-Y. Lu, ``Ai-accelerated discovery of high critical temperature superconductors,'' 2024. [Online]. Available: \url{https://arxiv.org/abs/2409.08065}
\BIBentrySTDinterwordspacing

\bibitem{xie2018crystal}
\BIBentryALTinterwordspacing
T.~Xie and J.~C. Grossman, ``Crystal graph convolutional neural networks for an accurate and interpretable prediction of material properties,'' \emph{Phys. Rev. Lett.}, vol. 120, p. 145301, Apr 2018. [Online]. Available: \url{https://link.aps.org/doi/10.1103/PhysRevLett.120.145301}
\BIBentrySTDinterwordspacing

\bibitem{MatAltMag}
\BIBentryALTinterwordspacing
Z.-F. Gao, S.~Qu, B.~Zeng, Y.~Liu, J.-R. Wen, H.~Sun, P.-J. Guo, and Z.-Y. Lu, ``Ai-accelerated discovery of altermagnetic materials,'' \emph{arXiv}, vol. 2311.04418, 2024. [Online]. Available: \url{https://arxiv.org/abs/2311.04418}
\BIBentrySTDinterwordspacing

\bibitem{chen2019graph}
C.~Chen, W.~Ye, Y.~Zuo, C.~Zheng, and S.~P. Ong, ``Graph networks as a universal machine learning framework for molecules and crystals,'' \emph{Chem. Mater}, vol.~31, no.~9, pp. 3564--3572, 2019.

\bibitem{Zhang_2023}
\BIBentryALTinterwordspacing
C.~Zhang, H.~Tang, C.~Pan, H.~Jiang, H.-J. Sun, K.-M. Ho, and C.-Z. Wang, ``Machine learning guided discovery of superconducting calcium borocarbides,'' \emph{Physical Review B}, vol. 108, no.~2, Jul. 2023. [Online]. Available: \url{http://dx.doi.org/10.1103/PhysRevB.108.024512}
\BIBentrySTDinterwordspacing

\bibitem{zhang2023dpa}
\BIBentryALTinterwordspacing
D.~Zhang, X.~Liu, X.~Zhang, C.~Zhang, C.~Cai, H.~Bi, Y.~Du, X.~Qin, J.~Huang, B.~Li \emph{et~al.}, ``Dpa-2: Towards a universal large atomic model for molecular and material simulation,'' \emph{arXiv}, vol. 2312.15492, 2023. [Online]. Available: \url{https://arxiv.org/abs/2312.15492}
\BIBentrySTDinterwordspacing

\bibitem{wines2023inverse}
D.~Wines, T.~Xie, and K.~Choudhary, ``Inverse design of next-generation superconductors using data-driven deep generative models,'' \emph{The Journal of Physical Chemistry Letters}, vol.~14, no.~29, pp. 6630--6638, 2023.

\bibitem{phononprediction_2024}
\BIBentryALTinterwordspacing
R.~Okabe, A.~Chotrattanapituk, A.~Boonkird, N.~Andrejevic, X.~Fu, T.~S. Jaakkola, Q.~Song, T.~Nguyen, N.~Drucker, S.~Mu, Y.~Wang, B.~Liao, Y.~Cheng, and M.~Li, ``Virtual node graph neural network for full phonon prediction,'' \emph{Nature Computational Science}, vol.~4, no.~7, pp. 522--531, jul 2024. [Online]. Available: \url{https://doi.org/10.1038/s43588-024-00661-0}
\BIBentrySTDinterwordspacing

\bibitem{Zhong_epc_2024}
\BIBentryALTinterwordspacing
Y.~Zhong, S.~Liu, B.~Zhang, Z.~Tao, Y.~Sun, W.~Chu, X.-G. Gong, J.-H. Yang, and H.~Xiang, ``Accelerating the calculation of electron–phonon coupling strength with machine learning,'' \emph{Nature Computational Science}, vol.~4, no.~8, pp. 615--625, aug 2024. [Online]. Available: \url{https://doi.org/10.1038/s43588-024-00668-7}
\BIBentrySTDinterwordspacing

\bibitem{scalingCuO2Tc}
O.~Chmaissem, J.~Jorgensen, S.~Short, A.~Knizhnik, Y.~Eckstein, and H.~Shaked, ``Scaling of transition temperature and cuo2 plane buckling in a high-temperature superconductor,'' \emph{Nature}, vol. 397, no. 6714, pp. 45--48, 1999.

\bibitem{lee2012relationship}
C.~Lee, K.~Kihou, A.~Iyo, H.~Kito, P.~Shirage, and H.~Eisaki, ``Relationship between crystal structure and superconductivity in iron-based superconductors,'' \emph{Solid State Communications}, vol. 152, no.~8, pp. 644--648, 2012.

\bibitem{mizuguchi2010anion}
Y.~Mizuguchi, Y.~Hara, K.~Deguchi, S.~Tsuda, T.~Yamaguchi, K.~Takeda, H.~Kotegawa, H.~Tou, and Y.~Takano, ``Anion height dependence of tc for the fe-based superconductor,'' \emph{Superconductor Science and Technology}, vol.~23, no.~5, p. 054013, 2010.

\bibitem{peng2017influence}
Y.~Peng, G.~Dellea, M.~Minola, M.~Conni, A.~Amorese, D.~Di~Castro, G.~De~Luca, K.~Kummer, M.~Salluzzo, X.~Sun \emph{et~al.}, ``Influence of apical oxygen on the extent of in-plane exchange interaction in cuprate superconductors,'' \emph{Nature Physics}, vol.~13, no.~12, pp. 1201--1206, 2017.

\bibitem{gu2024bond}
L.~Gu, Y.~Liu, P.~Chen, H.~Huang, N.~Chen, Y.~Li, T.~Lookman, Y.~Lu, and Y.~Su, ``Bond sensitive graph neural networks for predicting high temperature superconductors,'' \emph{Materials Genome Engineering Advances}, p. e48, 2024.

\bibitem{Pogue2023}
\BIBentryALTinterwordspacing
E.~A. Pogue, A.~New, K.~McElroy, N.~Q. Le, M.~J. Pekala, I.~McCue, E.~Gienger, J.~Domenico, E.~Hedrick, T.~M. McQueen, B.~Wilfong, C.~D. Piatko, C.~R. Ratto, A.~Lennon, C.~Chung, T.~Montalbano, G.~Bassen, and C.~D. Stiles, ``Closed-loop superconducting materials discovery,'' \emph{npj Computational Materials}, vol.~9, no.~1, p. 181, 2023. [Online]. Available: \url{https://doi.org/10.1038/s41524-023-01131-3}
\BIBentrySTDinterwordspacing

\bibitem{cullity2011introduction}
B.~D. Cullity and C.~D. Graham, \emph{Introduction to magnetic materials}.\hskip 1em plus 0.5em minus 0.4em\relax John Wiley \& Sons, 2011.

\bibitem{goldman2012handbook}
A.~Goldman, \emph{Handbook of modern ferromagnetic materials}.\hskip 1em plus 0.5em minus 0.4em\relax Springer Science \& Business Media, 2012, vol. 505.

\bibitem{jungwirth2016antiferromagnetic}
T.~Jungwirth, X.~Marti, P.~Wadley, and J.~Wunderlich, ``Antiferromagnetic spintronics,'' \emph{Nature nanotechnology}, vol.~11, no.~3, pp. 231--241, 2016.

\bibitem{PhysRevX.12.040501}
\BIBentryALTinterwordspacing
L.~\ifmmode~\check{S}\else \v{S}\fi{}mejkal, J.~Sinova, and T.~Jungwirth, ``Emerging research landscape of altermagnetism,'' \emph{Phys. Rev. X}, vol.~12, p. 040501, Dec 2022. [Online]. Available: \url{https://link.aps.org/doi/10.1103/PhysRevX.12.040501}
\BIBentrySTDinterwordspacing

\bibitem{PhysRevX.12.031042}
\BIBentryALTinterwordspacing
------, ``Beyond conventional ferromagnetism and antiferromagnetism: A phase with nonrelativistic spin and crystal rotation symmetry,'' \emph{Phys. Rev. X}, vol.~12, p. 031042, Sep 2022. [Online]. Available: \url{https://link.aps.org/doi/10.1103/PhysRevX.12.031042}
\BIBentrySTDinterwordspacing

\bibitem{PhysRevX.12.040002}
\BIBentryALTinterwordspacing
I.~Mazin, ``Editorial: Altermagnetism---a new punch line of fundamental magnetism,'' \emph{Phys. Rev. X}, vol.~12, p. 040002, Dec 2022. [Online]. Available: \url{https://link.aps.org/doi/10.1103/PhysRevX.12.040002}
\BIBentrySTDinterwordspacing

\bibitem{doi:10.7566/JPSJ.88.123702}
\BIBentryALTinterwordspacing
S.~Hayami, Y.~Yanagi, and H.~Kusunose, ``Momentum-dependent spin splitting by collinear antiferromagnetic ordering,'' \emph{Journal of the Physical Society of Japan}, vol.~88, no.~12, p. 123702, 2019. [Online]. Available: \url{https://doi.org/10.7566/JPSJ.88.123702}
\BIBentrySTDinterwordspacing

\bibitem{altermagnetism-2}
\BIBentryALTinterwordspacing
L.~$\check{\rm{S}}$mejkal, R.~Gonz$\acute{\rm{a}}$lez-Hern$\acute{\rm{a}}$ndez, T.~Jungwirth, and J.~Sinova, ``{Crystal time-reversal symmetry breaking and spontaneous Hall effect in collinear antiferromagnets},'' \emph{Sci. Adv.}, vol.~6, no.~23, p. eaaz8809, 2020. [Online]. Available: \url{https://www.science.org/doi/abs/10.1126/sciadv.aaz8809}
\BIBentrySTDinterwordspacing

\bibitem{altermagnetism-3}
\BIBentryALTinterwordspacing
L.-D. Yuan, Z.~Wang, J.-W. Luo, E.~I. Rashba, and A.~Zunger, ``{Giant momentum-dependent spin splitting in centrosymmetric low-$Z$ antiferromagnets},'' \emph{Phys. Rev. B}, vol. 102, p. 014422, Jul 2020. [Online]. Available: \url{https://link.aps.org/doi/10.1103/PhysRevB.102.014422}
\BIBentrySTDinterwordspacing

\bibitem{altermagnetism-4}
\BIBentryALTinterwordspacing
I.~I. Mazin, K.~Koepernik, M.~D. Johannes, R.~Gonz$\acute{\rm{a}}$lez-Hern$\acute{\rm{a}}$ndez, and L.~$\check{\rm{S}}$mejkal, ``{Prediction of unconventional magnetism in doped FeSb$_2$},'' \emph{Proc. Natl. Acad. Sci. U.S.A.}, vol. 118, no.~42, p. e2108924118, 2021. [Online]. Available: \url{https://www.pnas.org/doi/abs/10.1073/pnas.2108924118}
\BIBentrySTDinterwordspacing

\bibitem{jungwirth2024altermagnets}
T.~Jungwirth, R.~M. Fernandes, J.~Sinova, and L.~Smejkal, ``Altermagnets and beyond: Nodal magnetically-ordered phases,'' \emph{arXiv preprint arXiv:2409.10034}, 2024.

\bibitem{PhysRevB.103.245127}
\BIBentryALTinterwordspacing
A.~Bouhon, G.~F. Lange, and R.-J. Slager, ``Topological correspondence between magnetic space group representations and subdimensions,'' \emph{Phys. Rev. B}, vol. 103, p. 245127, Jun 2021. [Online]. Available: \url{https://link.aps.org/doi/10.1103/PhysRevB.103.245127}
\BIBentrySTDinterwordspacing

\bibitem{hubert2008magnetic}
A.~Hubert and R.~Schafer, \emph{Magnetic Domains: The Analysis of Magnetic Microstructures}, 1st~ed.\hskip 1em plus 0.5em minus 0.4em\relax Springer Berlin, Heidelberg, 1998.

\bibitem{GMR-PRX2022}
\BIBentryALTinterwordspacing
L.~$\check{\rm{S}}$mejkal, A.~B. Hellenes, R.~Gonz\'alez-Hern\'andez, J.~Sinova, and T.~Jungwirth, ``{Giant and Tunneling Magnetoresistance in Unconventional Collinear Antiferromagnets with Nonrelativistic Spin-Momentum Coupling},'' \emph{Phys. Rev. X}, vol.~12, p. 011028, Feb 2022. [Online]. Available: \url{https://link.aps.org/doi/10.1103/PhysRevX.12.011028}
\BIBentrySTDinterwordspacing

\bibitem{SC-AM}
\BIBentryALTinterwordspacing
D.~Zhu, Z.-Y. Zhuang, Z.~Wu, and Z.~Yan, ``Topological superconductivity in two-dimensional altermagnetic metals,'' \emph{Phys. Rev. B}, vol. 108, p. 184505, Nov 2023. [Online]. Available: \url{https://link.aps.org/doi/10.1103/PhysRevB.108.184505}
\BIBentrySTDinterwordspacing

\bibitem{MCM-liu2023}
\BIBentryALTinterwordspacing
Y.-X. Li and C.-C. Liu, ``{Majorana corner modes and tunable patterns in an altermagnet heterostructure},'' \emph{Phys. Rev. B}, vol. 108, p. 205410, Nov 2023. [Online]. Available: \url{https://link.aps.org/doi/10.1103/PhysRevB.108.205410}
\BIBentrySTDinterwordspacing

\bibitem{piezomagnetism-NC}
H.-Y. Ma, M.~Hu, N.~Li, J.~Liu, W.~Yao, J.-F. Jia, and J.~Liu, ``Multifunctional antiferromagnetic materials with giant piezomagnetism and noncollinear spin current,'' \emph{Nat. Commun.}, vol.~12, p. 2846, MAY 14 2021.

\bibitem{SST-PRL2021}
\BIBentryALTinterwordspacing
R.~Gonz\'alez-Hern\'andez, L.~$\check{\rm{S}}$mejkal, K.~V\'yborn\'y, Y.~Yahagi, J.~Sinova, T.~c.~v. Jungwirth, and J.~$\check{\rm{Z}}$elezn\'y, ``{Efficient Electrical Spin Splitter Based on Nonrelativistic Collinear Antiferromagnetism},'' \emph{Phys. Rev. Lett.}, vol. 126, p. 127701, Mar 2021. [Online]. Available: \url{https://link.aps.org/doi/10.1103/PhysRevLett.126.127701}
\BIBentrySTDinterwordspacing

\bibitem{SST-PRL2022}
\BIBentryALTinterwordspacing
H.~Bai, L.~Han, X.~Y. Feng, Y.~J. Zhou, R.~X. Su, Q.~Wang, L.~Y. Liao, W.~X. Zhu, X.~Z. Chen, F.~Pan, X.~L. Fan, and C.~Song, ``{Observation of Spin Splitting Torque in a Collinear Antiferromagnet ${\mathrm{RuO}}_{2}$},'' \emph{Phys. Rev. Lett.}, vol. 128, p. 197202, May 2022. [Online]. Available: \url{https://link.aps.org/doi/10.1103/PhysRevLett.128.197202}
\BIBentrySTDinterwordspacing

\bibitem{SST-PRL2022-2}
\BIBentryALTinterwordspacing
S.~Karube, T.~Tanaka, D.~Sugawara, N.~Kadoguchi, M.~Kohda, and J.~Nitta, ``{Observation of Spin-Splitter Torque in Collinear Antiferromagnetic ${\mathrm{RuO}}_{2}$},'' \emph{Phys. Rev. Lett.}, vol. 129, p. 137201, Sep 2022. [Online]. Available: \url{https://link.aps.org/doi/10.1103/PhysRevLett.129.137201}
\BIBentrySTDinterwordspacing

\bibitem{AHE-Sinova2022}
L.~$\check{\rm{S}}$mejkal, A.~H. MacDonald, J.~Sinova, S.~Nakatsuji, and T.~Jungwirth, ``{Anomalous Hall antiferromagnets},'' \emph{Nat. Rev. Mater.}, vol.~7, no.~6, pp. 482--496, JUN 2022.

\bibitem{AHE-RuO2-NE2022}
Z.~Feng, X.~Zhou, L.~$\check{\rm{S}}$mejkal, L.~Wu, Z.~Zhu, H.~Guo, R.~Gonz\'alez-Hern\'andez, X.~Wang, H.~Yan, P.~Qin, X.~Zhang, H.~Wu, H.~Chen, Z.~Meng, L.~Liu, Z.~Xia, J.~Sinova, T.~Jungwirth, and Z.~Liu, ``An anomalous hall effect in altermagnetic ruthenium dioxide,'' \emph{Nat. Electron.}, vol.~5, no.~11, p. 735, NOV 2022.

\bibitem{AHE-MnTe-PRL2023}
\BIBentryALTinterwordspacing
R.~D. Gonzalez~Betancourt, J.~Zub\'a\ifmmode~\check{c}\else \v{c}\fi{}, R.~Gonzalez-Hernandez, K.~Geishendorf, Z.~\ifmmode \check{S}\else \v{S}\fi{}ob\'a\ifmmode~\check{n}\else \v{n}\fi{}, G.~Springholz, K.~Olejn\'{\i}k, L.~\ifmmode~\check{S}\else \v{S}\fi{}mejkal, J.~Sinova, T.~Jungwirth, S.~T.~B. Goennenwein, A.~Thomas, H.~Reichlov\'a, J.~\ifmmode~\check{Z}\else \v{Z}\fi{}elezn\'y, and D.~Kriegner, ``{Spontaneous Anomalous Hall Effect Arising from an Unconventional Compensated Magnetic Phase in a Semiconductor},'' \emph{Phys. Rev. Lett.}, vol. 130, p. 036702, Jan 2023. [Online]. Available: \url{https://link.aps.org/doi/10.1103/PhysRevLett.130.036702}
\BIBentrySTDinterwordspacing

\bibitem{AHE-hou2023}
\BIBentryALTinterwordspacing
X.-Y. Hou, H.-C. Yang, Z.-X. Liu, P.-J. Guo, and Z.-Y. Lu, ``{Large intrinsic anomalous Hall effect in both Nb$_2$FeB$_2$ and Ta$_2$FeB$_2$ with collinear antiferromagnetism},'' \emph{Phys. Rev. B}, vol. 107, p. L161109, Apr 2023. [Online]. Available: \url{https://link.aps.org/doi/10.1103/PhysRevB.107.L161109}
\BIBentrySTDinterwordspacing

\bibitem{MOE-Yao2021}
\BIBentryALTinterwordspacing
X.~Zhou, W.~Feng, X.~Yang, G.-Y. Guo, and Y.~Yao, ``Crystal chirality magneto-optical effects in collinear antiferromagnets,'' \emph{Phys. Rev. B}, vol. 104, p. 024401, Jul 2021. [Online]. Available: \url{https://link.aps.org/doi/10.1103/PhysRevB.104.024401}
\BIBentrySTDinterwordspacing

\bibitem{CTHE-Yao2024}
\BIBentryALTinterwordspacing
X.~Zhou, W.~Feng, R.-W. Zhang, L.~$\check{\rm{S}}$mejkal, J.~Sinova, Y.~Mokrousov, and Y.~Yao, ``{Crystal Thermal Transport in Altermagnetic RuO$_2$},'' \emph{Phys. Rev. Lett.}, vol. 132, p. 056701, Jan 2024. [Online]. Available: \url{https://link.aps.org/doi/10.1103/PhysRevLett.132.056701}
\BIBentrySTDinterwordspacing

\bibitem{QAH-npj2023}
P.-J. Guo, Z.-X. Liu, and Z.-Y. Lu, ``{Quantum anomalous hall effect in collinear antiferromagnetism},'' \emph{npj Comput. Mater.}, vol.~9, p.~70, MAY 1 2023.

\bibitem{HighoT-liu2024}
\BIBentryALTinterwordspacing
Y.-X. Li, Y.~Liu, and C.-C. Liu, ``{Creation and manipulation of higher-order topological states by altermagnets},'' \emph{Phys. Rev. B}, vol. 109, p. L201109, May 2024. [Online]. Available: \url{https://link.aps.org/doi/10.1103/PhysRevB.109.L201109}
\BIBentrySTDinterwordspacing

\bibitem{LiFe2F6-guo2023}
P.-J. Guo, Y.~Gu, Z.-F. Gao, and Z.-Y. Lu, ``{Altermagnetic ferroelectric LiFe$_2$F$_6$ and spin-triplet excitonic insulator phase},'' 2023.

\bibitem{NiF3-qu2024}
S.~Qu, Z.-F. Gao, H.~Sun, K.~Liu, P.-J. Guo, and Z.-Y. Lu, ``{Extremely strong spin-orbit coupling effect in light element altermagnetic materials},'' 2024.

\bibitem{tan2024bipolarizedweylsemimetalsquantum}
\BIBentryALTinterwordspacing
C.-Y. Tan, Z.-F. Gao, H.-C. Yang, K.~Liu, P.-J. Guo, and Z.-Y. Lu, ``Bipolarized weyl semimetals and quantum crystal valley hall effect in two-dimensional altermagnetic materials,'' 2024. [Online]. Available: \url{https://arxiv.org/abs/2406.16603}
\BIBentrySTDinterwordspacing

\bibitem{itani2024northeast}
S.~Itani, Y.~Zhang, and J.~Zang, ``Northeast materials database (nemad): Enabling discovery of high transition temperature magnetic compounds,'' \emph{arXiv preprint arXiv:2409.15675}, 2024.

\bibitem{zhang2024gptarticleextractor}
Y.~Zhang, S.~Itani, K.~Khanal, E.~Okyere, G.~Smith, K.~Takahashi, and J.~Zang, ``Gptarticleextractor: An automated workflow for magnetic material database construction,'' \emph{Journal of Magnetism and Magnetic Materials}, vol. 597, p. 172001, 2024.

\bibitem{MPdata}
A.~Jain, J.~Montoya, S.~Dwaraknath, N.~E. Zimmermann, J.~Dagdelen, M.~Horton, P.~Huck, D.~Winston, S.~Cholia, S.~P. Ong \emph{et~al.}, ``The materials project: Accelerating materials design through theory-driven data and tools,'' \emph{Handbook of Materials Modeling: Methods: Theory and Modeling}, pp. 1751--1784, 2020.

\bibitem{rowe2018thermoelectrics}
D.~M. Rowe, \emph{Thermoelectrics handbook: macro to nano}.\hskip 1em plus 0.5em minus 0.4em\relax CRC press, 2018.

\bibitem{tan2016rationally}
G.~Tan, L.-D. Zhao, and M.~G. Kanatzidis, ``Rationally designing high-performance bulk thermoelectric materials,'' \emph{Chemical reviews}, vol. 116, no.~19, pp. 12\,123--12\,149, 2016.

\bibitem{mamur2021thermoelectric}
H.~Mamur, {\"O}.~F. Dilma{\c{c}}, J.~Begum, and M.~R.~A. Bhuiyan, ``Thermoelectric generators act as renewable energy sources,'' \emph{Cleaner Materials}, vol.~2, p. 100030, 2021.

\bibitem{vedernikov1998af}
M.~Vedernikov and E.~Iordanishvili, ``Af ioffe and origin of modern semiconductor thermoelectric energy conversion,'' in \emph{Seventeenth International Conference on Thermoelectrics. Proceedings ICT98 (Cat. No. 98TH8365)}.\hskip 1em plus 0.5em minus 0.4em\relax IEEE, 1998, pp. 37--42.

\bibitem{zevalkink2018practical}
A.~Zevalkink, D.~M. Smiadak, J.~L. Blackburn, A.~J. Ferguson, M.~L. Chabinyc, O.~Delaire, J.~Wang, K.~Kovnir, J.~Martin, L.~T. Schelhas \emph{et~al.}, ``A practical field guide to thermoelectrics: Fundamentals, synthesis, and characterization,'' \emph{Applied Physics Reviews}, vol.~5, no.~2, 2018.

\bibitem{gorai2017computationally}
P.~Gorai, V.~Stevanovi{\'c}, and E.~S. Toberer, ``Computationally guided discovery of thermoelectric materials,'' \emph{Nature Reviews Materials}, vol.~2, no.~9, pp. 1--16, 2017.

\bibitem{chen2022machine}
D.~Chen, F.~Jiang, L.~Fang, Y.-B. Zhu, C.-C. Ye, and W.-S. Liu, ``Machine learning assisted discovering of new m2x3-type thermoelectric materials,'' \emph{Rare Metals}, vol.~41, no.~5, pp. 1543--1553, 2022.

\bibitem{jia2022unsupervised}
X.~Jia, Y.~Deng, X.~Bao, H.~Yao, S.~Li, Z.~Li, C.~Chen, X.~Wang, J.~Mao, F.~Cao \emph{et~al.}, ``Unsupervised machine learning for discovery of promising half-heusler thermoelectric materials,'' \emph{npj Computational Materials}, vol.~8, no.~1, p.~34, 2022.

\bibitem{na2021predicting}
G.~S. Na, S.~Jang, and H.~Chang, ``Predicting thermoelectric properties from chemical formula with explicitly identifying dopant effects,'' \emph{npj Computational Materials}, vol.~7, no.~1, p. 106, 2021.

\bibitem{sasaki2020identifying}
M.~Sasaki, S.~Ju, Y.~Xu, J.~Shiomi, and M.~Goto, ``Identifying optimal strain in bismuth telluride thermoelectric film by combinatorial gradient thermal annealing and machine learning,'' \emph{ACS Combinatorial Science}, vol.~22, no.~12, pp. 782--790, 2020.

\bibitem{li2019high}
R.~Li, X.~Li, L.~Xi, J.~Yang, D.~J. Singh, and W.~Zhang, ``High-throughput screening for advanced thermoelectric materials: diamond-like abx2 compounds,'' \emph{ACS applied materials \& interfaces}, vol.~11, no.~28, pp. 24\,859--24\,866, 2019.

\bibitem{gan2021prediction}
Y.~Gan, G.~Wang, J.~Zhou, and Z.~Sun, ``Prediction of thermoelectric performance for layered iv-v-vi semiconductors by high-throughput ab initio calculations and machine learning,'' \emph{NPJ Computational Materials}, vol.~7, no.~1, p. 176, 2021.

\bibitem{islamov2023high}
M.~Islamov, H.~Babaei, R.~Anderson, K.~B. Sezginel, J.~R. Long, A.~J. McGaughey, D.~A. Gomez-Gualdron, and C.~E. Wilmer, ``High-throughput screening of hypothetical metal-organic frameworks for thermal conductivity,'' \emph{npj Computational Materials}, vol.~9, no.~1, p.~11, 2023.

\bibitem{luo2022high}
H.~Luo, X.~Li, Y.~Wang, Y.~Jin, M.~Yao, and J.~Yang, ``High-throughput screening of room temperature active peltier cooling materials in heusler compounds,'' \emph{npj Computational Materials}, vol.~8, no.~1, p. 199, 2022.

\bibitem{choubisa2023closed}
H.~Choubisa, M.~A. Haque, T.~Zhu, L.~Zeng, M.~Vafaie, D.~Baran, and E.~H. Sargent, ``Closed-loop error-correction learning accelerates experimental discovery of thermoelectric materials,'' \emph{Advanced Materials}, vol.~35, no.~40, p. 2302575, 2023.

\bibitem{iijima1991helical}
S.~Iijima, ``Helical microtubules of graphitic carbon,'' \emph{nature}, vol. 354, no. 6348, pp. 56--58, 1991.

\bibitem{geim2007rise}
A.~K. Geim and K.~S. Novoselov, ``The rise of graphene,'' \emph{Nature materials}, vol.~6, no.~3, pp. 183--191, 2007.

\bibitem{mendoza2018functionalization}
D.~Mendoza-Cach{\'u}, J.~L{\'o}pez-Miranda, C.~Mercado-Z{\'u}{\~n}iga, and G.~Rosas, ``Functionalization of mwcnts with ag-aunps by a green method and their catalytic properties,'' \emph{Diamond and Related Materials}, vol.~84, pp. 26--31, 2018.

\bibitem{javey2003ballistic}
A.~Javey, J.~Guo, Q.~Wang, M.~Lundstrom, and H.~Dai, ``Ballistic carbon nanotube field-effect transistors,'' \emph{nature}, vol. 424, no. 6949, pp. 654--657, 2003.

\bibitem{yu2000strength}
M.-F. Yu, O.~Lourie, M.~J. Dyer, K.~Moloni, T.~F. Kelly, and R.~S. Ruoff, ``Strength and breaking mechanism of multiwalled carbon nanotubes under tensile load,'' \emph{Science}, vol. 287, no. 5453, pp. 637--640, 2000.

\bibitem{an2001supercapacitors}
K.~H. An, W.~S. Kim, Y.~S. Park, Y.~C. Choi, S.~M. Lee, D.~C. Chung, D.~J. Bae, S.~C. Lim, and Y.~H. Lee, ``Supercapacitors using single-walled carbon nanotube electrodes,'' \emph{Advanced Materials}, vol.~13, no.~7, pp. 497--500, 2001.

\bibitem{zhang2024active}
D.~Zhang, P.~Yi, X.~Lai, L.~Peng, and H.~Li, ``Active machine learning model for the dynamic simulation and growth mechanisms of carbon on metal surface,'' \emph{Nature Communications}, vol.~15, no.~1, p. 344, 2024.

\bibitem{hedman2024dynamics}
D.~Hedman, B.~McLean, C.~Bichara, S.~Maruyama, J.~A. Larsson, and F.~Ding, ``Dynamics of growing carbon nanotube interfaces probed by machine learning-enabled molecular simulations,'' \emph{Nature Communications}, vol.~15, no.~1, p. 4076, 2024.

\bibitem{li2024transforming}
Y.~Li, S.~Wang, Z.~Lv, Z.~Wang, Y.~Zhao, Y.~Xie, Y.~Xu, L.~Qian, Y.~Yang, Z.~Zhao \emph{et~al.}, ``Transforming the synthesis of carbon nanotubes with machine learning models and automation,'' \emph{arXiv preprint arXiv:2404.01006}, 2024.

\bibitem{liu2022stochastic}
B.~Liu, N.~Vu-Bac, X.~Zhuang, X.~Fu, and T.~Rabczuk, ``Stochastic full-range multiscale modeling of thermal conductivity of polymeric carbon nanotubes composites: A machine learning approach,'' \emph{Composite Structures}, vol. 289, p. 115393, 2022.

\bibitem{elaskalany2024stochastic}
M.~Elaskalany and K.~Behdinan, ``Stochastic multiscale modeling of electrical conductivity of carbon nanotube polymer nanocomposites: An interpretable machine learning approach,'' \emph{Advanced Engineering Materials}, p. 2401233, 2024.

\bibitem{matos2019predictions}
M.~Matos, S.~Pinho, and V.~Tagarielli, ``Predictions of the electrical conductivity of composites of polymers and carbon nanotubes by an artificial neural network,'' \emph{Scripta Materialia}, vol. 166, pp. 117--121, 2019.

\bibitem{safavigerdini2023predicting}
K.~Safavigerdini, K.~Nouduri, R.~Surya, A.~Reinhard, Z.~Quinlan, F.~Bunyak, M.~R. Maschmann, and K.~Palaniappan, ``Predicting mechanical properties of carbon nanotube (cnt) images using multi-layer synthetic finite element model simulations,'' in \emph{2023 IEEE International Conference on Image Processing (ICIP)}.\hskip 1em plus 0.5em minus 0.4em\relax IEEE, 2023, pp. 3264--3268.

\bibitem{papadopoulos2018neural}
V.~Papadopoulos, G.~Soimiris, D.~Giovanis, and M.~Papadrakakis, ``A neural network-based surrogate model for carbon nanotubes with geometric nonlinearities,'' \emph{Computer Methods in Applied Mechanics and Engineering}, vol. 328, pp. 411--430, 2018.

\bibitem{fernandez2016geometrical}
M.~Fernandez, H.~Shi, and A.~S. Barnard, ``Geometrical features can predict electronic properties of graphene nanoflakes,'' \emph{Carbon}, vol. 103, pp. 142--150, 2016.

\bibitem{alred2018machine}
J.~M. Alred, K.~V. Bets, Y.~Xie, and B.~I. Yakobson, ``Machine learning electron density in sulfur crosslinked carbon nanotubes,'' \emph{Composites Science and Technology}, vol. 166, pp. 3--9, 2018.

\bibitem{singh2020artificial}
A.~V. Singh, D.~Rosenkranz, M.~H.~D. Ansari, R.~Singh, A.~Kanase, S.~P. Singh, B.~Johnston, J.~Tentschert, P.~Laux, and A.~Luch, ``Artificial intelligence and machine learning empower advanced biomedical material design to toxicity prediction,'' \emph{Advanced Intelligent Systems}, vol.~2, no.~12, p. 2000084, 2020.

\bibitem{novoselov20162d}
K.~Novoselov, A.~Mishchenko, A.~Carvalho, and A.~Castro~Neto, ``2d materials and van der waals heterostructures,'' \emph{Science}, vol. 353, no. 6298, p. aac9439, 2016.

\bibitem{Slager2013}
\BIBentryALTinterwordspacing
R.-J. Slager, A.~Mesaros, V.~Juričić, and J.~Zaanen, ``The space group classification of topological band-insulators,'' \emph{Nature Physics}, vol.~9, pp. 98--102, Feb 2013. [Online]. Available: \url{https://doi.org/10.1038/nphys2513}
\BIBentrySTDinterwordspacing

\bibitem{PhysRevX.7.041069}
\BIBentryALTinterwordspacing
J.~Kruthoff, J.~de~Boer, J.~van Wezel, C.~L. Kane, and R.-J. Slager, ``Topological classification of crystalline insulators through band structure combinatorics,'' \emph{Phys. Rev. X}, vol.~7, p. 041069, Dec 2017. [Online]. Available: \url{https://link.aps.org/doi/10.1103/PhysRevX.7.041069}
\BIBentrySTDinterwordspacing

\bibitem{po2017symmetry}
H.~C. Po, A.~Vishwanath, and H.~Watanabe, ``Symmetry-based indicators of band topology in the 230 space groups,'' \emph{Nature communications}, vol.~8, no.~1, p.~50, 2017.

\bibitem{bradlyn2017topological}
B.~Bradlyn, L.~Elcoro, J.~Cano, M.~G. Vergniory, Z.~Wang, C.~Felser, M.~I. Aroyo, and B.~A. Bernevig, ``Topological quantum chemistry,'' \emph{Nature}, vol. 547, no. 7663, pp. 298--305, 2017.

\bibitem{vergniory2019complete}
M.~Vergniory, L.~Elcoro, C.~Felser, N.~Regnault, B.~A. Bernevig, and Z.~Wang, ``A complete catalogue of high-quality topological materials,'' \emph{Nature}, vol. 566, no. 7745, pp. 480--485, 2019.

\bibitem{zhang2019catalogue}
T.~Zhang, Y.~Jiang, Z.~Song, H.~Huang, Y.~He, Z.~Fang, H.~Weng, and C.~Fang, ``Catalogue of topological electronic materials,'' \emph{Nature}, vol. 566, no. 7745, pp. 475--479, 2019.

\bibitem{chen2019high}
W.~Chen, J.~George, J.~B. Varley, G.-M. Rignanese, and G.~Hautier, ``High-throughput computational discovery of in2mn2o7 as a high curie temperature ferromagnetic semiconductor for spintronics,'' \emph{npj Computational Materials}, vol.~5, no.~1, p.~72, 2019.

\bibitem{horton2019high}
M.~K. Horton, J.~H. Montoya, M.~Liu, and K.~A. Persson, ``High-throughput prediction of the ground-state collinear magnetic order of inorganic materials using density functional theory,'' \emph{npj Computational Materials}, vol.~5, no.~1, p.~64, 2019.

\bibitem{geim2009graphene}
A.~K. Geim, ``Graphene: status and prospects,'' \emph{science}, vol. 324, no. 5934, pp. 1530--1534, 2009.

\bibitem{manzeli20172d}
S.~Manzeli, D.~Ovchinnikov, D.~Pasquier, O.~V. Yazyev, and A.~Kis, ``2d transition metal dichalcogenides,'' \emph{Nature Reviews Materials}, vol.~2, no.~8, pp. 1--15, 2017.

\bibitem{pei2022interlayer}
S.~Pei, Z.~Wang, and J.~Xia, ``Interlayer coupling: an additional degree of freedom in two-dimensional materials,'' \emph{ACS nano}, vol.~16, no.~8, pp. 11\,498--11\,503, 2022.

\bibitem{das2019topological}
S.~K. Das, B.~Yan, J.~van~den Brink, and I.~C. Fulga, ``Topological crystalline insulators from stacked graphene layers,'' \emph{Physical Review B}, vol.~99, no.~16, p. 165418, 2019.

\bibitem{yi2015review}
M.~Yi and Z.~Shen, ``A review on mechanical exfoliation for the scalable production of graphene,'' \emph{Journal of Materials Chemistry A}, vol.~3, no.~22, pp. 11\,700--11\,715, 2015.

\bibitem{zhang2013review}
Y.~Zhang, L.~Zhang, and C.~Zhou, ``Review of chemical vapor deposition of graphene and related applications,'' \emph{Accounts of chemical research}, vol.~46, no.~10, pp. 2329--2339, 2013.

\bibitem{lu2022machine}
M.~Lu, H.~Ji, Y.~Zhao, Y.~Chen, J.~Tao, Y.~Ou, Y.~Wang, Y.~Huang, J.~Wang, and G.~Hao, ``Machine learning-assisted synthesis of two-dimensional materials,'' \emph{ACS Applied Materials \& Interfaces}, vol.~15, no.~1, pp. 1871--1878, 2022.

\bibitem{ryu2022understanding}
B.~Ryu, L.~Wang, H.~Pu, M.~K. Chan, and J.~Chen, ``Understanding, discovery, and synthesis of 2d materials enabled by machine learning,'' \emph{Chemical Society Reviews}, vol.~51, no.~6, pp. 1899--1925, 2022.

\bibitem{hasan2010colloquium}
M.~Z. Hasan and C.~L. Kane, ``Colloquium: topological insulators,'' \emph{Reviews of modern physics}, vol.~82, no.~4, pp. 3045--3067, 2010.

\bibitem{ando2013topological}
Y.~Ando, ``Topological insulator materials,'' \emph{Journal of the Physical Society of Japan}, vol.~82, no.~10, p. 102001, 2013.

\bibitem{schleder2021machine}
G.~R. Schleder, B.~Focassio, and A.~Fazzio, ``Machine learning for materials discovery: Two-dimensional topological insulators,'' \emph{Applied Physics Reviews}, vol.~8, no.~3, 2021.

\bibitem{choudhary2017high}
K.~Choudhary, I.~Kalish, R.~Beams, and F.~Tavazza, ``High-throughput identification and characterization of two-dimensional materials using density functional theory,'' \emph{Scientific reports}, vol.~7, no.~1, p. 5179, 2017.

\bibitem{mounet2018two}
N.~Mounet, M.~Gibertini, P.~Schwaller, D.~Campi, A.~Merkys, A.~Marrazzo, T.~Sohier, I.~E. Castelli, A.~Cepellotti, G.~Pizzi \emph{et~al.}, ``Two-dimensional materials from high-throughput computational exfoliation of experimentally known compounds,'' \emph{Nature nanotechnology}, vol.~13, no.~3, pp. 246--252, 2018.

\bibitem{choudharyefficient}
K.~Choudhary, K.~Garrity, G.~Pilania, and F.~Tavazza, ``Efficient computational design of 2d van der waals heterostructures: Band-alignment,'' \emph{Lattice-Mismatch, Web-app Generation and Machinelearning. arXiv}, 2004.

\bibitem{harris2024autonomous}
S.~B. Harris, A.~Biswas, S.~J. Yun, K.~M. Roccapriore, C.~M. Rouleau, A.~A. Puretzky, R.~K. Vasudevan, D.~B. Geohegan, and K.~Xiao, ``Autonomous synthesis of thin film materials with pulsed laser deposition enabled by in situ spectroscopy and automation,'' \emph{Small Methods}, p. 2301763, 2024.

\bibitem{wang2021unsupervised}
Z.~Wang, J.~Cai, Q.~Wang, S.~Wu, and J.~Li, ``Unsupervised discovery of thin-film photovoltaic materials from unlabeled data,'' \emph{npj Computational Materials}, vol.~7, no.~1, p. 128, 2021.

\bibitem{frey2019prediction}
N.~C. Frey, J.~Wang, G.~I. Vega~Bellido, B.~Anasori, Y.~Gogotsi, and V.~B. Shenoy, ``Prediction of synthesis of 2d metal carbides and nitrides (mxenes) and their precursors with positive and unlabeled machine learning,'' \emph{ACS nano}, vol.~13, no.~3, pp. 3031--3041, 2019.

\bibitem{li2024reinforced}
Z.~Li, F.~Yao, and H.~Sun, ``Reinforced active learning for cvd-grown two-dimensional materials characterization,'' \emph{IISE Transactions}, vol.~56, no.~8, pp. 811--823, 2024.

\bibitem{rajak2020quantum}
P.~Rajak, A.~Krishnamoorthy, R.~Kalia, A.~Nakano, and P.~Vashishta, ``Quantum material synthesis by reinforcement learning,'' in \emph{Proceedings of NeurIPS Workshop on Machine Learning and the Physical Sciences}, 2020, p. 170.

\bibitem{xu2021machine}
M.~Xu, B.~Tang, Y.~Lu, C.~Zhu, Q.~Lu, C.~Zhu, L.~Zheng, J.~Zhang, N.~Han, W.~Fang \emph{et~al.}, ``Machine learning driven synthesis of few-layered wte2 with geometrical control,'' \emph{Journal of the American Chemical Society}, vol. 143, no.~43, pp. 18\,103--18\,113, 2021.

\bibitem{tang2020machine}
B.~Tang, Y.~Lu, J.~Zhou, T.~Chouhan, H.~Wang, P.~Golani, M.~Xu, Q.~Xu, C.~Guan, and Z.~Liu, ``Machine learning-guided synthesis of advanced inorganic materials,'' \emph{Materials Today}, vol.~41, pp. 72--80, 2020.

\bibitem{haastrup2018computational}
S.~Haastrup, M.~Strange, M.~Pandey, T.~Deilmann, P.~S. Schmidt, N.~F. Hinsche, M.~N. Gjerding, D.~Torelli, P.~M. Larsen, A.~C. Riis-Jensen \emph{et~al.}, ``The computational 2d materials database: high-throughput modeling and discovery of atomically thin crystals,'' \emph{2D Materials}, vol.~5, no.~4, p. 042002, 2018.

\bibitem{gjerding2021recent}
M.~N. Gjerding, A.~Taghizadeh, A.~Rasmussen, S.~Ali, F.~Bertoldo, T.~Deilmann, N.~R. Kn{\o}sgaard, M.~Kruse, A.~H. Larsen, S.~Manti \emph{et~al.}, ``Recent progress of the computational 2d materials database (c2db),'' \emph{2D Materials}, vol.~8, no.~4, p. 044002, 2021.

\bibitem{zhou20192dmatpedia}
J.~Zhou, L.~Shen, M.~D. Costa, K.~A. Persson, S.~P. Ong, P.~Huck, Y.~Lu, X.~Ma, Y.~Chen, H.~Tang \emph{et~al.}, ``2dmatpedia, an open computational database of two-dimensional materials from top-down and bottom-up approaches,'' \emph{Scientific data}, vol.~6, no.~1, p.~86, 2019.

\bibitem{talirz2020materials}
L.~Talirz, S.~Kumbhar, E.~Passaro, A.~V. Yakutovich, V.~Granata, F.~Gargiulo, M.~Borelli, M.~Uhrin, S.~P. Huber, S.~Zoupanos \emph{et~al.}, ``Materials cloud, a platform for open computational science,'' \emph{Scientific data}, vol.~7, no.~1, p. 299, 2020.

\bibitem{campi2023expansion}
D.~Campi, N.~Mounet, M.~Gibertini, G.~Pizzi, and N.~Marzari, ``Expansion of the materials cloud 2d database,'' \emph{ACS nano}, vol.~17, no.~12, pp. 11\,268--11\,278, 2023.

\bibitem{marchenko2020database}
E.~I. Marchenko, S.~A. Fateev, A.~A. Petrov, V.~V. Korolev, A.~Mitrofanov, A.~V. Petrov, E.~A. Goodilin, and A.~B. Tarasov, ``Database of two-dimensional hybrid perovskite materials: open-access collection of crystal structures, band gaps, and atomic partial charges predicted by machine learning,'' \emph{Chemistry of materials}, vol.~32, no.~17, pp. 7383--7388, 2020.

\bibitem{frey2020machine}
N.~C. Frey, D.~Akinwande, D.~Jariwala, and V.~B. Shenoy, ``Machine learning-enabled design of point defects in 2d materials for quantum and neuromorphic information processingfrey2020machine,'' \emph{ACS nano}, vol.~14, no.~10, pp. 13\,406--13\,417, 2020.

\bibitem{xu2023small}
P.~Xu, X.~Ji, M.~Li, and W.~Lu, ``Small data machine learning in materials science,'' \emph{npj Computational Materials}, vol.~9, no.~1, p.~42, 2023.

\bibitem{brito2023network}
A.~C.~M. Brito, M.~C.~F. Oliveira, O.~N. Oliveira~Jr, F.~N. Silva, and D.~R. Amancio, ``Network analysis and natural language processing to obtain a landscape of the scientific literature on materials applications,'' \emph{ACS Applied Materials \& Interfaces}, vol.~15, no.~23, pp. 27\,437--27\,446, 2023.

\bibitem{green2024solar}
M.~A. Green, E.~D. Dunlop, M.~Yoshita, N.~Kopidakis, K.~Bothe, G.~Siefer, D.~Hinken, M.~Rauer, J.~Hohl-Ebinger, and X.~Hao, ``Solar cell efficiency tables (version 64),'' \emph{Progress in Photovoltaics: Research and Applications}, vol.~32, no.~7, pp. 425--441, 2024.

\bibitem{polman2016photovoltaic}
A.~Polman, M.~Knight, E.~C. Garnett, B.~Ehrler, and W.~C. Sinke, ``Photovoltaic materials: Present efficiencies and future challenges,'' \emph{Science}, vol. 352, no. 6283, p. aad4424, 2016.

\bibitem{nelson2003physics}
J.~Nelson, ``The physics of solar cells,'' \emph{Imperial College Press google schola}, vol.~2, pp. 62--68, 2003.

\bibitem{philipps2019photovoltaics}
S.~Philipps and W.~Warmuth, ``Photovoltaics report fraunhofer institute for solar energy systems,'' \emph{ISE with Support of PSE GmbH November 14th; Fraunhofer ISE: Freiburg, Germany}, 2019.

\bibitem{streetman2000solid}
B.~G. Streetman, S.~Banerjee \emph{et~al.}, \emph{Solid state electronic devices}.\hskip 1em plus 0.5em minus 0.4em\relax Prentice hall New Jersey, 2000, vol.~4.

\bibitem{nakamura2013blue}
S.~Nakamura and G.~Fasol, \emph{The blue laser diode: GaN based light emitters and lasers}.\hskip 1em plus 0.5em minus 0.4em\relax Springer Science \& Business Media, 2013.

\bibitem{levinshtein2001properties}
M.~E. Levinshtein, S.~L. Rumyantsev, and M.~S. Shur, \emph{Properties of Advanced Semiconductor Materials: GaN, AIN, InN, BN, SiC, SiGe}.\hskip 1em plus 0.5em minus 0.4em\relax John Wiley \& Sons, 2001.

\bibitem{li2021brief}
J.~Li, A.~Aierken, Y.~Liu, Y.~Zhuang, X.~Yang, J.~Mo, R.~Fan, Q.~Chen, S.~Zhang, Y.~Huang \emph{et~al.}, ``A brief review of high efficiency iii-v solar cells for space application,'' \emph{Frontiers in Physics}, vol.~8, p. 631925, 2021.

\bibitem{dimroth2007high}
F.~Dimroth and S.~Kurtz, ``High-efficiency multijunction solar cells,'' \emph{MRS bulletin}, vol.~32, no.~3, pp. 230--235, 2007.

\bibitem{kojima2009organometal}
A.~Kojima, K.~Teshima, Y.~Shirai, and T.~Miyasaka, ``Organometal halide perovskites as visible-light sensitizers for photovoltaic cells,'' \emph{Journal of the american chemical society}, vol. 131, no.~17, pp. 6050--6051, 2009.

\bibitem{snaith2013perovskites}
H.~J. Snaith, ``Perovskites: the emergence of a new era for low-cost, high-efficiency solar cells,'' \emph{The journal of physical chemistry letters}, vol.~4, no.~21, pp. 3623--3630, 2013.

\bibitem{tilley2016perovskites}
R.~J. Tilley, \emph{Perovskites: structure-property relationships}.\hskip 1em plus 0.5em minus 0.4em\relax John Wiley \& Sons, 2016.

\bibitem{tao2021machine}
Q.~Tao, P.~Xu, M.~Li, and W.~Lu, ``Machine learning for perovskite materials design and discovery,'' \emph{Npj computational materials}, vol.~7, no.~1, p.~23, 2021.

\bibitem{royperovskite2022}
\BIBentryALTinterwordspacing
P.~Roy, A.~Ghosh, F.~Barclay, A.~Khare, and E.~Cuce, ``\BIBforeignlanguage{en}{Perovskite {Solar} {Cells}: {A} {Review} of the {Recent} {Advances}},'' \emph{\BIBforeignlanguage{en}{Coatings}}, vol.~12, no.~8, p. 1089, Jul. 2022. [Online]. Available: \url{https://www.mdpi.com/2079-6412/12/8/1089}
\BIBentrySTDinterwordspacing

\bibitem{liumachine2024}
\BIBentryALTinterwordspacing
Y.~Liu, X.~Tan, P.~Xiang, Y.~Tu, T.~Shao, Y.~Zang, X.~Li, and W.~Yan, ``\BIBforeignlanguage{en}{Machine learning as a characterization method for analysis and design of perovskite solar cells},'' \emph{\BIBforeignlanguage{en}{Materials Today Physics}}, vol.~42, p. 101359, Mar. 2024. [Online]. Available: \url{https://linkinghub.elsevier.com/retrieve/pii/S254252932400035X}
\BIBentrySTDinterwordspacing

\bibitem{mohantycomprehensive2023}
\BIBentryALTinterwordspacing
D.~Mohanty and A.~K. Palai, ``\BIBforeignlanguage{en}{Comprehensive {Machine} {Learning} {Pipeline} for {Prediction} of {Power} {Conversion} {Efficiency} in {Perovskite} {Solar} {Cells}},'' \emph{\BIBforeignlanguage{en}{Advanced Theory and Simulations}}, vol.~6, no.~12, p. 2300309, Dec. 2023. [Online]. Available: \url{https://onlinelibrary.wiley.com/doi/10.1002/adts.202300309}
\BIBentrySTDinterwordspacing

\bibitem{akbarunveiling2024}
\BIBentryALTinterwordspacing
B.~Akbar, H.~Tayara, and K.~T. Chong, ``\BIBforeignlanguage{en}{Unveiling dominant recombination loss in perovskite solar cells with a {XGBoost}-based machine learning approach},'' \emph{\BIBforeignlanguage{en}{iScience}}, vol.~27, no.~3, p. 109200, Mar. 2024. [Online]. Available: \url{https://linkinghub.elsevier.com/retrieve/pii/S2589004224004218}
\BIBentrySTDinterwordspacing

\bibitem{karimitariaccurate2024}
\BIBentryALTinterwordspacing
N.~Karimitari, W.~J. Baldwin, E.~W. Muller, Z.~J.~L. Bare, W.~J. Kennedy, G.~Csányi, and C.~Sutton, ``\BIBforeignlanguage{en}{Accurate {Crystal} {Structure} {Prediction} of {New} {2D} {Hybrid} {Organic}–{Inorganic} {Perovskites}},'' \emph{\BIBforeignlanguage{en}{Journal of the American Chemical Society}}, vol. 146, no.~40, pp. 27\,392--27\,404, Oct. 2024. [Online]. Available: \url{https://pubs.acs.org/doi/10.1021/jacs.4c06549}
\BIBentrySTDinterwordspacing

\bibitem{ranhalide2023}
\BIBentryALTinterwordspacing
J.~Ran, B.~Wang, Y.~Wu, D.~Liu, C.~Mora~Perez, A.~S. Vasenko, and O.~V. Prezhdo, ``\BIBforeignlanguage{en}{Halide {Vacancies} {Create} {No} {Charge} {Traps} on {Lead} {Halide} {Perovskite} {Surfaces} but {Can} {Generate} {Deep} {Traps} in the {Bulk}},'' \emph{\BIBforeignlanguage{en}{The Journal of Physical Chemistry Letters}}, vol.~14, no.~26, pp. 6028--6036, Jul. 2023. [Online]. Available: \url{https://pubs.acs.org/doi/10.1021/acs.jpclett.3c01231}
\BIBentrySTDinterwordspacing

\bibitem{biinfluence2024}
\BIBentryALTinterwordspacing
H.~Bi, M.~Wang, L.~Liu, J.~Yan, R.~Zeng, Z.~Xu, and J.~Wang, ``\BIBforeignlanguage{en}{The influence of perovskite crystal structure on its stability},'' \emph{\BIBforeignlanguage{en}{Journal of Materials Chemistry A}}, vol.~12, no.~21, pp. 12\,744--12\,751, 2024. [Online]. Available: \url{https://xlink.rsc.org/?DOI=D3TA07457A}
\BIBentrySTDinterwordspacing

\bibitem{wuuniversal2024}
Y.~Wu, C.-F. Wang, M.-G. Ju, Q.~Jia, Q.~Zhou, S.~Lu, X.~Gao, Y.~Zhang, and J.~Wang, ``Universal machine learning aided synthesis approach of two-dimensional perovskites in a typical laboratory,'' \emph{Nature Communications}, vol.~15, no.~1, p. 138, 2024.

\bibitem{maidata2024}
H.~Mai, X.~Wen, X.~Li, N.~S. Dissanayake, X.~Sun, Y.~Lu, T.~C. Le, S.~P. Russo, D.~Chen, D.~A. Winkler \emph{et~al.}, ``Data driven high quantum yield halide perovskite phosphors design and fabrication,'' \emph{Materials Today}, vol.~74, pp. 12--21, 2024.

\bibitem{lirational2023}
\BIBentryALTinterwordspacing
X.~Li, H.~Mai, J.~Lu, X.~Wen, T.~C. Le, S.~P. Russo, D.~A. Winkler, D.~Chen, and R.~A. Caruso, ``\BIBforeignlanguage{en}{Rational {Atom} {Substitution} to {Obtain} {Efficient}, {Lead}‐{Free} {Photocatalytic} {Perovskites} {Assisted} by {Machine} {Learning} and {DFT} {Calculations}},'' \emph{\BIBforeignlanguage{en}{Angewandte Chemie International Edition}}, vol.~62, no.~52, p. e202315002, Dec. 2023. [Online]. Available: \url{https://onlinelibrary.wiley.com/doi/10.1002/anie.202315002}
\BIBentrySTDinterwordspacing

\bibitem{choubisainterpretable2023}
\BIBentryALTinterwordspacing
H.~Choubisa, P.~Todorović, J.~M. Pina, D.~H. Parmar, Z.~Li, O.~Voznyy, I.~Tamblyn, and E.~H. Sargent, ``\BIBforeignlanguage{en}{Interpretable discovery of semiconductors with machine learning},'' \emph{\BIBforeignlanguage{en}{npj Computational Materials}}, vol.~9, no.~1, p. 117, Jun. 2023. [Online]. Available: \url{https://www.nature.com/articles/s41524-023-01066-9}
\BIBentrySTDinterwordspacing

\bibitem{selvaratnaminterpretable2023}
\BIBentryALTinterwordspacing
B.~Selvaratnam, A.~O. Oliynyk, and A.~Mar, ``\BIBforeignlanguage{en}{Interpretable {Machine} {Learning} in {Solid}-{State} {Chemistry}, with {Applications} to {Perovskites}, {Spinels}, and {Rare}-{Earth} {Intermetallics}: {Finding} {Descriptors} {Using} {Decision} {Trees}},'' \emph{\BIBforeignlanguage{en}{Inorganic Chemistry}}, vol.~62, no.~28, pp. 10\,865--10\,875, Jul. 2023. [Online]. Available: \url{https://pubs.acs.org/doi/10.1021/acs.inorgchem.3c01153}
\BIBentrySTDinterwordspacing

\bibitem{solak2023advances}
E.~K. Solak and E.~Irmak, ``Advances in organic photovoltaic cells: A comprehensive review of materials, technologies, and performance,'' \emph{RSC advances}, vol.~13, no.~18, pp. 12\,244--12\,269, 2023.

\bibitem{brabec2010polymer}
C.~J. Brabec, S.~Gowrisanker, J.~J. Halls, D.~Laird, S.~Jia, and S.~P. Williams, ``Polymer--fullerene bulk-heterojunction solar cells,'' \emph{Advanced Materials}, vol.~22, no.~34, pp. 3839--3856, 2010.

\bibitem{sariciftciphotoinduced1992}
\BIBentryALTinterwordspacing
N.~S. Sariciftci, L.~Smilowitz, A.~J. Heeger, and F.~Wudl, ``\BIBforeignlanguage{en}{Photoinduced {Electron} {Transfer} from a {Conducting} {Polymer} to {Buckminsterfullerene}},'' \emph{\BIBforeignlanguage{en}{Science}}, vol. 258, no. 5087, pp. 1474--1476, Nov. 1992. [Online]. Available: \url{https://www.science.org/doi/10.1126/science.258.5087.1474}
\BIBentrySTDinterwordspacing

\bibitem{li2012polymer}
G.~Li, R.~Zhu, and Y.~Yang, ``Polymer solar cells,'' \emph{Nature photonics}, vol.~6, no.~3, pp. 153--161, 2012.

\bibitem{yuansinglejunction2019}
\BIBentryALTinterwordspacing
J.~Yuan, Y.~Zhang, L.~Zhou, G.~Zhang, H.-L. Yip, T.-K. Lau, X.~Lu, C.~Zhu, H.~Peng, P.~A. Johnson, M.~Leclerc, Y.~Cao, J.~Ulanski, Y.~Li, and Y.~Zou, ``\BIBforeignlanguage{en}{Single-{Junction} {Organic} {Solar} {Cell} with over 15\% {Efficiency} {Using} {Fused}-{Ring} {Acceptor} with {Electron}-{Deficient} {Core}},'' \emph{\BIBforeignlanguage{en}{Joule}}, vol.~3, no.~4, pp. 1140--1151, Apr. 2019. [Online]. Available: \url{https://linkinghub.elsevier.com/retrieve/pii/S2542435119300327}
\BIBentrySTDinterwordspacing

\bibitem{solakadvances2023}
\BIBentryALTinterwordspacing
E.~K. Solak and E.~Irmak, ``\BIBforeignlanguage{en}{Advances in organic photovoltaic cells: a comprehensive review of materials, technologies, and performance},'' \emph{\BIBforeignlanguage{en}{RSC Advances}}, vol.~13, no.~18, pp. 12\,244--12\,269, 2023. [Online]. Available: \url{https://xlink.rsc.org/?DOI=D3RA01454A}
\BIBentrySTDinterwordspacing

\bibitem{zhanghighefficiency2022}
\BIBentryALTinterwordspacing
Q.~Zhang, Y.~J. Zheng, W.~Sun, Z.~Ou, O.~Odunmbaku, M.~Li, S.~Chen, Y.~Zhou, J.~Li, B.~Qin, and K.~Sun, ``\BIBforeignlanguage{en}{High‐{Efficiency} {Non}‐{Fullerene} {Acceptors} {Developed} by {Machine} {Learning} and {Quantum} {Chemistry}},'' \emph{\BIBforeignlanguage{en}{Advanced Science}}, vol.~9, no.~6, p. 2104742, Feb. 2022. [Online]. Available: \url{https://onlinelibrary.wiley.com/doi/10.1002/advs.202104742}
\BIBentrySTDinterwordspacing

\bibitem{morishitamachine2024}
\BIBentryALTinterwordspacing
Y.~Morishita, M.~Yarimizu, M.~Kaneko, and A.~Muraoka, ``\BIBforeignlanguage{en}{Machine learning approach for predicting high {JSC} donor molecules in fullerene-typed organic solar cells},'' \emph{\BIBforeignlanguage{en}{Chemical Physics Letters}}, p. 141719, Oct. 2024. [Online]. Available: \url{https://linkinghub.elsevier.com/retrieve/pii/S0009261424006614}
\BIBentrySTDinterwordspacing

\bibitem{liuaccelerating2022}
\BIBentryALTinterwordspacing
X.~Liu, Y.~Shao, T.~Lu, D.~Chang, M.~Li, and W.~Lu, ``\BIBforeignlanguage{en}{Accelerating the discovery of high-performance donor/acceptor pairs in photovoltaic materials via machine learning and density functional theory},'' \emph{\BIBforeignlanguage{en}{Materials \& Design}}, vol. 216, p. 110561, Apr. 2022. [Online]. Available: \url{https://linkinghub.elsevier.com/retrieve/pii/S0264127522001824}
\BIBentrySTDinterwordspacing

\bibitem{sutharmachine2023}
\BIBentryALTinterwordspacing
R.~Suthar, T.~Abhijith, P.~Sharma, and S.~Karak, ``\BIBforeignlanguage{en}{Machine learning framework for the analysis and prediction of energy loss for non-fullerene organic solar cells},'' \emph{\BIBforeignlanguage{en}{Solar Energy}}, vol. 250, pp. 119--127, Jan. 2023. [Online]. Available: \url{https://linkinghub.elsevier.com/retrieve/pii/S0038092X22009008}
\BIBentrySTDinterwordspacing

\bibitem{limachine2024}
\BIBentryALTinterwordspacing
M.~Li, C.~Zhang, M.~Zhang, J.~Gong, X.~Liu, Y.~Chen, Z.~Liu, Y.~Wu, and H.~Chen, ``\BIBforeignlanguage{en}{Machine {Learning} {Study} on the {Virtual} {Screening} of {Donor}–{Acceptor} {Pairs} for {Organic} {Solar} {Cells}},'' \emph{\BIBforeignlanguage{en}{physica status solidi (a)}}, vol. 221, no.~9, p. 2400008, May 2024. [Online]. Available: \url{https://onlinelibrary.wiley.com/doi/10.1002/pssa.202400008}
\BIBentrySTDinterwordspacing

\bibitem{sutharmachinelearningguided2023}
\BIBentryALTinterwordspacing
R.~Suthar, A.~T, and S.~Karak, ``\BIBforeignlanguage{en}{Machine-learning-guided prediction of photovoltaic performance of non-fullerene organic solar cells using novel molecular and structural descriptors},'' \emph{\BIBforeignlanguage{en}{Journal of Materials Chemistry A}}, vol.~11, no.~41, pp. 22\,248--22\,258, 2023. [Online]. Available: \url{https://xlink.rsc.org/?DOI=D3TA04603F}
\BIBentrySTDinterwordspacing

\bibitem{miyakeimproved2022}
\BIBentryALTinterwordspacing
Y.~Miyake, K.~Kranthiraja, F.~Ishiwari, and A.~Saeki, ``\BIBforeignlanguage{en}{Improved {Predictions} of {Organic} {Photovoltaic} {Performance} through {Machine} {Learning} {Models} {Empowered} by {Artificially} {Generated} {Failure} {Data}},'' \emph{\BIBforeignlanguage{en}{Chemistry of Materials}}, vol.~34, no.~15, pp. 6912--6920, Aug. 2022. [Online]. Available: \url{https://pubs.acs.org/doi/10.1021/acs.chemmater.2c01294}
\BIBentrySTDinterwordspacing

\bibitem{wangefficient2023}
\BIBentryALTinterwordspacing
H.~Wang, J.~Feng, Z.~Dong, L.~Jin, M.~Li, J.~Yuan, and Y.~Li, ``\BIBforeignlanguage{en}{Efficient screening framework for organic solar cells with deep learning and ensemble learning},'' \emph{\BIBforeignlanguage{en}{npj Computational Materials}}, vol.~9, no.~1, p. 200, Oct. 2023. [Online]. Available: \url{https://www.nature.com/articles/s41524-023-01155-9}
\BIBentrySTDinterwordspacing

\bibitem{shettyaccelerating2024}
\BIBentryALTinterwordspacing
P.~Shetty, A.~Adeboye, S.~Gupta, C.~Zhang, and R.~Ramprasad, ``\BIBforeignlanguage{en}{Accelerating {Materials} {Discovery} for {Polymer} {Solar} {Cells}: {Data}-{Driven} {Insights} {Enabled} by {Natural} {Language} {Processing}},'' \emph{\BIBforeignlanguage{en}{Chemistry of Materials}}, p. acs.chemmater.4c00709, Aug. 2024. [Online]. Available: \url{https://pubs.acs.org/doi/10.1021/acs.chemmater.4c00709}
\BIBentrySTDinterwordspacing

\bibitem{mou2023machine}
L.-H. Mou, T.~Han, P.~E. Smith, E.~Sharman, and J.~Jiang, ``Machine learning descriptors for data-driven catalysis study,'' \emph{Advanced Science}, vol.~10, no.~22, p. 2301020, 2023.

\bibitem{guo2021machine}
Y.~Guo, X.~He, Y.~Su, Y.~Dai, M.~Xie, S.~Yang, J.~Chen, K.~Wang, D.~Zhou, and C.~Wang, ``Machine-learning-guided discovery and optimization of additives in preparing cu catalysts for co2 reduction,'' \emph{Journal of the American Chemical Society}, vol. 143, no.~15, pp. 5755--5762, 2021.

\bibitem{ishioka2022designing}
S.~Ishioka, A.~Fujiwara, S.~Nakanowatari, L.~Takahashi, T.~Taniike, and K.~Takahashi, ``Designing catalyst descriptors for machine learning in oxidative coupling of methane,'' \emph{ACS Catalysis}, vol.~12, no.~19, pp. 11\,541--11\,546, 2022.

\bibitem{andersen2019beyond}
M.~Andersen, S.~V. Levchenko, M.~Scheffler, and K.~Reuter, ``Beyond scaling relations for the description of catalytic materials,'' \emph{Acs Catalysis}, vol.~9, no.~4, pp. 2752--2759, 2019.

\bibitem{ouyang2018sisso}
R.~Ouyang, S.~Curtarolo, E.~Ahmetcik, M.~Scheffler, and L.~M. Ghiringhelli, ``Sisso: A compressed-sensing method for identifying the best low-dimensional descriptor in an immensity of offered candidates,'' \emph{Physical Review Materials}, vol.~2, no.~8, p. 083802, 2018.

\bibitem{bartel2019new}
C.~J. Bartel, C.~Sutton, B.~R. Goldsmith, R.~Ouyang, C.~B. Musgrave, L.~M. Ghiringhelli, and M.~Scheffler, ``New tolerance factor to predict the stability of perovskite oxides and halides,'' \emph{Science advances}, vol.~5, no.~2, p. eaav0693, 2019.

\bibitem{xu2020data}
W.~Xu, M.~Andersen, and K.~Reuter, ``Data-driven descriptor engineering and refined scaling relations for predicting transition metal oxide reactivity,'' \emph{ACS Catalysis}, vol.~11, no.~2, pp. 734--742, 2020.

\bibitem{mou2023bridging}
T.~Mou, H.~S. Pillai, S.~Wang, M.~Wan, X.~Han, N.~M. Schweitzer, F.~Che, and H.~Xin, ``Bridging the complexity gap in computational heterogeneous catalysis with machine learning,'' \emph{Nature Catalysis}, vol.~6, no.~2, pp. 122--136, 2023.

\bibitem{ren2022universal}
C.~Ren, S.~Lu, Y.~Wu, Y.~Ouyang, Y.~Zhang, Q.~Li, C.~Ling, and J.~Wang, ``A universal descriptor for complicated interfacial effects on electrochemical reduction reactions,'' \emph{Journal of the American Chemical Society}, vol. 144, no.~28, pp. 12\,874--12\,883, 2022.

\bibitem{li2017high}
Z.~Li, S.~Wang, W.~S. Chin, L.~E. Achenie, and H.~Xin, ``High-throughput screening of bimetallic catalysts enabled by machine learning,'' \emph{Journal of Materials Chemistry A}, vol.~5, no.~46, pp. 24\,131--24\,138, 2017.

\bibitem{li2020adaptive}
Z.~Li, L.~E. Achenie, and H.~Xin, ``An adaptive machine learning strategy for accelerating discovery of perovskite electrocatalysts,'' \emph{ACS Catalysis}, vol.~10, no.~7, pp. 4377--4384, 2020.

\bibitem{lu2020neural}
Z.~Lu, Z.~W. Chen, and C.~V. Singh, ``Neural network-assisted development of high-entropy alloy catalysts: decoupling ligand and coordination effects,'' \emph{Matter}, vol.~3, no.~4, pp. 1318--1333, 2020.

\bibitem{wang2020electric}
X.~Wang, S.~Ye, W.~Hu, E.~Sharman, R.~Liu, Y.~Liu, Y.~Luo, and J.~Jiang, ``Electric dipole descriptor for machine learning prediction of catalyst surface--molecular adsorbate interactions,'' \emph{Journal of the American Chemical Society}, vol. 142, no.~17, pp. 7737--7743, 2020.

\bibitem{zhong2021electronic}
W.~Zhong, Y.~Qiu, H.~Shen, X.~Wang, J.~Yuan, C.~Jia, S.~Bi, and J.~Jiang, ``Electronic spin moment as a catalytic descriptor for fe single-atom catalysts supported on c2n,'' \emph{Journal of the American Chemical Society}, vol. 143, no.~11, pp. 4405--4413, 2021.

\bibitem{gu2022nitrogen}
Y.~Gu, Q.~Zhu, Z.~Liu, C.~Fu, J.~Wu, Q.~Zhu, Q.~Jia, and J.~Ma, ``Nitrogen reduction reaction energy and pathways in metal-zeolites: deep learning and explainable machine learning with local acidity and hydrogen bonding features,'' \emph{Journal of Materials Chemistry A}, vol.~10, no.~28, pp. 14\,976--14\,988, 2022.

\bibitem{williams2019enabling}
T.~Williams, K.~McCullough, and J.~A. Lauterbach, ``Enabling catalyst discovery through machine learning and high-throughput experimentation,'' \emph{Chemistry of Materials}, vol.~32, no.~1, pp. 157--165, 2019.

\bibitem{karim2020coupling}
M.~R. Karim, M.~Ferrandon, S.~Medina, E.~Sture, N.~Kariuki, D.~J. Myers, E.~F. Holby, P.~Zelenay, and T.~Ahmed, ``Coupling high-throughput experiments and regression algorithms to optimize pgm-free orr electrocatalyst synthesis,'' \emph{ACS Applied Energy Materials}, vol.~3, no.~9, pp. 9083--9088, 2020.

\bibitem{artrith2020predicting}
N.~Artrith, Z.~Lin, and J.~G. Chen, ``Predicting the activity and selectivity of bimetallic metal catalysts for ethanol reforming using machine learning,'' \emph{Acs Catalysis}, vol.~10, no.~16, pp. 9438--9444, 2020.

\bibitem{zhu2022all}
Q.~Zhu, F.~Zhang, Y.~Huang, H.~Xiao, L.~Zhao, X.~Zhang, T.~Song, X.~Tang, X.~Li, G.~He \emph{et~al.}, ``An all-round ai-chemist with a scientific mind,'' \emph{National Science Review}, vol.~9, no.~10, p. nwac190, 2022.

\bibitem{zhong2020accelerated}
M.~Zhong, K.~Tran, Y.~Min, C.~Wang, Z.~Wang, C.-T. Dinh, P.~De~Luna, Z.~Yu, A.~S. Rasouli, P.~Brodersen \emph{et~al.}, ``Accelerated discovery of co2 electrocatalysts using active machine learning,'' \emph{Nature}, vol. 581, no. 7807, pp. 178--183, 2020.

\bibitem{han2021single}
Z.-K. Han, D.~Sarker, R.~Ouyang, A.~Mazheika, Y.~Gao, and S.~V. Levchenko, ``Single-atom alloy catalysts designed by first-principles calculations and artificial intelligence,'' \emph{Nature communications}, vol.~12, no.~1, p. 1833, 2021.

\bibitem{yin2024machine}
P.~Yin, X.~Niu, S.-B. Li, K.~Chen, X.~Zhang, M.~Zuo, L.~Zhang, and H.-W. Liang, ``Machine-learning-accelerated design of high-performance platinum intermetallic nanoparticle fuel cell catalysts,'' \emph{Nature Communications}, vol.~15, no.~1, p. 415, 2024.

\bibitem{chen2022universal}
L.~Chen, Y.~Tian, X.~Hu, S.~Yao, Z.~Lu, S.~Chen, X.~Zhang, and Z.~Zhou, ``A universal machine learning framework for electrocatalyst innovation: a case study of discovering alloys for hydrogen evolution reaction,'' \emph{Advanced Functional Materials}, vol.~32, no.~47, p. 2208418, 2022.

\bibitem{mok2024generativelanguagemodelcatalyst}
\BIBentryALTinterwordspacing
D.~H. Mok and S.~Back, ``Generative language model for catalyst discovery,'' 2024. [Online]. Available: \url{https://arxiv.org/abs/2407.14040}
\BIBentrySTDinterwordspacing

\bibitem{yeh2004nanostructured}
J.-W. Yeh, S.-K. Chen, S.-J. Lin, J.-Y. Gan, T.-S. Chin, T.-T. Shun, C.-H. Tsau, and S.-Y. Chang, ``Nanostructured high-entropy alloys with multiple principal elements: novel alloy design concepts and outcomes,'' \emph{Advanced engineering materials}, vol.~6, no.~5, pp. 299--303, 2004.

\bibitem{CANTOR2004213}
\BIBentryALTinterwordspacing
B.~Cantor, I.~Chang, P.~Knight, and A.~Vincent, ``Microstructural development in equiatomic multicomponent alloys,'' \emph{Materials Science and Engineering: A}, vol. 375-377, pp. 213--218, 2004. [Online]. Available: \url{https://www.sciencedirect.com/science/article/pii/S0921509303009936}
\BIBentrySTDinterwordspacing

\bibitem{george2019high}
E.~P. George, D.~Raabe, and R.~O. Ritchie, ``High-entropy alloys,'' \emph{Nature reviews materials}, vol.~4, no.~8, pp. 515--534, 2019.

\bibitem{OTTO20132628}
\BIBentryALTinterwordspacing
F.~Otto, Y.~Yang, H.~Bei, and E.~George, ``Relative effects of enthalpy and entropy on the phase stability of equiatomic high-entropy alloys,'' \emph{Acta Materialia}, vol.~61, no.~7, pp. 2628--2638, 2013. [Online]. Available: \url{https://www.sciencedirect.com/science/article/pii/S1359645413000694}
\BIBentrySTDinterwordspacing

\bibitem{rao2022machine}
Z.~Rao, P.-Y. Tung, R.~Xie, Y.~Wei, H.~Zhang, A.~Ferrari, T.~Klaver, F.~K{\"o}rmann, P.~T. Sukumar, A.~Kwiatkowski~da Silva \emph{et~al.}, ``Machine learning--enabled high-entropy alloy discovery,'' \emph{Science}, vol. 378, no. 6615, pp. 78--85, 2022.

\bibitem{liu2023machine}
X.~Liu, J.~Zhang, and Z.~Pei, ``Machine learning for high-entropy alloys: Progress, challenges and opportunities,'' \emph{Progress in Materials Science}, vol. 131, p. 101018, 2023.

\bibitem{wang2023neural}
J.~Wang, H.~Kwon, H.~S. Kim, and B.-J. Lee, ``A neural network model for high entropy alloy design,'' \emph{npj Computational Materials}, vol.~9, no.~1, p.~60, 2023.

\bibitem{chau2023support}
N.~H. Chau, M.~Kubo, L.~V. Hai, and T.~Yamamoto, ``Support vector machine-based phase prediction of multi-principal element alloys,'' \emph{Vietnam Journal of Computer Science}, vol.~10, no.~01, pp. 101--116, 2023.

\bibitem{HUANG2019225}
\BIBentryALTinterwordspacing
W.~Huang, P.~Martin, and H.~L. Zhuang, ``Machine-learning phase prediction of high-entropy alloys,'' \emph{Acta Materialia}, vol. 169, pp. 225--236, 2019. [Online]. Available: \url{https://www.sciencedirect.com/science/article/pii/S1359645419301454}
\BIBentrySTDinterwordspacing

\bibitem{krishna2021machine}
Y.~V. Krishna, U.~K. Jaiswal, and M.~Rahul, ``Machine learning approach to predict new multiphase high entropy alloys,'' \emph{Scripta Materialia}, vol. 197, p. 113804, 2021.

\bibitem{li2022towards}
H.~Li, R.~Yuan, H.~Liang, W.~Y. Wang, J.~Li, and J.~Wang, ``Towards high entropy alloy with enhanced strength and ductility using domain knowledge constrained active learning,'' \emph{Materials \& Design}, vol. 223, p. 111186, 2022.

\bibitem{zhu2022phase}
W.~Zhu, W.~Huo, S.~Wang, X.~Wang, K.~Ren, S.~Tan, F.~Fang, Z.~Xie, and J.~Jiang, ``Phase formation prediction of high-entropy alloys: a deep learning study,'' \emph{journal of materials research and technology}, vol.~18, pp. 800--809, 2022.

\bibitem{wang2022element}
X.~Wang, N.-D. Tran, S.~Zeng, C.~Hou, Y.~Chen, and J.~Ni, ``Element-wise representations with ecnet for material property prediction and applications in high-entropy alloys,'' \emph{npj Computational Materials}, vol.~8, no.~1, p. 253, 2022.

\bibitem{pei2021machine}
Z.~Pei, K.~A. Rozman, {\"O}.~N. Do{\u{g}}an, Y.~Wen, N.~Gao, E.~A. Holm, J.~A. Hawk, D.~E. Alman, and M.~C. Gao, ``Machine-learning microstructure for inverse material design,'' \emph{Advanced Science}, vol.~8, no.~23, p. 2101207, 2021.

\bibitem{slater2015function}
A.~G. Slater and A.~I. Cooper, ``Function-led design of new porous materials,'' \emph{Science}, vol. 348, no. 6238, p. aaa8075, 2015.

\bibitem{thomas2020much}
A.~Thomas, ``Much ado about nothing--a decade of porous materials research,'' \emph{Nature Communications}, vol.~11, no.~1, p. 4985, 2020.

\bibitem{park2024inverse}
J.~Park, A.~P.~S. Gill, S.~M. Moosavi, and J.~Kim, ``Inverse design of porous materials: a diffusion model approach,'' \emph{Journal of Materials Chemistry A}, vol.~12, no.~11, pp. 6507--6514, 2024.

\bibitem{yao2021inverse}
Z.~Yao, B.~S{\'a}nchez-Lengeling, N.~S. Bobbitt, B.~J. Bucior, S.~G.~H. Kumar, S.~P. Collins, T.~Burns, T.~K. Woo, O.~K. Farha, R.~Q. Snurr \emph{et~al.}, ``Inverse design of nanoporous crystalline reticular materials with deep generative models,'' \emph{Nature Machine Intelligence}, vol.~3, no.~1, pp. 76--86, 2021.

\bibitem{park2024generative}
H.~Park, X.~Yan, R.~Zhu, E.~A. Huerta, S.~Chaudhuri, D.~Cooper, I.~Foster, and E.~Tajkhorshid, ``A generative artificial intelligence framework based on a molecular diffusion model for the design of metal-organic frameworks for carbon capture,'' \emph{Communications Chemistry}, vol.~7, no.~1, p.~21, 2024.

\bibitem{igashov2024equivariant}
I.~Igashov, H.~St{\"a}rk, C.~Vignac, A.~Schneuing, V.~G. Satorras, P.~Frossard, M.~Welling, M.~Bronstein, and B.~Correia, ``Equivariant 3d-conditional diffusion model for molecular linker design,'' \emph{Nature Machine Intelligence}, pp. 1--11, 2024.

\bibitem{fu2023mofdiff}
X.~Fu, T.~Xie, A.~S. Rosen, T.~Jaakkola, and J.~Smith, ``Mofdiff: Coarse-grained diffusion for metal-organic framework design,'' \emph{arXiv preprint arXiv:2310.10732}, 2023.

\bibitem{park2024multi}
J.~Park, Y.~Lee, and J.~Kim, ``Multi-modal conditioning for metal-organic frameworks generation using 3d modeling techniques,'' \emph{ChemRxiv, DOI: 10.26434/chemrxiv-2024-w8fps.}, 2024.

\bibitem{wang2023machine}
T.~Wang, R.~Pan, M.~L. Martins, J.~Cui, Z.~Huang, B.~P. Thapaliya, C.-L. Do-Thanh, M.~Zhou, J.~Fan, Z.~Yang \emph{et~al.}, ``Machine-learning-assisted material discovery of oxygen-rich highly porous carbon active materials for aqueous supercapacitors,'' \emph{Nature communications}, vol.~14, no.~1, p. 4607, 2023.

\bibitem{kumar2021synthesis}
S.~Kumar, G.~Ignacz, and G.~Szekely, ``Synthesis of covalent organic frameworks using sustainable solvents and machine learning,'' \emph{Green Chemistry}, vol.~23, no.~22, pp. 8932--8939, 2021.

\bibitem{delpisheh2024leveraging}
M.~Delpisheh, B.~Ebrahimpour, A.~Fattahi, M.~Siavashi, H.~Mir, H.~Mashhadimoslem, M.~A. Abdol, M.~Ghorbani, J.~Shokri, D.~Niblett \emph{et~al.}, ``Leveraging machine learning in porous media,'' \emph{Journal of Materials Chemistry A}, 2024.

\bibitem{d2022machine}
M.~D'Elia, H.~Deng, C.~Fraces, K.~Garikipati, L.~Graham-Brady, A.~Howard, G.~Karniadakis, V.~Keshavarzzadeh, R.~M. Kirby, N.~Kutz \emph{et~al.}, ``Machine learning in heterogeneous porous materials,'' \emph{arXiv preprint arXiv:2202.04137}, 2022.

\bibitem{breiman2001random}
L.~Breiman, ``Random forests,'' \emph{Machine learning}, vol.~45, pp. 5--32, 2001.

\bibitem{quinlan1986induction}
J.~R. Quinlan, ``Induction of decision trees,'' \emph{Machine learning}, vol.~1, pp. 81--106, 1986.

\bibitem{cortes1995support}
C.~Cortes, ``Support-vector networks,'' \emph{Machine Learning}, 1995.

\bibitem{rumelhart1986learning}
D.~E. Rumelhart, G.~E. Hinton, and R.~J. Williams, ``Learning representations by back-propagating errors,'' \emph{nature}, vol. 323, no. 6088, pp. 533--536, 1986.

\bibitem{goodfellow2016deep}
I.~Goodfellow, ``Deep learning,'' 2016.

\bibitem{deng2022inverse}
B.~Deng, A.~Zareei, X.~Ding, J.~C. Weaver, C.~H. Rycroft, and K.~Bertoldi, ``Inverse design of mechanical metamaterials with target nonlinear response via a neural accelerated evolution strategy,'' \emph{Advanced Materials}, vol.~34, no.~41, p. 2206238, 2022.

\bibitem{geng2022data}
X.~Geng, Z.~Cheng, S.~Wang, C.~Peng, A.~Ullah, H.~Wang, and G.~Wu, ``A data-driven machine learning approach to predict the hardenability curve of boron steels and assist alloy design,'' \emph{Journal of Materials Science}, vol.~57, no.~23, pp. 10\,755--10\,768, 2022.

\bibitem{ijaz2023machine}
S.~Ijaz, S.~Noureen, B.~Rehman, O.~Aldaghri, H.~Cabrera, K.~H. Ibnaouf, N.~Madkhali, and M.~Q. Mehmood, ``Machine-learning-driven accelerated design-method for meta-devices,'' \emph{Materials Today Communications}, vol.~37, p. 106951, 2023.

\bibitem{mcdonald2023applied}
S.~M. McDonald, E.~K. Augustine, Q.~Lanners, C.~Rudin, L.~Catherine~Brinson, and M.~L. Becker, ``Applied machine learning as a driver for polymeric biomaterials design,'' \emph{Nature Communications}, vol.~14, no.~1, p. 4838, 2023.

\bibitem{lei2019machine}
X.~Lei, C.~Liu, Z.~Du, W.~Zhang, and X.~Guo, ``Machine learning-driven real-time topology optimization under moving morphable component-based framework,'' \emph{Journal of Applied Mechanics}, vol.~86, no.~1, p. 011004, 2019.

\bibitem{lecun2015deep}
Y.~LeCun, Y.~Bengio, and G.~Hinton, ``Deep learning,'' \emph{nature}, vol. 521, no. 7553, pp. 436--444, 2015.

\bibitem{he2016deep}
K.~He, X.~Zhang, S.~Ren, and J.~Sun, ``Deep residual learning for image recognition,'' in \emph{Proceedings of the IEEE conference on computer vision and pattern recognition}, 2016, pp. 770--778.

\bibitem{abueidda2020topology}
D.~W. Abueidda, S.~Koric, and N.~A. Sobh, ``Topology optimization of 2d structures with nonlinearities using deep learning,'' \emph{Computers \& Structures}, vol. 237, p. 106283, 2020.

\bibitem{wei2022use}
X.~Wei, S.~van~der Zwaag, Z.~Jia, C.~Wang, and W.~Xu, ``On the use of transfer modeling to design new steels with excellent rotating bending fatigue resistance even in the case of very small calibration datasets,'' \emph{Acta Materialia}, vol. 235, p. 118103, 2022.

\bibitem{gu2018novo}
G.~X. Gu, C.-T. Chen, and M.~J. Buehler, ``De novo composite design based on machine learning algorithm,'' \emph{Extreme Mechanics Letters}, vol.~18, pp. 19--28, 2018.

\bibitem{gilmerneural2017}
\BIBentryALTinterwordspacing
J.~Gilmer, S.~S. Schoenholz, P.~F. Riley, O.~Vinyals, and G.~E. Dahl, ``Neural {Message} {Passing} for {Quantum} {Chemistry},'' in \emph{Proceedings of the 34th {International} {Conference} on {Machine} {Learning}}, ser. Proceedings of {Machine} {Learning} {Research}, D.~Precup and Y.~W. Teh, Eds., vol.~70.\hskip 1em plus 0.5em minus 0.4em\relax PMLR, Aug. 2017, pp. 1263--1272. [Online]. Available: \url{https://proceedings.mlr.press/v70/gilmer17a.html}
\BIBentrySTDinterwordspacing

\bibitem{schuttschnet2018}
\BIBentryALTinterwordspacing
K.~T. Sch{\"u}tt, H.~E. Sauceda, P.-J. Kindermans, A.~Tkatchenko, and K.-R. Müller, ``\BIBforeignlanguage{en}{{SchNet} – {A} deep learning architecture for molecules and materials},'' \emph{\BIBforeignlanguage{en}{The Journal of Chemical Physics}}, vol. 148, no.~24, p. 241722, Jun. 2018. [Online]. Available: \url{https://pubs.aip.org/jcp/article/148/24/241722/962591/SchNet-A-deep-learning-architecture-for-molecules}
\BIBentrySTDinterwordspacing

\bibitem{schuttquantumchemical2017}
\BIBentryALTinterwordspacing
K.~T. Schutt, F.~Arbabzadah, S.~Chmiela, K.~R. Muller, and A.~Tkatchenko, ``\BIBforeignlanguage{en}{Quantum-chemical insights from deep tensor neural networks},'' \emph{\BIBforeignlanguage{en}{Nature Communications}}, vol.~8, no.~1, p. 13890, 2017. [Online]. Available: \url{https://www.nature.com/articles/ncomms13890}
\BIBentrySTDinterwordspacing

\bibitem{huforcenet2021}
\BIBentryALTinterwordspacing
W.~Hu, M.~Shuaibi, A.~Das, S.~Goyal, A.~Sriram, J.~Leskovec, D.~Parikh, and C.~L. Zitnick, ``{ForceNet}: {A} {Graph} {Neural} {Network} for {Large}-{Scale} {Quantum} {Calculations},'' 2021, version Number: 1. [Online]. Available: \url{https://arxiv.org/abs/2103.01436}
\BIBentrySTDinterwordspacing

\bibitem{xiecrystal2018}
\BIBentryALTinterwordspacing
T.~Xie and J.~C. Grossman, ``\BIBforeignlanguage{en}{Crystal {Graph} {Convolutional} {Neural} {Networks} for an {Accurate} and {Interpretable} {Prediction} of {Material} {Properties}},'' \emph{\BIBforeignlanguage{en}{Physical Review Letters}}, vol. 120, no.~14, p. 145301, Apr. 2018. [Online]. Available: \url{https://link.aps.org/doi/10.1103/PhysRevLett.120.145301}
\BIBentrySTDinterwordspacing

\bibitem{unkephysnet2019}
\BIBentryALTinterwordspacing
O.~T. Unke and M.~Meuwly, ``\BIBforeignlanguage{en}{{PhysNet}: {A} {Neural} {Network} for {Predicting} {Energies}, {Forces}, {Dipole} {Moments}, and {Partial} {Charges}},'' \emph{\BIBforeignlanguage{en}{Journal of Chemical Theory and Computation}}, vol.~15, no.~6, pp. 3678--3693, Jun. 2019. [Online]. Available: \url{https://pubs.acs.org/doi/10.1021/acs.jctc.9b00181}
\BIBentrySTDinterwordspacing

\bibitem{gasteigerdirectional2020}
\BIBentryALTinterwordspacing
J.~Gasteiger, J.~Groß, and S.~Günnemann, ``Directional {Message} {Passing} for {Molecular} {Graphs},'' 2020, version Number: 2. [Online]. Available: \url{https://arxiv.org/abs/2003.03123}
\BIBentrySTDinterwordspacing

\bibitem{liuspherical2022}
\BIBentryALTinterwordspacing
Y.~Liu, L.~Wang, M.~Liu, Y.~Lin, X.~Zhang, B.~Oztekin, and S.~Ji, ``Spherical {Message} {Passing} for {3D} {Molecular} {Graphs},'' in \emph{International {Conference} on {Learning} {Representations}}, 2022. [Online]. Available: \url{https://openreview.net/forum?id=givsRXsOt9r}
\BIBentrySTDinterwordspacing

\bibitem{yueplugandplay2024}
\BIBentryALTinterwordspacing
A.~Yue, D.~Luo, and H.~Xu, ``A {Plug}-and-{Play} {Quaternion} {Message}-{Passing} {Module} for {Molecular} {Conformation} {Representation},'' \emph{Proceedings of the AAAI Conference on Artificial Intelligence}, vol.~38, no.~15, pp. 16\,633--16\,641, Mar. 2024. [Online]. Available: \url{https://ojs.aaai.org/index.php/AAAI/article/view/29602}
\BIBentrySTDinterwordspacing

\bibitem{klicperagemnet2024}
J.~Klicpera, F.~Becker, and S.~Günnemann, ``{GemNet}: universal directional graph neural networks for molecules,'' in \emph{Proceedings of the 35th {International} {Conference} on {Neural} {Information} {Processing} {Systems}}, ser. {NIPS} '21.\hskip 1em plus 0.5em minus 0.4em\relax Red Hook, NY, USA: Curran Associates Inc., 2024.

\bibitem{wangcomenet2022}
\BIBentryALTinterwordspacing
L.~Wang, Y.~Liu, Y.~Lin, H.~Liu, and S.~Ji, ``{ComENet}: {Towards} {Complete} and {Efficient} {Message} {Passing} for {3D} {Molecular} {Graphs},'' in \emph{Advances in {Neural} {Information} {Processing} {Systems}}, S.~Koyejo, S.~Mohamed, A.~Agarwal, D.~Belgrave, K.~Cho, and A.~Oh, Eds., vol.~35.\hskip 1em plus 0.5em minus 0.4em\relax Curran Associates, Inc., 2022, pp. 650--664. [Online]. Available: \url{https://proceedings.neurips.cc/paper_files/paper/2022/file/0418973e545b932939302cb605d06f43-Paper-Conference.pdf}
\BIBentrySTDinterwordspacing

\bibitem{egnn}
\BIBentryALTinterwordspacing
V.~G. Satorras, E.~Hoogeboom, and M.~Welling, ``E(n) equivariant graph neural networks,'' 2022. [Online]. Available: \url{https://arxiv.org/abs/2102.09844}
\BIBentrySTDinterwordspacing

\bibitem{schutt2017schnet}
K.~Sch{\"u}tt, P.-J. Kindermans, H.~E. Sauceda~Felix, S.~Chmiela, A.~Tkatchenko, and K.-R. M{\"u}ller, ``Schnet: A continuous-filter convolutional neural network for modeling quantum interactions,'' \emph{Advances in neural information processing systems}, vol.~30, 2017.

\bibitem{kohler2019equivariant}
J.~K{\"o}hler, L.~Klein, and F.~No{\'e}, ``Equivariant flows: sampling configurations for multi-body systems with symmetric energies,'' \emph{arXiv preprint arXiv:1910.00753}, 2019.

\bibitem{huang2022equivariant}
W.~Huang, J.~Han, Y.~Rong, T.~Xu, F.~Sun, and J.~Huang, ``Equivariant graph mechanics networks with constraints,'' in \emph{ICLR}, 2022.

\bibitem{du2022se}
W.~Du, H.~Zhang, Y.~Du, Q.~Meng, W.~Chen, N.~Zheng, B.~Shao, and T.-Y. Liu, ``{SE}(3) equivariant graph neural networks with complete local frames,'' in \emph{Proceedings of the 39th International Conference on Machine Learning}, ser. Proceedings of Machine Learning Research, vol. 162.\hskip 1em plus 0.5em minus 0.4em\relax PMLR, 17--23 Jul 2022, pp. 5583--5608.

\bibitem{kofinas2021rototranslated}
M.~Kofinas, N.~S. Nagaraja, and E.~Gavves, ``Roto-translated local coordinate frames for interacting dynamical systems,'' in \emph{Advances in Neural Information Processing Systems}, 2021.

\bibitem{kofinas2023latent}
M.~Kofinas, E.~J. Bekkers, N.~S. Nagaraja, and E.~Gavves, ``Latent field discovery in interacting dynamical systems with neural fields,'' in \emph{Thirty-seventh Conference on Neural Information Processing Systems}, 2023.

\bibitem{gvpgnn}
B.~Jing, S.~Eismann, P.~Suriana, R.~J.~L. Townshend, and R.~Dror, ``Learning from protein structure with geometric vector perceptrons,'' in \emph{ICLR}, 2021.

\bibitem{du2024new}
Y.~Du, L.~Wang, D.~Feng, G.~Wang, S.~Ji, C.~P. Gomes, Z.-M. Ma \emph{et~al.}, ``A new perspective on building efficient and expressive 3d equivariant graph neural networks,'' \emph{Advances in Neural Information Processing Systems}, vol.~36, 2024.

\bibitem{puny2021frame}
O.~Puny, M.~Atzmon, E.~J. Smith, I.~Misra, A.~Grover, H.~Ben-Hamu, and Y.~Lipman, ``Frame averaging for invariant and equivariant network design,'' in \emph{International Conference on Learning Representations}, 2021.

\bibitem{han2022equivariant}
J.~Han, W.~Huang, T.~Xu, and Y.~Rong, ``Equivariant graph hierarchy-based neural networks,'' in \emph{NeurIPS}, 2022.

\bibitem{TFN2018}
\BIBentryALTinterwordspacing
N.~Thomas, T.~Smidt, S.~Kearnes, L.~Yang, L.~Li, K.~Kohlhoff, and P.~Riley, ``Tensor field networks: Rotation- and translation-equivariant neural networks for 3d point clouds,'' 2018. [Online]. Available: \url{https://arxiv.org/abs/1802.08219}
\BIBentrySTDinterwordspacing

\bibitem{e3nn}
\BIBentryALTinterwordspacing
M.~Geiger and T.~Smidt, ``e3nn: Euclidean neural networks,'' 2022. [Online]. Available: \url{https://arxiv.org/abs/2207.09453}
\BIBentrySTDinterwordspacing

\bibitem{segnns}
\BIBentryALTinterwordspacing
J.~Brandstetter, R.~Hesselink, E.~van~der Pol, E.~J. Bekkers, and M.~Welling, ``Geometric and physical quantities improve e(3) equivariant message passing,'' 2022. [Online]. Available: \url{https://arxiv.org/abs/2110.02905}
\BIBentrySTDinterwordspacing

\bibitem{NequIP}
\BIBentryALTinterwordspacing
S.~Batzner, A.~Musaelian, L.~Sun, M.~Geiger, J.~P. Mailoa, M.~Kornbluth, N.~Molinari, T.~E. Smidt, and B.~Kozinsky, ``E(3)-equivariant graph neural networks for data-efficient and accurate interatomic potentials,'' \emph{Nature Communications}, vol.~13, no.~1, p. 2453, 2022. [Online]. Available: \url{https://doi.org/10.1038/s41467-022-29939-5}
\BIBentrySTDinterwordspacing

\bibitem{DimeNet}
\BIBentryALTinterwordspacing
J.~Gasteiger, J.~Groß, and S.~Günnemann, ``Directional message passing for molecular graphs,'' 2022. [Online]. Available: \url{https://arxiv.org/abs/2003.03123}
\BIBentrySTDinterwordspacing

\bibitem{scn2022}
\BIBentryALTinterwordspacing
C.~L. Zitnick, A.~Das, A.~Kolluru, J.~Lan, M.~Shuaibi, A.~Sriram, Z.~Ulissi, and B.~Wood, ``Spherical channels for modeling atomic interactions,'' 2022. [Online]. Available: \url{https://arxiv.org/abs/2206.14331}
\BIBentrySTDinterwordspacing

\bibitem{eSCN}
\BIBentryALTinterwordspacing
S.~Passaro and C.~L. Zitnick, ``Reducing so(3) convolutions to so(2) for efficient equivariant gnns,'' 2023. [Online]. Available: \url{https://arxiv.org/abs/2302.03655}
\BIBentrySTDinterwordspacing

\bibitem{MACE2023}
\BIBentryALTinterwordspacing
I.~Batatia, D.~P. Kovács, G.~N.~C. Simm, C.~Ortner, and G.~Csányi, ``Mace: Higher order equivariant message passing neural networks for fast and accurate force fields,'' 2023. [Online]. Available: \url{https://arxiv.org/abs/2206.07697}
\BIBentrySTDinterwordspacing

\bibitem{PaiNN}
\BIBentryALTinterwordspacing
K.~T. Schütt, O.~T. Unke, and M.~Gastegger, ``Equivariant message passing for the prediction of tensorial properties and molecular spectra,'' 2021. [Online]. Available: \url{https://arxiv.org/abs/2102.03150}
\BIBentrySTDinterwordspacing

\bibitem{Allegro}
A.~Musaelian, S.~Batzner, A.~Johansson, L.~Sun, C.~J. Owen, M.~Kornbluth, and B.~Kozinsky, ``Learning local equivariant representations for large-scale atomistic dynamics,'' \emph{Nature Communications}, vol.~14, no.~1, p. 579, 2023.

\bibitem{equiformerV1}
\BIBentryALTinterwordspacing
Y.-L. Liao and T.~Smidt, ``Equiformer: Equivariant graph attention transformer for 3d atomistic graphs,'' 2023. [Online]. Available: \url{https://arxiv.org/abs/2206.11990}
\BIBentrySTDinterwordspacing

\bibitem{equiformerV2}
\BIBentryALTinterwordspacing
Y.-L. Liao, B.~Wood, A.~Das, and T.~Smidt, ``Equiformerv2: Improved equivariant transformer for scaling to higher-degree representations,'' 2024. [Online]. Available: \url{https://arxiv.org/abs/2306.12059}
\BIBentrySTDinterwordspacing

\bibitem{graphoformer}
\BIBentryALTinterwordspacing
C.~Ying, T.~Cai, S.~Luo, S.~Zheng, G.~Ke, D.~He, Y.~Shen, and T.-Y. Liu, ``Do transformers really perform bad for graph representation?'' 2021. [Online]. Available: \url{https://arxiv.org/abs/2106.05234}
\BIBentrySTDinterwordspacing

\bibitem{fuchs2020se}
F.~Fuchs, D.~Worrall, V.~Fischer, and M.~Welling, ``Se(3)-transformers: 3d roto-translation equivariant attention networks,'' in \emph{NeurIPS}, vol.~33, 2020.

\bibitem{TorchMD-NET}
\BIBentryALTinterwordspacing
P.~Thölke and G.~D. Fabritiis, ``Torchmd-net: Equivariant transformers for neural network based molecular potentials,'' 2022. [Online]. Available: \url{https://arxiv.org/abs/2202.02541}
\BIBentrySTDinterwordspacing

\bibitem{hutchinson2021lietransformer}
M.~J. Hutchinson, C.~Le~Lan, S.~Zaidi, E.~Dupont, Y.~W. Teh, and H.~Kim, ``Lietransformer: equivariant self-attention for lie groups,'' in \emph{ICML}, 2021.

\bibitem{ramakrishnan2014quantum}
R.~Ramakrishnan, P.~O. Dral, M.~Rupp, and O.~A. Von~Lilienfeld, ``Quantum chemistry structures and properties of 134 kilo molecules,'' \emph{Scientific data}, vol.~1, no.~1, pp. 1--7, 2014.

\bibitem{choudhary2021atomistic}
K.~Choudhary and B.~DeCost, ``Atomistic line graph neural network for improved materials property predictions,'' \emph{npj Computational Materials}, vol.~7, no.~1, p. 185, 2021.

\bibitem{karamad2020orbital}
M.~Karamad, R.~Magar, Y.~Shi, S.~Siahrostami, I.~D. Gates, and A.~Barati~Farimani, ``Orbital graph convolutional neural network for material property prediction,'' \emph{Physical Review Materials}, vol.~4, no.~9, p. 093801, 2020.

\bibitem{lam2017machine}
T.~Lam~Pham, H.~Kino, K.~Terakura, T.~Miyake, K.~Tsuda, I.~Takigawa, and H.~Chi~Dam, ``Machine learning reveals orbital interaction in materials,'' \emph{Science and technology of advanced materials}, vol.~18, no.~1, pp. 756--765, 2017.

\bibitem{kaba2022equivariant}
O.~Kaba and S.~Ravanbakhsh, ``Equivariant networks for crystal structures,'' \emph{Advances in Neural Information Processing Systems}, vol.~35, pp. 4150--4164, 2022.

\bibitem{yan2022periodic}
K.~Yan, Y.~Liu, Y.~Lin, and S.~Ji, ``Periodic graph transformers for crystal material property prediction,'' \emph{Advances in Neural Information Processing Systems}, vol.~35, pp. 15\,066--15\,080, 2022.

\bibitem{magar2022crystal}
R.~Magar, Y.~Wang, and A.~Barati~Farimani, ``Crystal twins: self-supervised learning for crystalline material property prediction,'' \emph{npj Computational Materials}, vol.~8, no.~1, p. 231, 2022.

\bibitem{zbontar2021barlow}
J.~Zbontar, L.~Jing, I.~Misra, Y.~LeCun, and S.~Deny, ``Barlow twins: Self-supervised learning via redundancy reduction,'' in \emph{International conference on machine learning}.\hskip 1em plus 0.5em minus 0.4em\relax PMLR, 2021, pp. 12\,310--12\,320.

\bibitem{chen2020simple}
T.~Chen, S.~Kornblith, M.~Norouzi, and G.~Hinton, ``A simple framework for contrastive learning of visual representations,'' in \emph{International conference on machine learning}.\hskip 1em plus 0.5em minus 0.4em\relax PMLR, 2020, pp. 1597--1607.

\bibitem{yu2023crystal}
H.~Yu, Y.~Song, J.~Hu, C.~Guo, and B.~Yang, ``A crystal-specific pre-training framework for crystal material property prediction,'' \emph{arXiv preprint arXiv:2306.05344}, 2023.

\bibitem{devlin2018bert}
\BIBentryALTinterwordspacing
J.~Devlin, M.~Chang, K.~Lee, and K.~Toutanova, ``{BERT:} pre-training of deep bidirectional transformers for language understanding,'' \emph{CoRR}, vol. abs/1810.04805, 2018. [Online]. Available: \url{http://arxiv.org/abs/1810.04805}
\BIBentrySTDinterwordspacing

\bibitem{kirklin2015open}
S.~Kirklin, J.~E. Saal, B.~Meredig, A.~Thompson, J.~W. Doak, M.~Aykol, S.~R{\"u}hl, and C.~Wolverton, ``The open quantum materials database (oqmd): assessing the accuracy of dft formation energies,'' \emph{npj Computational Materials}, vol.~1, no.~1, pp. 1--15, 2015.

\bibitem{gasteiger2020fast}
J.~Gasteiger, S.~Giri, J.~T. Margraf, and S.~G{\"u}nnemann, ``Fast and uncertainty-aware directional message passing for non-equilibrium molecules,'' \emph{arXiv preprint arXiv:2011.14115}, 2020.

\bibitem{rombach2021highresolution}
R.~Rombach, A.~Blattmann, D.~Lorenz, P.~Esser, and B.~Ommer, ``High-resolution image synthesis with latent diffusion models,'' 2021.

\bibitem{ho2020denoising}
J.~Ho, A.~Jain, and P.~Abbeel, ``Denoising diffusion probabilistic models,'' \emph{Advances in neural information processing systems}, vol.~33, pp. 6840--6851, 2020.

\bibitem{song2021Score-based}
\BIBentryALTinterwordspacing
Y.~Song, J.~Sohl{-}Dickstein, D.~P. Kingma, A.~Kumar, S.~Ermon, and B.~Poole, ``Score-based generative modeling through stochastic differential equations,'' in \emph{9th International Conference on Learning Representations, {ICLR} 2021, Virtual Event, Austria, May 3-7, 2021}.\hskip 1em plus 0.5em minus 0.4em\relax OpenReview.net, 2021. [Online]. Available: \url{https://openreview.net/forum?id=PxTIG12RRHS}
\BIBentrySTDinterwordspacing

\bibitem{abramson2024accurate}
J.~Abramson, J.~Adler, J.~Dunger, R.~Evans, T.~Green, A.~Pritzel, O.~Ronneberger, L.~Willmore, A.~J. Ballard, J.~Bambrick \emph{et~al.}, ``Accurate structure prediction of biomolecular interactions with alphafold 3,'' \emph{Nature}, pp. 1--3, 2024.

\bibitem{gabriele2023diffdock}
\BIBentryALTinterwordspacing
G.~Corso, H.~St{\"{a}}rk, B.~Jing, R.~Barzilay, and T.~S. Jaakkola, ``Diffdock: Diffusion steps, twists, and turns for molecular docking,'' in \emph{The Eleventh International Conference on Learning Representations, {ICLR} 2023, Kigali, Rwanda, May 1-5, 2023}.\hskip 1em plus 0.5em minus 0.4em\relax OpenReview.net, 2023. [Online]. Available: \url{https://openreview.net/forum?id=kKF8\_K-mBbS}
\BIBentrySTDinterwordspacing

\bibitem{jing2022torsional}
\BIBentryALTinterwordspacing
B.~Jing, G.~Corso, J.~Chang, R.~Barzilay, and T.~S. Jaakkola, ``Torsional diffusion for molecular conformer generation,'' in \emph{Advances in Neural Information Processing Systems 35: Annual Conference on Neural Information Processing Systems 2022, NeurIPS 2022, New Orleans, LA, USA, November 28 - December 9, 2022}, S.~Koyejo, S.~Mohamed, A.~Agarwal, D.~Belgrave, K.~Cho, and A.~Oh, Eds., 2022. [Online]. Available: \url{http://papers.nips.cc/paper\_files/paper/2022/hash/994545b2308bbbbc97e3e687ea9e464f-Abstract-Conference.html}
\BIBentrySTDinterwordspacing

\bibitem{CrystalGAN}
\BIBentryALTinterwordspacing
A.~Nouira, N.~Sokolovska, and J.-C. Crivello, ``Crystalgan: Learning to discover crystallographic structures with generative adversarial networks,'' 2019. [Online]. Available: \url{https://arxiv.org/abs/1810.11203}
\BIBentrySTDinterwordspacing

\bibitem{FlowMM}
\BIBentryALTinterwordspacing
B.~K. Miller, R.~T.~Q. Chen, A.~Sriram, and B.~M. Wood, ``Flowmm: Generating materials with riemannian flow matching,'' 2024. [Online]. Available: \url{https://arxiv.org/abs/2406.04713}
\BIBentrySTDinterwordspacing

\bibitem{jiao2024space}
R.~Jiao, W.~Huang, Y.~Liu, D.~Zhao, and Y.~Liu, ``Space group constrained crystal generation,'' \emph{arXiv preprint arXiv:2402.03992}, 2024.

\bibitem{xie2021crystal}
T.~Xie, X.~Fu, O.-E. Ganea, R.~Barzilay, and T.~Jaakkola, ``Crystal diffusion variational autoencoder for periodic material generation,'' \emph{arXiv preprint arXiv:2110.06197}, 2021.

\bibitem{YE2024100003}
\BIBentryALTinterwordspacing
C.-Y. Ye, H.-M. Weng, and Q.-S. Wu, ``Con-cdvae: A method for the conditional generation of crystal structures,'' \emph{Computational Materials Today}, vol.~1, p. 100003, 2024. [Online]. Available: \url{https://www.sciencedirect.com/science/article/pii/S2950463524000036}
\BIBentrySTDinterwordspacing

\bibitem{DALL-E2}
\BIBentryALTinterwordspacing
A.~Ramesh, P.~Dhariwal, A.~Nichol, C.~Chu, and M.~Chen, ``Hierarchical text-conditional image generation with clip latents,'' 2022. [Online]. Available: \url{https://arxiv.org/abs/2204.06125}
\BIBentrySTDinterwordspacing

\bibitem{Luo2024}
\BIBentryALTinterwordspacing
X.~Luo, Z.~Wang, P.~Gao, J.~Lv, Y.~Wang, C.~Chen, and Y.~Ma, ``Deep learning generative model for crystal structure prediction,'' \emph{npj Computational Materials}, vol.~10, p. 254, Nov 2024. [Online]. Available: \url{https://doi.org/10.1038/s41524-024-01443-y}
\BIBentrySTDinterwordspacing

\bibitem{zeni2023mattergen}
C.~Zeni, R.~Pinsler, D.~Z{\"u}gner, A.~Fowler, M.~Horton, X.~Fu, S.~Shysheya, J.~Crabb{\'e}, L.~Sun, J.~Smith \emph{et~al.}, ``Mattergen: a generative model for inorganic materials design,'' \emph{arXiv preprint arXiv:2312.03687}, 2023.

\bibitem{joshi2023expressive}
C.~K. Joshi, C.~Bodnar, S.~V. Mathis, T.~Cohen, and P.~Lio, ``On the expressive power of geometric graph neural networks,'' in \emph{International conference on machine learning}.\hskip 1em plus 0.5em minus 0.4em\relax PMLR, 2023, pp. 15\,330--15\,355.

\bibitem{hall2013lie}
B.~C. Hall, \emph{Lie groups, Lie algebras, and representations}.\hskip 1em plus 0.5em minus 0.4em\relax Springer, 2013.

\bibitem{cao2024CrystalFormer}
\BIBentryALTinterwordspacing
Z.~Cao, X.~Luo, J.~Lv, and L.~Wang, ``Space group informed transformer for crystalline materials generation,'' 2024. [Online]. Available: \url{https://arxiv.org/abs/2403.15734}
\BIBentrySTDinterwordspacing

\bibitem{brooks1998markov}
S.~Brooks, ``Markov chain monte carlo method and its application,'' \emph{Journal of the royal statistical society: series D (the Statistician)}, vol.~47, no.~1, pp. 69--100, 1998.

\bibitem{reynolds2009gaussian}
D.~A. Reynolds \emph{et~al.}, ``Gaussian mixture models.'' \emph{Encyclopedia of biometrics}, vol. 741, no. 659-663, 2009.

\bibitem{FlowLLM2024}
\BIBentryALTinterwordspacing
A.~Sriram, B.~K. Miller, R.~T.~Q. Chen, and B.~M. Wood, ``Flowllm: Flow matching for material generation with large language models as base distributions,'' 2024. [Online]. Available: \url{https://arxiv.org/abs/2410.23405}
\BIBentrySTDinterwordspacing

\bibitem{brown2020language}
T.~B. Brown, ``Language models are few-shot learners,'' \emph{arXiv preprint arXiv:2005.14165}, 2020.

\bibitem{Boiko2023}
\BIBentryALTinterwordspacing
D.~A. Boiko, R.~MacKnight, B.~Kline, and G.~Gomes, ``Autonomous chemical research with large language models,'' \emph{Nature}, vol. 624, no. 7992, pp. 570--578, 2023. [Online]. Available: \url{https://doi.org/10.1038/s41586-023-06792-0}
\BIBentrySTDinterwordspacing

\bibitem{beltagy2019scibert}
I.~Beltagy, K.~Lo, and A.~Cohan, ``Scibert: A pretrained language model for scientific text,'' \emph{arXiv preprint arXiv:1903.10676}, 2019.

\bibitem{gupta2021matscibert}
T.~Gupta, M.~Zaki, N.~Krishnan \emph{et~al.}, ``Matscibert: A materials domain language model for text mining and information extraction,'' \emph{arXiv preprint arXiv:2109.15290}, 2021.

\bibitem{volker2024LLMs}
C.~V{\"o}lker, T.~Rug, K.~M. Jablonka, and S.~Kruschwitz, ``Llms can design sustainable concrete--a systematic benchmark,'' 2024.

\bibitem{zhao2024potential}
S.~Zhao, S.~Chen, J.~Zhou, C.~Li, T.~Tang, S.~J. Harris, Y.~Liu, J.~Wan, and X.~Li, ``Potential to transform words to watts with large language models in battery research,'' \emph{Cell Reports Physical Science}, vol.~5, no.~3, 2024.

\bibitem{satpute2024exploring}
P.~Satpute, S.~Tiwari, M.~Gupta, and S.~Ghosh, ``Exploring large language models for microstructure evolution in materials,'' \emph{Materials Today Communications}, p. 109583, 2024.

\bibitem{lei2024materials}
G.~Lei, R.~Docherty, and S.~J. Cooper, ``Materials science in the era of large language models: a perspective,'' \emph{Digital Discovery}, 2024.

\bibitem{gupta2022matscibert}
T.~Gupta, M.~Zaki, N.~A. Krishnan, and Mausam, ``Matscibert: A materials domain language model for text mining and information extraction,'' \emph{npj Computational Materials}, vol.~8, no.~1, p. 102, 2022.

\bibitem{blum2009970}
L.~C. Blum and J.-L. Reymond, ``970 million druglike small molecules for virtual screening in the chemical universe database gdb-13,'' \emph{Journal of the American Chemical Society}, vol. 131, no.~25, pp. 8732--8733, 2009.

\bibitem{chmiela2017machine}
S.~Chmiela, A.~Tkatchenko, H.~E. Sauceda, I.~Poltavsky, K.~T. Sch{\"u}tt, and K.-R. M{\"u}ller, ``Machine learning of accurate energy-conserving molecular force fields,'' \emph{Science advances}, vol.~3, no.~5, p. e1603015, 2017.

\bibitem{chmiela2018towards}
S.~Chmiela, H.~E. Sauceda, K.-R. M{\"u}ller, and A.~Tkatchenko, ``Towards exact molecular dynamics simulations with machine-learned force fields,'' \emph{Nature communications}, vol.~9, no.~1, p. 3887, 2018.

\bibitem{jain2013commentary}
A.~Jain, S.~P. Ong, G.~Hautier, W.~Chen, W.~D. Richards, S.~Dacek, S.~Cholia, D.~Gunter, D.~Skinner, G.~Ceder \emph{et~al.}, ``Commentary: The materials project: A materials genome approach to accelerating materials innovation,'' \emph{APL materials}, vol.~1, no.~1, 2013.

\bibitem{chanussot2021open}
L.~Chanussot, A.~Das, S.~Goyal, T.~Lavril, M.~Shuaibi, M.~Riviere, K.~Tran, J.~Heras-Domingo, C.~Ho, W.~Hu \emph{et~al.}, ``Open catalyst 2020 (oc20) dataset and community challenges,'' \emph{Acs Catalysis}, vol.~11, no.~10, pp. 6059--6072, 2021.

\bibitem{choudhary2020joint}
K.~Choudhary, K.~F. Garrity, A.~C. Reid, B.~DeCost, A.~J. Biacchi, A.~R. Hight~Walker, Z.~Trautt, J.~Hattrick-Simpers, A.~G. Kusne, A.~Centrone \emph{et~al.}, ``The joint automated repository for various integrated simulations (jarvis) for data-driven materials design,'' \emph{npj computational materials}, vol.~6, no.~1, p. 173, 2020.

\bibitem{Nakata2023}
\BIBentryALTinterwordspacing
M.~Nakata and T.~Maeda, ``Pubchemqc b3lyp/6-31g*//pm6 data set: The electronic structures of 86 million molecules using b3lyp/6-31g* calculations,'' \emph{Journal of Chemical Information and Modeling}, vol.~63, pp. 5734--5754, 2023. [Online]. Available: \url{https://doi.org/10.1021/acs.jcim.3c00899}
\BIBentrySTDinterwordspacing

\bibitem{ullah2024molecularquantumchemicaldata}
\BIBentryALTinterwordspacing
A.~Ullah, Y.~Chen, and P.~O. Dral, ``Molecular quantum chemical data sets and databases for machine learning potentials,'' 2024. [Online]. Available: \url{https://arxiv.org/abs/2408.12058}
\BIBentrySTDinterwordspacing

\bibitem{jia2024llmatdesign}
S.~Jia, C.~Zhang, and V.~Fung, ``Llmatdesign: Autonomous materials discovery with large language models,'' \emph{arXiv preprint arXiv:2406.13163}, 2024.

\bibitem{zhang2024honeycomb}
H.~Zhang, Y.~Song, Z.~Hou, S.~Miret, and B.~Liu, ``Honeycomb: A flexible llm-based agent system for materials science,'' \emph{arXiv preprint arXiv:2409.00135}, 2024.

\bibitem{miret2024llms}
S.~Miret and N.~Krishnan, ``Are llms ready for real-world materials discovery?'' \emph{arXiv preprint arXiv:2402.05200}, 2024.

\bibitem{nie2024active}
S.~Nie, Y.~Xiang, L.~Wu, G.~Lin, Q.~Liu, S.~Chu, and X.~Wang, ``Active learning guided discovery of high entropy oxides featuring high h2-production,'' \emph{Journal of the American Chemical Society}, 2024.

\bibitem{qu2023leveraging}
J.~Qu, Y.~R. Xie, K.~M. Ciesielski, C.~E. Porter, E.~S. Toberer, and E.~Ertekin, ``Leveraging language representation for material recommendation, ranking, and exploration,'' \emph{arXiv preprint arXiv:2305.01101}, 2023.

\bibitem{zhong2023accelerating}
Y.~Zhong, Z.~Tao, W.~Chu, X.~Gong, and H.~Xiang, ``Accelerating the calculation of electron-phonon coupling by machine learning methods,'' \emph{arXiv preprint arXiv:2302.00439}, 2023.

\bibitem{gibson2024accelerating}
J.~B. Gibson, A.~C. Hire, P.~M. Dee, O.~Barrera, B.~Geisler, P.~J. Hirschfeld, and R.~G. Hennig, ``Accelerating superconductor discovery through tempered deep learning of the electron-phonon spectral function,'' \emph{arXiv preprint arXiv:2401.16611}, 2024.

\bibitem{haldar2024machine}
A.~Haldar, Q.~Clark, M.~Zacharias, F.~Giustino, and S.~Sharifzadeh, ``Machine learning electron-phonon interactions in two-dimensional semiconducting materials: The case of zero-point renormalization,'' \emph{Physical Review Materials}, vol.~8, no.~10, p. L101001, 2024.

\bibitem{okabe2024virtual}
R.~Okabe, A.~Chotrattanapituk, A.~Boonkird, N.~Andrejevic, X.~Fu, T.~S. Jaakkola, Q.~Song, T.~Nguyen, N.~Drucker, S.~Mu \emph{et~al.}, ``Virtual node graph neural network for full phonon prediction,'' \emph{Nature Computational Science}, vol.~4, no.~7, pp. 522--531, 2024.

\bibitem{fang2024phonon}
S.~Fang, M.~Geiger, J.~G. Checkelsky, and T.~Smidt, ``Phonon predictions with e (3)-equivariant graph neural networks,'' \emph{arXiv preprint arXiv:2403.11347}, 2024.

\bibitem{zhang2024overcoming}
H.~Zhang, S.~Liu, J.~You, C.~Liu, S.~Zheng, Z.~Lu, T.~Wang, N.~Zheng, and B.~Shao, ``Overcoming the barrier of orbital-free density functional theory for molecular systems using deep learning,'' \emph{Nature Computational Science}, vol.~4, no.~3, pp. 210--223, 2024.

\bibitem{li2022deep}
H.~Li, Z.~Wang, N.~Zou, M.~Ye, R.~Xu, X.~Gong, W.~Duan, and Y.~Xu, ``Deep-learning density functional theory hamiltonian for efficient ab initio electronic-structure calculation,'' \emph{Nature Computational Science}, vol.~2, no.~6, pp. 367--377, 2022.

\bibitem{gong2023general}
X.~Gong, H.~Li, N.~Zou, R.~Xu, W.~Duan, and Y.~Xu, ``General framework for e (3)-equivariant neural network representation of density functional theory hamiltonian,'' \emph{Nature Communications}, vol.~14, no.~1, p. 2848, 2023.

\bibitem{li2024deep}
H.~Li, Z.~Tang, J.~Fu, W.-H. Dong, N.~Zou, X.~Gong, W.~Duan, and Y.~Xu, ``Deep-learning density functional perturbation theory,'' \emph{Physical Review Letters}, vol. 132, no.~9, p. 096401, 2024.

\bibitem{li2023deep}
H.~Li, Z.~Tang, X.~Gong, N.~Zou, W.~Duan, and Y.~Xu, ``Deep-learning electronic-structure calculation of magnetic superstructures,'' \emph{Nature Computational Science}, vol.~3, no.~4, pp. 321--327, 2023.

\bibitem{tang2024deep}
Z.~Tang, H.~Li, P.~Lin, X.~Gong, G.~Jin, L.~He, H.~Jiang, X.~Ren, W.~Duan, and Y.~Xu, ``A deep equivariant neural network approach for efficient hybrid density functional calculations,'' \emph{Nature Communications}, vol.~15, no.~1, p. 8815, 2024.

\bibitem{gong2024generalizing}
X.~Gong, S.~G. Louie, W.~Duan, and Y.~Xu, ``Generalizing deep learning electronic structure calculation to the plane-wave basis,'' \emph{Nature computational science}, pp. 1--9, 2024.

\bibitem{tang2024improving}
Z.~Tang, N.~Zou, H.~Li, Y.~Wang, Z.~Yuan, H.~Tao, Y.~Li, Z.~Chen, B.~Zhao, M.~Sun \emph{et~al.}, ``Improving density matrix electronic structure method by deep learning,'' \emph{arXiv preprint arXiv:2406.17561}, 2024.

\bibitem{wang2024deeph}
Y.~Wang, H.~Li, Z.~Tang, H.~Tao, Y.~Wang, Z.~Yuan, Z.~Chen, W.~Duan, and Y.~Xu, ``Deeph-2: Enhancing deep-learning electronic structure via an equivariant local-coordinate transformer,'' \emph{arXiv preprint arXiv:2401.17015}, 2024.

\bibitem{li2024neural}
Y.~Li, Z.~Tang, Z.~Chen, M.~Sun, B.~Zhao, H.~Li, H.~Tao, Z.~Yuan, W.~Duan, and Y.~Xu, ``Neural-network density functional theory based on variational energy minimization,'' \emph{Physical Review Letters}, vol. 133, no.~7, p. 076401, 2024.

\bibitem{wang2024universal}
Y.~Wang, Y.~Li, Z.~Tang, H.~Li, Z.~Yuan, H.~Tao, N.~Zou, T.~Bao, X.~Liang, Z.~Chen \emph{et~al.}, ``Universal materials model of deep-learning density functional theory hamiltonian,'' \emph{Science Bulletin}, 2024.

\bibitem{zhong2023transferable}
Y.~Zhong, H.~Yu, M.~Su, X.~Gong, and H.~Xiang, ``Transferable equivariant graph neural networks for the hamiltonians of molecules and solids,'' \emph{npj Computational Materials}, vol.~9, no.~1, p. 182, 2023.

\bibitem{zhong2024universal}
Y.~Zhong, H.~Yu, J.~Yang, X.~Guo, H.~Xiang, and X.~Gong, ``Universal machine learning kohn-sham hamiltonian for materials,'' \emph{Chinese Physics Letters}, 2024.

\bibitem{su2023efficient}
M.~Su, J.-H. Yang, H.-J. Xiang, and X.-G. Gong, ``Efficient determination of the hamiltonian and electronic properties using graph neural network with complete local coordinates,'' \emph{Machine Learning: Science and Technology}, vol.~4, no.~3, p. 035010, 2023.

\end{thebibliography}

\end{document}